\documentclass[useAMS, usenatbib]{mn2e}

\usepackage{aasmacros}
\usepackage{amssymb} \usepackage{graphicx}
\usepackage{subcaption} \usepackage{float}
\usepackage{booktabs} \usepackage{multirow} \usepackage{amsfonts}
\usepackage{lmodern}
\usepackage{amsmath}
\usepackage{xcolor}
\usepackage{soul,color}

\let\oldbullet\bullet \renewcommand{\bullet}[1][0pt]{% 
\mathrel{\raisebox{#1}{$\oldbullet$}}% 
}
\title{Feedback from Active Galactic Nuclei: Energy- versus momentum-driving} \author[Costa et al.]{Tiago Costa\footnotemark[1], Debora
  Sijacki and Martin G. Haehnelt \\
  Institute of Astronomy and Kavli Institute for Cosmology,
  University of Cambridge, Madingley Road, Cambridge CB$3$ $0$HA}

\begin{document}
 
\maketitle

\begin{abstract}
We employ hydrodynamical simulations using the moving-mesh code {\sc AREPO} to investigate the role of energy and momentum input from Active Galactic Nuclei (AGN) in driving large-scale galactic outflows. 
We start by reproducing analytic solutions for both energy- and momentum-driven outflowing shells in simulations of a spherical isolated dark matter potential with gas in hydrostatic equilibrium and with no radiative cooling. 
We confirm that for this simplified setup, galactic outflows driven by a momentum input rate of order $L_{\rm Edd}/c$ can establish an $M_{\rm BH}- \sigma$ relation with slope and normalisation similar to that observed. 
We show that momentum input at a rate of $L_{\rm Edd}/c$ is however insufficient to drive efficient outflows once cooling and gas inflows as predicted by cosmological simulations at resolved scales are taken into account.
We argue that observed large-scale AGN-driven outflows are instead likely to be energy-driven and show that such outflows can reach momentum fluxes exceeding $10 L_{\rm Edd}/c$ within the innermost $10 \, \rm kpc$ of the galaxy.
The outflows are highly anisotropic, with outflow rates and a velocity structure found to be inadequately described by spherical outflow models.
We verify that the hot energy-driven outflowing gas is expected to be strongly affected by metal-line cooling, leading to significant amounts ($\gtrsim 10^9 \, \rm M_\odot$) of entrained cold gas.
\end{abstract}


\section{Introduction}
\renewcommand{\thefootnote}{\fnsymbol{footnote}}
\footnotetext[1]{E-mail: taf34@ast.cam.ac.uk}

There is now mounting evidence that Active Galactic Nucleus (AGN) activity can drive powerful large-scale outflows.
Observational signatures of outflows in the form of broadened (Doppler shifted) emission and absorption lines have been detected for a number of galaxies hosting an AGN.
Detected outflows have characteristic speeds $\gtrsim 1000 \, \rm km \, s^{-1}$, spatial scales of $\approx 1 \-- 10 \, \rm kpc$ and often appear to consist of a complex multi-phase medium.
This typically comprises a hot ionised component that can travel at speeds as high as $\approx 3000 \, \rm km \, s^{-1}$ \citep{Humphrey:10, Greene:11, Nesvadba:11, Westmoquette:12, Rupke:13, Liu:13, Mullaney:13, Forster-Schreiber:13, RodriguezZaurin:13, VillarMartin:14, Arribas:14, Harrison:14, Genzel:14}, a partially overlapping neutral atomic component at speeds not much exceeding $\approx 1000 \, \rm km \, s^{-1}$ \citep{Krug:10, Sturm:11, Rupke:13} and a substantial portion of cold molecular gas, as revealed by spatially resolved CO-emission and as OH-emission/absorption features \citep{Feruglio:10, Sturm:11, Alatalo:11, Aalto:12, Cicone:12, Veilleux:13, Rupke:13b, Combes:13, Cicone:14, Sakamoto:14}.
Molecular outflows detected in ultra-luminous infrared galaxies (ULIRGs) and quasars (QSOs) possess speeds of the order of $\approx 1000 \, \rm km \, s^{-1}$ and high mass outflow rates up to $\approx 1000 \, \rm M_\odot \, yr^{-1}$ \citep[see e.g.][]{Sturm:11, Cicone:12}. 
Large-scale outflows at scales of $\approx 10 \, \rm kpc$ and speeds $\gtrsim 1000 \, \rm km \, s^{-1}$ have also been detected in luminous high redshift QSOs \citep{Cano-Diaz:12, Maiolino:12}, where the estimated outflow rates ($200 \-- 3500 \, \rm M_\odot \, yr^{-1}$) exceed the inferred star formation rates of the host galaxies.
While it is difficult to disentangle the combined effects of both supernova and AGN feedback effects in driving outflows, theoretical models of supernova-driven feedback exclude outflow speeds much exceeding $\approx 600 \, \rm km \, s^{-1}$ on energetic grounds \citep{Martin:05, Sharma:13}.
Most crucially, several studies indicate that the presence of an AGN boosts the properties of detected outflows.
Using CO observations of a sample of local ULIRGs and QSOs, \citet{Cicone:14} have shown that the mass outflow rates, the kinetic luminosity, the momentum flux and the spatial extension of detected outflows all correlate with the fraction of the host galaxy's bolometric luminosity attributed to a central AGN.
The depletion times of molecular gas of $\approx 10^6 \, \rm yr$ estimated for outflows in \citet{Cicone:14} are lower than the depletion time for star formation, providing tentative evidence that AGN-driven outflows can quench the host galaxy.

Observed AGN-driven outflows thus appear to strongly affect their host galaxies, potentially realising the negative AGN feedback effect invoked by galaxy formation models in order to address a series of open questions concerning the properties of massive galaxies. AGN feedback may account for how massive galaxies move away from the `main-sequence' of star forming galaxies to become `red and dead' \citep{DiMatteo:05, Springel:05c, Sijacki:06, Okamoto:08, Hopkins:08, Dubois:13b, Martizzi:14}, explaining the shape of the otherwise overpopulated high-mass end of the galactic stellar mass function \citep{Scannapieco:04, Churazov:05, Kawata:05, Croton:06, Bower:06, Sijacki:07, Somerville:08, Booth:09, Puchwein:13, Vogelsberger:14}.
Other potential roles of AGN feedback include helping to drive the morphological transformation between spirals and ellipticals, leading to metal enrichment of the intracluster (ICM) and intergalactic medium (IGM) and thereby shaping their thermodynamic properties \citep{Sijacki:07, Puchwein:08, McCarthy:10, Gaspari:11, Teyssier:11, Martizzi:12, Planelles:14}.
It has also been argued that AGN-driven outflows could provide a `positive feedback' effect whereby star formation occurs in compressed outflowing gas \citep{Silk:05, Zubovas:13, Silk:13} or even that it may drive the apparent size evolution of elliptical galaxies \citep{Ishibashi:13, Ishibashi:14}.
Finally, AGN feedback also offers a compelling mechanism for the origin of the correlations observed between the mass of the supermassive black hole and the mass and velocity dispersion of the bulge of the host galaxy \citep[e.g.][]{Magorrian:98, Tremaine:02, Marconi:03, Ferrarese:05, Gultekin:09, McConnell:13, Kormendy:13}.
Balancing the energy or momentum released by the AGN with the gravitational weight of its surrounding interstellar material has been argued to reproduce the scaling and normalisation of observed correlations \citep{Silk:98, Haehnelt:98, Fabian:99, King:03, Murray:05}.

The efficiency of AGN feedback is ultimately dependent on the detailed hydrodynamics of AGN-driven outflows, which in turn is sensitive to a wide range of physical effects including the efficiency of (in-shock) cooling and star formation, the geometry, dynamical and thermodynamic state of the gaseous medium through which the outflow propagates.
In Section~\ref{enemomout}, we review the current theoretical understanding of AGN-driven outflows as envisaged in models in which the energy and/or momentum deposited by the AGN on galactic scales couples hydrodynamically (as opposed to radiatively) with the interstellar medium (ISM) via a fast inner AGN wind. 
We clarify the nature of energy- and momentum-driven shells of outflowing material in spherically symmetric galactic potentials and discuss important distinctions between these two limiting cases.
In Section~\ref{secnumimp}, we discuss the implementation of energy- and momentum-driven outflow feedback into numerical simulations.
In Section~\ref{seccomp}, using simulations of isolated dark matter potentials with gas in hydrostatic equilibrium and no radiative cooling, we reproduce analytical solutions for energy- and momentum-driven shells. 
Using simulations including radiative cooling, we show that under the right circumstances, large amounts of outflowing molecular gas are likely to form due to the cooling of shock-heated outflowing material.
Finally, in Section~\ref{seccosmo}, we implement the same AGN energy- and momentum-driven outflow models in fully cosmological hydrodynamic simulations and present several important departures from the idealised picture envisaged in analytic models. We place particular emphasis on the origin of an $M_{\rm BH} \-- \sigma$ relation in the context of such simulations.
We summarise our conclusions in Section~\ref{secconc}.
Tests of our numerical implementations of energy- and momentum-driven outflows against changes in numerical resolution and the energy/momentum injection procedure are discussed in Appendix~\ref{appendixa}. Possible future research avenues in the context of AGN feedback are discussed in Appendix~\ref{appendixb}.

\section{Energy and Momentum-driven outflows}
\label{enemomout}

\subsection{Accretion as source of energy}

Energy and momentum deposited by AGN into the ISM of galaxies as well as into their surrounding IGM/ICM is initially released by accretion of baryonic matter on to supermassive black holes \citep{Lynden-Bell:69}. 
A significant fraction of the accreted rest mass energy is radiated away close to the event horizon of the black hole, leading to an accretion luminosity $L$ given by

\begin{equation}
L  \,=\,  \eta \dot{M}_{\rm BH} c^2 \, ,
\label{acclum}
\end{equation}

where $\dot{M}_{\rm BH}$ is the black hole accretion rate, $c$ is the speed of light in vacuum and $\eta$ is the radiative efficiency of the black hole, which typically lies in the range $0.05 \-- 0.42$ depending on the spin of the black hole.
Studies comparing observed AGN luminosities and the inferred black hole mass density suggest that, on average, $\eta \,=\, 0.1 \-- 0.2$ \citep{Soltan:82, Fabian:99b, Yu:02}, a value consistent with moderately spinning black holes (a recent study by \citet{Ueda:14} however gives $\eta \,\approx\, 0.05$). 

A point of much debate is the fraction of the luminous energy $L$ (Eq.~\ref{acclum}) that can couple to surrounding interstellar gas and drive an outflow.
Observed AGN show signatures of outflows down to the smallest resolvable scales. 
These inner outflows take the form of highly collimated jets directly observable by their radio emission as well as wide-angle accretion disc winds inferred from very broad absorption lines in so called `broad absorption line' (BAL) QSOs \citep[e.g.][]{Pounds:03, Ganguly:07, Reeves:09, Tombesi:10}. Inferred outflow rates are typically similar or higher than the black hole accretion rate. 
The luminosity of AGN is thereby often assumed to be limited to the Eddington luminosity, which is given by

\begin{equation}
 L_{\rm Edd} \,=\, \frac{4 \pi GM_{\rm BH} c}{\kappa} \, ,
\end{equation}

where  $\kappa$ represents the opacity, $M_{\rm BH}$ the mass of the accreting black hole and $G$ the gravitational constant.

Here, we focus on the direct hydrodynamical coupling of an outflow launched from the inner parts of the accretion disc to interstellar gas at $\rm pc$ scales. 
In Section $2.5$, we briefly discuss the alternative suggestion of a coupling of radiation escaping from the AGN to dusty interstellar material at galactic ($\ga \rm kpc$) scales. 
We next review simple analytical models addressing the interaction between the inner AGN wind and the ISM without cooling.

\subsection{Spherical wind models}
\label{kingsmodel}

We start by investigating the simplified model proposed by \citet{King:03, King:05} describing the interaction of a sub-relativistic accretion disc wind, from now on referred to as `inner wind' for brevity, with the ISM surrounding the accreting black hole, which is assumed to lie at the centre of the galactic halo. 
In line with observed nuclear (ultra-)fast outflows \citep{Pounds:03, Pounds:09, Reeves:09, Tombesi:10, Tombesi:11, Tombesi:12, Tombesi:13, Pounds:13}, \citet{King:03, King:05} assumes the inner wind to have simple properties.
It is taken to have a high covering fraction with opening angle $\Omega_{\rm w}$ such that $\Omega_{\rm w}/4 \pi \,\approx\, 1$ and its mass outflow rate is taken to be equal to the black hole's Eddington accretion rate $\dot{M}_{\rm Edd}$. 
When combined, these two properties contribute to a high opacity in the inner wind and an optical depth to electron scattering of $\tau \,\approx\, 1$, supporting the view that such an inner wind can indeed be driven via radiation pressure from the AGN \citep{King:03b}.
Consequently, the inner wind initially transports momentum at a rate $\dot{p}_{\rm w}$ comparable with the momentum injection rate of the AGN as given by

\begin{equation}
	\dot{p}_{\rm w} \,=\, \dot{M}_{\rm w} v_{\rm w} \,\approx\, \frac{L_{\rm Edd}}{c} \, ,
\label{pwind}	
\end{equation}

where $v_{\rm w}$ is the speed of the inner wind and $\dot{M}_{\rm w}$ is the mass outflow rate of the inner wind, which is, as already mentioned, assumed to be equal to the Eddington accretion rate. From Eq.~\ref{acclum}, it thus follows that

\begin{equation}
	\dot{M}_{\rm w} \,=\, \dot{M}_{\rm Edd} \,=\, \frac{L_{\rm Edd}}{\eta c^2} \, .
\label{mwind}
\end{equation}

Substitution of $\dot{M}_{\rm w}$ as given in Eq.~\ref{mwind} into Eq.~\ref{pwind} then fixes the velocity $v_{\rm w}$ of the inner wind to

\begin{equation}
	v_{\rm w} \,=\, \eta c \, .
\label{vwind}
\end{equation}

Mass continuity implies that the mass density of the inner wind must in turn be given by

\begin{equation}
\rho_{\rm w} \,=\, \frac{\dot{M}_{\rm w}}{4 \pi R^2 v_{\rm w}} \,=\, \frac{GM_{\rm BH}}{\kappa \eta^2 c^2 R^2} \, .
\label{rhowind}
\end{equation}

Finally, the inner wind's kinetic luminosity $\dot{E}_{\rm k,\,w}$ is given by

\begin{equation}
	\dot{E}_{\rm k,\,w} \,=\, \frac{1}{2} \dot{M}_{\rm w} v_{\rm w}^2 \,=\, \frac{\eta}{2} L_{\rm Edd} \, .
\label{kinwind}
\end{equation}

If for the radiative efficiency, the canonical $\eta \,=\, 0.1$ is adopted, the inner wind speed is $v_{\rm w} \,=\, 0.1 c$ in agreement with observed blueshifted absorption lines in QSO spectra \citep[see e.g.][]{Pounds:13} and the kinetic luminosity is $\dot{E}_{\rm k,\,w} \,=\, \epsilon L_{\rm Edd}$ with $\epsilon = 0.05$.
For fixed momentum flux, note that Eqs.~\ref{pwind} and \ref{kinwind} however imply that a lower $v_{\rm w}$ results in higher $\dot{M}_{\rm w}$ and lower $\dot{E}_{\rm k,\,w}$, respectively.

\begin{figure}
\centering \includegraphics[scale = 0.7]{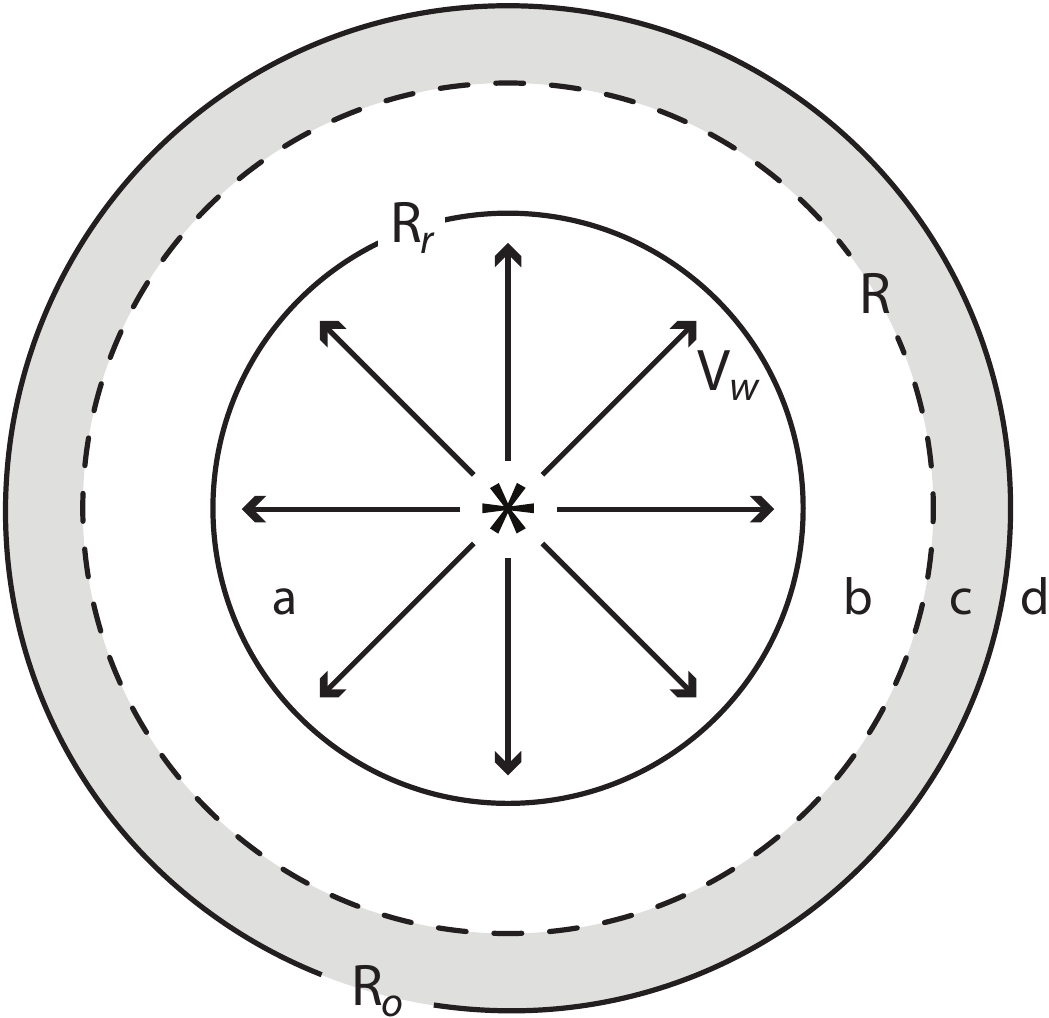}
\caption{The shock pattern resulting from the collision of a fast inner AGN wind with the surrounding ISM \citep[see also][]{Weaver:77, Dyson:97, Zubovas:12, Faucher-Giguere:12}. The flow is divided into four distinct regions, which are labelled with letters in order of increasing distance to the AGN. Region (a) is filled with the unshocked inner wind which flows out with a speed $v_{\rm w} \,=\, \eta c$. Since the inner wind acts as a piston pushing into the ISM at highly supersonic speeds, a shock must naturally form. However, in the reference frame of the inner wind itself, it is the ISM that moves against it highly supersonically, giving rise to a \emph{reverse shock} (region (b)). Neglecting relativistic effects, gas behind the reverse shock has a temperature $T_{\rm r} \,=\, 10^{10 - 11} \rm K$ (see text). On the outside, region (c) is bounded by a \emph{forward shock} that sweeps up interstellar gas into an expanding shell. Regions (b) and (c) are separated by a contact discontinuity (dashed circle). Finally, region (d) is occupied by the undisturbed ISM. It is the cooling of region (b) that determines whether the outflowing shell in region (c) is energy- or momentum-driven.}
\label{outflowstructure}
\end{figure}

With a speed of $v_{\rm w} \,\approx\, 0.1 c$, the inner wind is highly supersonic and must drive a strong shock into the ISM. The structure of the shock pattern that ensues is analogous to that resulting from the interaction between a fast stellar wind and the surrounding interstellar gas \citep[see e.g.][]{Weaver:77, Dyson:97}.
The inner wind drives a \emph{forward shock} wave that thrusts into the ambient ISM, while the inner wind itself must decelerate strongly in an inner \emph{reverse shock} facing towards the black hole. Thus, in order of increasing distance from the AGN, the flow pattern consists of four zones (see Fig.~\ref{outflowstructure}): (a) the unshocked highly supersonic inner wind; (b) the shocked inner wind material that has crossed the reverse shock, also often termed `wind shock' in the literature \citep[e.g.][]{Zubovas:12, Faucher-Giguere:12}; (c) a shell of interstellar gas swept up by the forward shock and (d) the unperturbed ambient ISM. 

If the forward shock is strong, the pressure from the ambient ISM (region d) can be neglected and the dynamics of the shell of swept up gas (region c) is fully determined by the difference between the outward pressure force $P$ exerted on it by the shocked inner wind (region b) and the inward pull due to gravity on the shell.
The latter depends on the total mass (dark matter and black hole)\footnote{Note that the contribution of a stellar component is not explicitly treated in \citet{King:03}.} enclosed by the radius of the shell $M_{\rm tot}(<R) \,\equiv\, M_{\rm halo}(<R) + M_{\rm BH}$ and on the mass of the swept up shell itself $M_{\rm sh}(R) \,=\, f_{\rm gas} M_{\rm halo}(<R)$, where $f_{\rm gas}$ is the gas mass fraction of the halo.
Balancing the two forces gives the equation of motion of the outflowing shell as

\begin{figure*}
\centering 
\includegraphics[scale = 0.15, trim = 0.1mm 0mm 0mm 0mm]{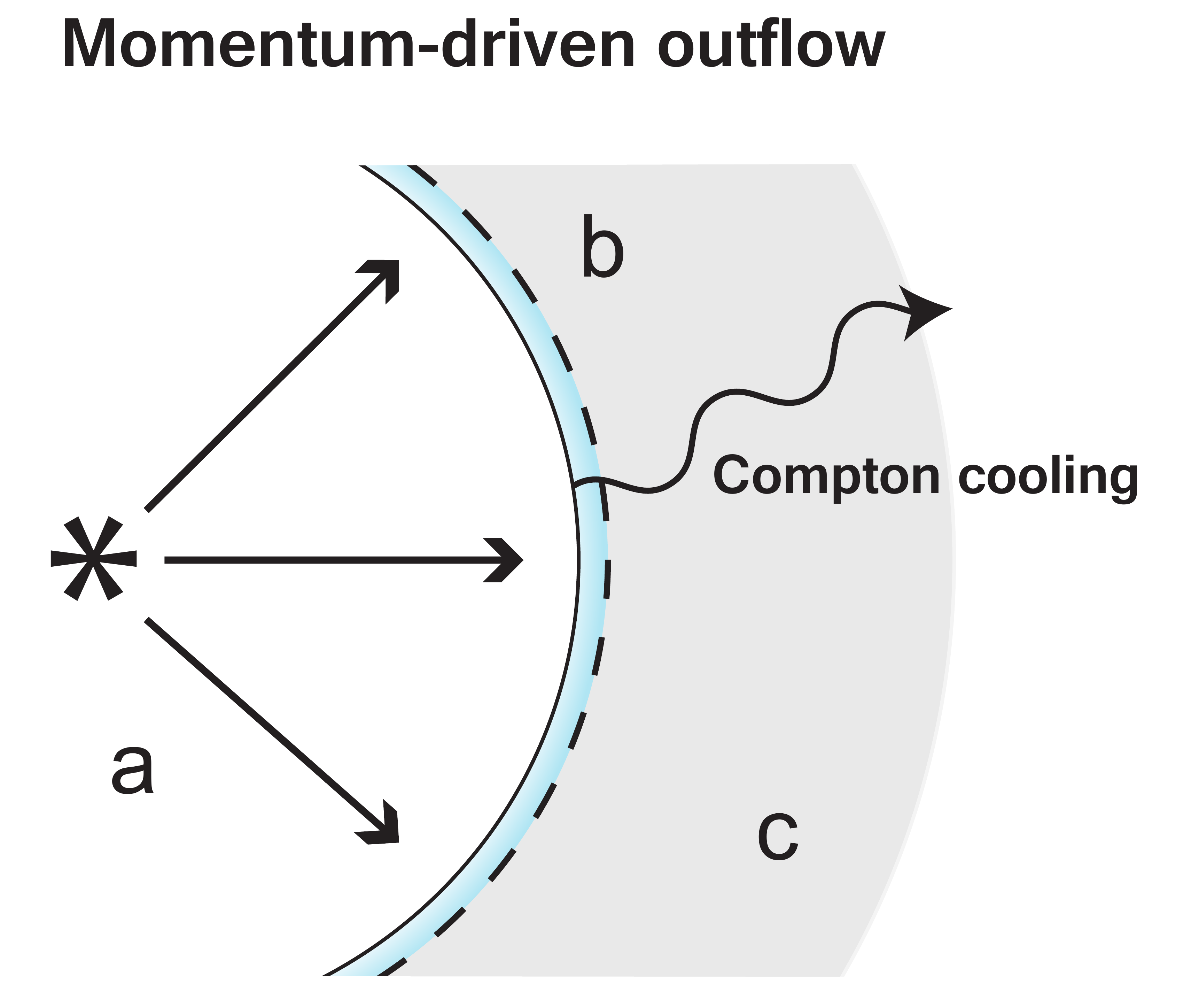}
\hspace{0.7in}
\includegraphics[scale = 0.15]{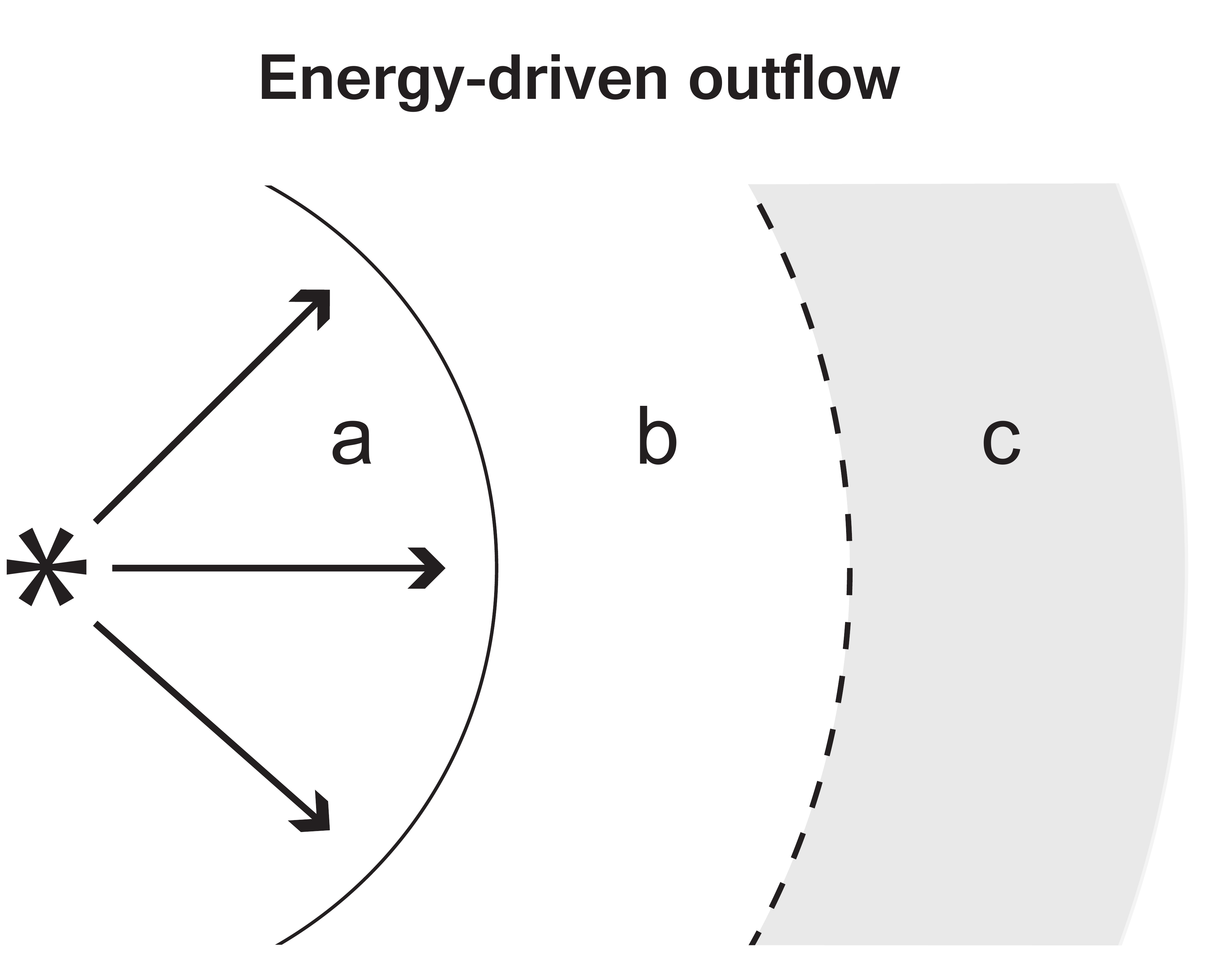}
\caption{Momentum and energy-driven AGN outflows. A momentum-driven outflow (shown on the left) occurs when the inner wind that shock-heats as it passes the reverse shock (region (b)) cools efficiently. The resulting (isothermal) shock consists of an initially adiabatic shock followed by a region where radiative cooling dominates shown in light blue. The loss of thermal energy results in a drop of pressure support and causes region (b) to be very thin. The post-shock pressure at the contact discontinuity in this case is $P \,=\, \rho_{\rm w} v_{\rm w}^2$. An energy-driven outflow (shown on the right) results when the energy injected by the inner wind (region (a)) is fully conserved throughout the outflow, i.e. no substantial cooling losses are sustained. Region (c) is driven out as the now hot and thick region (b) expands adiabatically. The separate question of how efficiently region (c) cools does not determine whether the outflow is energy- or momentum-driven. We revisit this question in Section~\ref{seccooling} and for now represent region (c) as thick, while mindful it can collapse into a very thin shell if cooling is efficient.}
\label{outflowstructure_momene}
\end{figure*}

\begin{equation}
 \frac{d \left[ M_{\rm sh}(R) \dot{R} \right]}{dt}  \,=\, 4 \pi R^2 P - \frac{GM_{\rm sh}(R)M_{\rm tot}(<R)}{R^2} \, .
\label{shelleq}
\end{equation}  

The properties of the fluid right across the reverse shock at $R_{\rm r}$, where the subscript $r$ denotes `reverse shock', can be obtained from the Rankine-Hugoniot jump conditions for a strong adiabatic shock.
The number density of the shocked inner wind $n_{\rm r}$ across the reverse shock is accordingly given by

\begin{equation}
\begin{split}
n_{\rm r} & \,=\, 4\frac{\rho_{\rm w}}{\mu m_{\rm p}} \\ 
        & \,\approx\, 10^{-3} \left(\frac{\eta}{0.1}\right)^{-2} \left(\frac{\mu}{0.59}\right)^{-1}  \left(\frac{M_{\rm BH}}{10^8 \mathrm{M_\odot}}\right) \left(\frac{R}{1 \mathrm{kpc}}\right)^{-2}\rm cm^{-3} \, ,
\label{nrankine}
\end{split}
\end{equation}

using the result from Eq.~\ref{rhowind} to substitute for $\rho_{\rm w}$.

Assuming the reverse shock to have a speed $\ll v_{\rm w}$, the temperature $T_{\rm r}$ just behind the shock is

\begin{equation}
  T_{\rm r} \,=\, \frac{3}{16} \frac{\mu m_{\rm p}}{k_{\rm }} v_{\rm w}^2 \,\approx\, 1.2 \times 10^{10} \left( \frac{\mu}{0.59}\right)\left( \frac{v_{\rm w}}{0.1 c} \right)^2 \rm K \, .
\label{trankine}
\end{equation}

Note that such high temperatures are only possible to achieve due to the high speed of $0.1 c$ of the inner (unshocked) AGN wind and by neglecting relativistic effects in the hot plasma.
The thermal energy in the shocked inner wind can be either transferred to the ISM or lost via radiative cooling.
The dynamics of the resulting (out)flow pattern depends critically on which of these is the fate of the thermal energy contained in the $\approx\, 10^{10 - 11} \rm K$ hot shocked inner wind. 

\subsection{Momentum-driven limit}
\label{secmom}

If the reverse shock cools efficiently, shocked inner wind gas in region (b) traverses a radiative region with a size of the order of the cooling length $l_{\rm cool}$ until its temperature returns to its pre-shock value at a radius $\approx R_{\rm r} + l_{\rm cool}$.
This sequence of an initially adiabatic shock and a cooling region constitutes an \emph{isothermal shock}.
Since cooling is efficient, $l_{\rm cool} << R_{\rm r}$, it follows that the shock is approximately planar, i.e. region (b) is very thin (see Fig.~\ref{outflowstructure_momene}).
For a strong isothermal shock, the Rankine-Hugoniot jump conditions yield a post-shock pressure $P$ given by

\begin{equation}
P \,=\, \rho_{\rm w} v_{\rm w}^2 \, .
\end{equation}

The post-shock gas then drives an outer shock into the unperturbed ISM, thrusting with a pressure equal to the ram-pressure of the pre-shock inner wind, i.e. $P \,=\, L_{\rm Edd}/(4 \pi c R^2)$. 
This result becomes physically intuitive when recalling that region (b) is very thin. 
The inner wind can then be imagined to collide with the ISM directly, transferring its momentum fully in the interaction. 
In this limit, the outflow is said to be \emph{momentum-driven} since the driving force is equal to the momentum flux of the inner wind.
Eq.~\ref{shelleq} can then be simplified to

\begin{equation}
 \frac{d \left[ M_{\rm sh}(R) \dot{R} \right]}{dt}  \,=\, \frac{L_{\rm Edd}}{c} - \frac{GM_{\rm sh}(R)M_{\rm tot} (<R)}{R^2} \, .
\label{shellmom1}
\end{equation}  

For an isothermal halo with velocity dispersion $\sigma$ \citep[see][]{King:05, McQuillin:12, Ishibashi:12}, this can be recast into the form

\begin{equation}
  \frac{d \left[ R \dot{R} \right]}{dt} \,=\, -2 \sigma^2 \left( 1 - \frac{M_{\rm BH}}{M_{\rm \sigma}} \right) - \frac{GM_{\rm BH}}{R} \, ,
\label{shellmom2}
\end{equation}  

where $M_{\rm \sigma} \,=\, \frac{f_{\rm gas} \kappa}{\pi G^2} \sigma^4$.
Neglecting the last term on the right hand side of Eq.~\ref{shellmom2} (valid if the launching speed of the shell is higher than the escape speed from the black hole's gravitational potential), the differential equation admits unbound solutions only if $M_{\rm BH} \geq M_{\rm \sigma}$.
If AGN outflows are momentum-driven, they can in principle escape to arbitrarily high radii only when the central black hole grows beyond a critical mass, at which the outflowing shell becomes unbound.
At this critical value, further accretion on to the black hole is expected to become inefficient and the black hole mass is limited to

\begin{equation}
M_{\rm final}^{\rm m} \,=\, M_{\rm \sigma} \,=\, \frac{f_{\rm gas} \kappa}{\pi G^2} \sigma^4 \, ,
\label{msgimaeq}
\end{equation}

where the superscript `$\rm m$' denotes `momentum-driven'.
Remarkably, the normalisation and the $M_{\rm final}^{\rm m} \,\propto\, \sigma^4$ scaling are in close agreement with the observed $M_{\rm BH} \,\propto\, \sigma^{4.3}$ relation \citep{Ferrarese:00, Gebhardt:00, Tremaine:02, Gultekin:09, McConnell:13, Kormendy:13}.

\subsection{Energy-driven limit}
\label{subsecene}

If the reverse shock is instead unable to radiate away its thermal energy, the shell is driven by the adiabatic expansion of the hot shocked wind bubble (see Fig.~\ref{outflowstructure_momene}).
In the limit where the full energy of the inner wind is conserved, the resulting outflow pattern is termed \emph{energy-driven}. 

In an energy-driven outflow, the expansion rate of the shell of shocked interstellar gas is equal to the rate at which the shocked wind does `PdV' work on its surroundings.
By balancing energy losses due to `PdV' work and work done for overcoming gravity with energy injection by the inner wind, we can write an energy equation that reads

\begin{equation}
\begin{split}
	\dot{E}    & \,=\, (\gamma - 1) \frac{d (PV)}{dt} \\
                   & \,=\, \epsilon L_{\rm Edd} - P \dot{V} - G \frac{M_{\rm sh}(R) M_{\rm tot}(<R)}{R^2}\dot{R} \, ,   
\label{shellene1}
\end{split}
\end{equation}

where the first equality simply follows from the ideal gas law and $\gamma \,=\, \frac{5}{3}$ is the adiabatic index of the gas. As mentioned, $\epsilon \,=\, \eta/2$ if $\dot{M}_{\rm w} \,=\, \dot{M}_{\rm Edd}$.
Using Eq.~\ref{shelleq} to eliminate $P$ from Eq.~\ref{shellene1}, assuming an isothermal halo such as in \citet{King:05},  \citet{King:11} and \citet{McQuillin:13} and that the shell's initial speed is sufficient for it to escape from the black hole potential yields the equation of motion for the shell

\begin{equation}
\begin{split}
  \epsilon L_{\rm Edd} \,=\, \frac{2 f_{\rm gas} \sigma^2}{G} & \left( \frac{1}{2} R^2 \dddot{R} + 3R\dot{R}\ddot{R} + \frac{3}{2} \dot{R}^3 \right) \\ 
                                                                                                      & \quad + 10 f_{\rm gas} \frac{\sigma^4}{G} \dot{R} \, .
\label{shellene2}
\end{split}
\end{equation}

Eq.~\ref{shellene2} also admits unbound solutions. 
Assuming a solution of the type $R \propto t$ \citep{King:11} and taking $\dot{R} \,=\, 2 \sigma$ into Eq.~\ref{shellene2} gives

\begin{equation}
\epsilon L_{\rm Edd} \,=\, \frac{44f_{\rm gas}}{G} \sigma^5 \, .
\label{msigmaene}
\end{equation}

When this condition is satisfied, the energy-driven outflow can clear the halo off its gas and terminate black hole accretion. 
In this case, the black hole mass is fixed to a value given by

\begin{equation}
M_{\rm final}^{\rm e} \,=\, \frac{11 f_{\rm gas} \kappa}{\epsilon \pi G^2 c} \sigma^5 \, ,
\label{msigmaeq2}
\end{equation}

using `$\rm e$' to denote `energy-driven' \citep[see also][]{Silk:98, Haehnelt:98, King:10, Fabian:12, McQuillin:13}.
Energy-driven outflows therefore yield $M_{\rm final}^{\rm e} \,\propto\, \sigma^5$, somewhat steeper than the observed $\,\propto\, \sigma^{4.3}$ scaling \citep[see e.g.][]{Kormendy:13}.

\subsection{Energy- versus momentum-driving and the role of Compton cooling/heating}
\label{secenevsmom}

\subsubsection{The efficiency of energy- vs. momentum-driving}

Using Eqs.~\ref{msgimaeq} and ~\ref{msigmaeq2}, the ratio of the final black hole masses for energy- and momentum-driven outflows is given as

\begin{equation}
\frac{M_{\rm final}^{\rm e}}{M_{\rm final}^{\rm m}} \,\approx\, 10^{-1} \left( \frac{\epsilon}{0.05} \right)^{-1} \left( \frac{\eta}{0.1} \right)^{-1}  \left( \frac{\sigma}{200 \, \mathrm{km \, s^{-1}}} \right) \, ,
\end{equation}

suggesting that lower black holes masses are required to drive a large-scale outflow for energy- than for momentum-driven solutions.
In this section, we clarify the reason for the relative efficiency of energy- over momentum-driven outflows.

Whether an outflow is energy- or momentum-driven, its kinetic luminosity can be parametrised in terms of a kinetic conversion efficiency $\epsilon_{\rm kin}$ that quantifies the fraction of the AGN luminosity which is converted into kinetic energy of the outflow on galactic scales. The outflow's kinetic luminosity is thus given by

\begin{equation}
\dot{E}_{\rm k,\,sh} \,=\, \epsilon_{\rm kin} L \, .
\end{equation}

The momentum flux of the outflow can, in turn, be written as

\begin{equation}
\dot{p}_{\rm sh} \,=\, \frac{2 \dot{E}_{\rm k, sh}}{v_{\rm sh}} \,=\, 2 \epsilon_{\rm kin} \frac{L}{c} \frac{c}{v_{\rm sh}} \, ,
\label{pshell1}
\end{equation}

again, irrespective of whether it is energy- or momentum-driven.

If the outflow is energy-driven, then $\epsilon_{\rm kin}$ is related to $\epsilon$ via

\begin{equation}
\epsilon_{\rm kin}^{\rm e} \,=\, \epsilon_{\rm th-kin} \times \epsilon \,=\, 0.05 \epsilon_{\rm th-kin} \left( \frac{\epsilon}{0.05} \right) \, ,
\label{ekinene}
\end{equation}

where $\epsilon_{\rm th-kin}$ is the efficiency in converting the thermal energy injected by the AGN into kinetic energy of the outflow on galactic scales.

We now evaluate the ratio of the momentum flux $\dot{p}_{\rm sh}$ of the outflow to that of the inner wind, a ratio often termed `momentum boost-factor'. 
Substituting Eq.~\ref{ekinene} into Eq.~\ref{pshell1} and using $\dot{p}_{\rm w} \,=\, L/c$ (Eq.~\ref{pwind}) gives

\begin{equation}
\frac{\dot{p}_{\rm sh}}{\dot{p}_{\rm w}} \bigg|^{\rm e} \,=\, 2 \epsilon_{\rm kin}^{\rm e} \frac{c}{v_{\rm sh}} \,=\, 30 \epsilon_{\rm th-kin} \left(\frac{\epsilon}{0.05}\right) \left(\frac{v_{\rm sh}}{1000 \mathrm{km \, s^{-1}}}\right)^{-1}\, .
\label{momboost}
\end{equation}

Energy-driven shells can therefore acquire momentum fluxes substantially higher than the original momentum flux of the inner wind \citep[see also][]{Zubovas:12, Faucher-Giguere:12}.

Conversely, if the outflow is momentum-driven, momentum conservation across the isothermal shock implies that

\begin{equation}
\dot{p}_{\rm sh} \,=\, \dot{p}_{\rm w} \,=\, \frac{L}{c} \, ,
\label{pshell2}
\end{equation}

and hence, by combining Eq.~\ref{pshell1} and Eq.~\ref{pshell2}

\begin{equation}
\epsilon_{\rm kin}^{\rm m} \,=\, \frac{1}{2} \frac{v_{\rm sh}}{c} \,=\, 6 \times 10^{-4} \left( \frac{v_{\rm sh}}{1000 \mathrm{km \, s^{-1}}} \right) \, .
\label{ekinmom}
\end{equation}

As expected, the boost factor in the  momentum-driven regime is given by

\begin{equation}
\frac{\dot{p}_{\rm sh}}{\dot{p}_{\rm w}} \bigg|^{\rm m} \,=\, 2 \epsilon_{\rm kin}^{\rm m} \frac{c}{v_{\rm sh}} \,=\, 1 \, ,
\end{equation}

in contrast with the energy-driven case, where high boost factors can apply.

The reason for the higher relative efficiency of energy-driven outflows stems from the inequality  $\epsilon_{\rm kin}^{\rm e} \,>\, \epsilon_{\rm kin}^{\rm m}$, which applies for typical values for $\epsilon_{\rm th-kin}$ ($\approx 0.5$).
Energy-driven shells propagate more rapidly than momentum-driven outflows and their higher  momentum flux means they can more vigorously remove gas from the halo.

\subsubsection{The role of Bremsstrahlung and inverse Compton cooling/heating}

In practice, if AGN-driven outflows arise due to an inner wind shock, they are likely to be a mixture of the energy- and momentum-driven limits discussed above.
Whether, however, the outflows are more suitably approximated by their energy- or momentum-driven limit depends on the relative locations of the reverse shock radius $R_{\rm r}$ and the cooling radius $R_{\rm cool}$, since it is the cooling of the shocked wind that discriminates between the two cases.
If $R_{\rm r} > R_{\rm cool}$, the hot inner wind shock cannot cool and the outflow falls towards its energy-driven limit.
Conversely, if $R_{\rm r} < R_{\rm cool}$, the reverse shock cools efficiently and the outflow is more accurately represented by its momentum-driven limit. 

For gas of temperatures in the range $10^{10} \-- 10^{11} \, \rm K$ (see Eq.~\ref{trankine}), the only significant cooling processes are free-free emission and inverse Compton scattering.
We begin by estimating the timescale for free-free (thermal Bremsstrahlung) emission $t_{\rm ff}$ \citep[see e.g.][]{Mo:10} for the shocked inner wind gas

\begin{equation}
\begin{split}
t_{\rm ff} & \, \approx \,  7.9 \times 10^{11}  \left(\frac{T}{10^{10} \mathrm{K}}\right)^{1/2} \left(\frac{n_{\rm e}}{10^{-3} \mathrm{cm^{-3}}}\right)^{-1} \, \rm yr  \\
                    & \, \approx \, 4.1 \times 10^{11} \left( \frac{T}{10^{10} \, \mathrm{K}}\right)^{1/2} \left( \frac{R}{1 \, \mathrm{kpc}} \right)^2 \, \rm \mathrm{yr} \, ,
\end{split}
\label{freefree}
\end{equation}

where we have taken the electron number density as $n_{\rm e} \,=\, \frac{27}{14} n_{\rm r}$ (for a fully ionised primordial gas) and assumed the fiducial values given in Eq.~$9$\footnote{Implicit in this calculation is the approximation that the density and temperature of the inner wind shock bubble is uniform \citep[see][]{Faucher-Giguere:12}.}.

The outflow time $t_{\rm flow}\,=\, \frac{R}{v_{\rm sh}}$ of the shocked inner wind is given by

\begin{equation}
t_{\rm flow} \,\approx\, 10^6 \left( \frac{R}{1 \mathrm{\,kpc}} \right) \left( \frac{v_{\rm sh}}{1000 \mathrm{ \,km \, s^{-1}}} \right)^{-1} \, \rm yr \, .
\label{tflow}
\end{equation}

Setting $t_{\rm ff} \,=\, t_{\rm flow}$ gives a cooling radius of $R_{\rm cool} \,\approx\, 2 \times 10^{-3} \left( \frac{v_{\rm sh}}{1000 \, \mathrm{km s^{-1}}} \right)^{-1} \left( \frac{T}{10^{10} \, \mathrm{K}} \right)^{-1/2} \, \rm pc$, i.e. the shocked inner wind cannot cool via free-free emission.

A much more efficient cooling mechanism arises from the energy exchange that takes place as photons in the radiation field of the AGN and electrons in the hot shocked inner wind gas undergo Compton scattering. 
In this process, high energy photons can transfer energy to (non-relativistic) electrons if $h\nu \,>\, 4k_{\rm B} T$, thereby heating the gas, while low energy photons with $h\nu \,<\, 4k_{\rm B} T$ gain energy, effectively cooling the gas.
The temperature at which Compton heating and cooling balance out, the `Compton temperature' $T_{\rm C}$, depends on the spectral energy density of the emitting quasar and will therefore vary from source to source and with accretion rate \citep[see][for a recent discussion on the AGN feedback effects of a varying Compton temperature]{Gan:14}. 
For the average QSO, \citet{Sazonov:04} estimate $T_{\rm C} \,\approx\, 2 \times 10^7 \rm \, K$. 
Since $T_{\rm C} \,\ll\, T \,\approx\, 10^{10} \, \rm K$ (as given in Eq.~\ref{trankine}), the hot shocked wind cools \citep[see][for the potential observational imprint of inverse Compton cooling of the inner wind shocked bubble]{Bourne:13}.
Using the estimated cooling function for Compton heating/cooling of a gas exposed to a QSO continuum with $T_{\rm C} \,=\, 1.9 \times 10^7 \rm \, K$ provided in \citet{Sazonov:05} gives a cooling time

\begin{equation}
t_{\rm C} \,\approx\, 10^8 \left( \frac{R}{1 \mathrm{kpc}} \right)^2 \left( \frac{M_{\rm BH}}{10^8 \mathrm{M_\odot}} \right)^{-1}  \left( \frac{L}{L_{\rm Edd}}\right)^{-1} \rm \, yr \, .
\label{tcmpt}
\end{equation}

In the ultra-relativistic regime (for $T \,\gtrsim\, 10^9 \, \rm K$), note that the inverse Compton cooling rate however scales as $\propto T^2$, while the gas thermal energy still scales as $T$, resulting in even lower cooling times than suggested by Eq.~$29$ (which only holds for non-relativistic electrons) \citep[see e.g.][]{Pozdnyakov:83}.
For mildly relativistic electrons, as is the case for $T \,\approx\, 10^{10} \, \rm K$, the cooling time given in Eq. $29$ is however likely overestimated by a just a factor of a few.
Setting $t_{\rm flow} \,=\, t_{\rm C}$ gives the cooling radius for inverse Compton cooling

\begin{equation}
R_{\rm cool} \, \approx \,  \left( \frac{v_{\rm sh}}{1000 \mathrm{ \,km \, s^{-1}}} \right)^{-1} \left( \frac{M_{\rm BH}}{10^8 \mathrm{M_\odot}} \right) \left( \frac{L}{L_{\rm Edd}} \right) \, \rm kpc \, .
\label{coolingradius}
\end{equation}

In the simulations presented in this study, we do not explicitly model Compton cooling/heating because neither the inner wind nor the reverse shock are resolved.
Efficient Compton cooling is implicit in our momentum-driven model however.
Note also that once gas cools below $T \,\lesssim\, T_{\rm C}$, Compton scattering can instead heat gas and can provide an important source of feedback in its own right as shown by \citet{Ciotti:97, Ciotti:01} and \citet{Sazonov:05}.

According to the estimate of Eq.~\ref{coolingradius}, the shocked inner wind cools very efficiently below a radius $R_{\rm cool} \, \approx \, 1 \, \rm kpc$. 
In this regime, the outflow will be momentum-driven \citep{King:03}.
Once the momentum-driven shell moves beyond $R_{\rm cool}$, the outflow becomes energy-driven due to the relative inefficiency of inverse Compton cooling in the shocked inner wind.
Note, however that if the electrons and protons in the hot shocked wind plasma are thermally decoupled, as is known to occur in very strong shocks, \citep[e.g. in supernova remnants, ][]{Ghavamian:07}, the cooling radius will be reduced to too close to the sphere of influence of the black hole for meaningful cooling to occur \citep{Faucher-Giguere:12}. 

\subsection{Radiation pressure-driven outflows}
\label{secrad}

A separate class of models for AGN feedback envisages that AGN-driven outflows are driven by direct radiation pressure on dust grains embedded in the ISM of the galaxy \citep{Fabian:99, Murray:05, Debuhr:11}.
In this scenario, interstellar gas is pressurised by the radiation field of the AGN rather than by a shocked AGN inner wind as in \citet{King:03, King:05}.
Observational support that has been invoked in favour of this picture 
is for instance  the observed lack of AGN with simultaneously high luminosities and high hydrogen column densities \citep{Raimundo:10, Fabian:12}, where radiation pressure would be most efficient at driving outflows.
It has been argued that radiation pressure can lead to large-scale outflows and to momentum-boosts, as IR reprocessed photons are repeatedly absorbed \citep[see e.g.][]{Roth:12}. Since their dynamics can be described by an equation essentially identical to Eq.~\ref{shellmom1} in the optically thick regime, radiation pressure-driven outflows have also been termed `momentum-driven' in the literature. These outflows are however not to be confused with the momentum-driven outflows discussed in this study, which have a fundamentally different physical origin.
We postpone a more detailed assessment  of the plausibility of IR radiation-pressure wind models and their implementation in numerical simulations to  future work. Note, however, that such radiation pressure driving requires large spatially coherent IR optical depth which is increasingly difficult to realise on galactic  ($\ga$ kpc) scales.  See also \citet{Krumholz:13} and \citet{Davis:14} for recent discussions on the possibility of launching  IR radiation pressure-driven winds.

\section{Numerical implementation of the galactic wind models}
\label{secnumimp}

\subsection{Preliminaries}

Models of AGN-driven outflows (see Section~\ref{enemomout}) involve a large disparity of physical scales; ranging from $\sim 0.001 \, \rm pc$ where energy is released from the accretion disc that feeds the supermassive black hole, all the way to galactic ($\rm kpc$) scales at which the large-scale outflow sweeps across the galaxy and the surrounding halo.
The \emph{ab initio} treatment of the interaction between an AGN and its host galaxy in numerical simulations requires currently infeasible resolution and dynamical range.
In the context of the inner wind model of AGN feedback (Section~\ref{enemomout}), lack of resolution means that it is currently neither possible to resolve the inner wind nor the reverse shock in numerical simulations.
Numerical implementations of AGN feedback therefore always rely on `subgrid' models that attempt to mimic the effects of unresolved outflow physics at resolved scales \citep[see e.g.][for a comparison of several widely used AGN feedback models]{Wurster:13}.
The relation of popularly adopted `subgrid' models with simple hydrodynamical models such as summarised in Section~\ref{kingsmodel} however often remains unclear.

Our strategy in this study is to mimic the presence of a reverse shock and its behaviour in both the momentum- and energy-driven (limiting) regimes using `subgrid' models \citep[see also][]{Nayakshin:10, Power:11, Nayakshin:12}. 
Uncertainties in the properties of the inner wind are bypassed in numerical simulations by assuming that either energy or momentum (scaled to the luminosity of the QSO) is injected at galactic scales ($0.1 \-- 1 \, \rm kpc$).  
Injection of energy is normally straightforward to implement and has the advantage that a fixed amount of energy is added to the wind.
It realises efficient feedback as long as cooling is not overwhelming, e.g. at large distances from the dense galactic centre or at high energy input rates. 
It probably, however, underestimates the effect of AGN feedback when cooling times are short compared to outflow times and too little of the thermal energy is converted into kinetic energy. 
Injection of momentum, on the other hand, is usually motivated by arguing that cooling is efficient in the inner outflow, where the optical depth is expected to be high such that a momentum flux $L/c$ should be injected at large scales. 
This approach has the advantage that galactic winds have greater impact in rather dense environments and therefore smaller radii where cooling is important. 
It has, however, the disadvantage that the corresponding injected energy scales linearly with the outflow velocity of the gas into which the momentum is injected and thus depends strongly on the physical and dynamical state of the ISM.  
Like energy injection, it is also very sensitive to the resolved distance to the AGN.    
It is important to note that momentum or thermal energy injection in numerical simulations do not necessarily respectively translate to what is meant by energy- or momentum-driven outflows in the classical sense.
Particles/cells receiving too large a momentum/velocity kick will shock and thermalise rapidly, driving an outflow via adiabatic expansion rather than by ram-pressure. 
Such an outflow rather resembles the energy-driven case and cannot be considered `momentum-driven' even if it relies on momentum injection.
Conversely, if thermal energy is added to gas which cools very efficiently so that most of the injected energy is radiated away, gas may be driven out due to residual bulk motion and ram-pressure rather than via adiabatic expansion. Such an outflow can no longer be described as `energy-driven'.
Such uncertainties imply that the physical interpretation of any `subgrid' model is dependent on the included physics (e.g. cooling) and the spatial resolution of the simulation.
Bearing these difficulties in mind, we now describe implementations of energy- and momentum-driven outflows that successfully reproduce the predictions of the analytical models of \citet{King:03, King:05} in their idealised setting.

\subsection{Numerical setup}
\label{hernhalo}

For our numerical simulations, we employ the moving-mesh code {\sc AREPO} \citep{Springel:10}, which uses a second-order accurate finite-volume method on an unstructured Voronoi mesh that is allowed to move with the flow in order to model gas hydrodynamics.
For our simulations of an isolated halo, we adopt a static Hernquist dark matter potential \citep{Hernquist:90, Sijacki:12}.
Even though an isothermal profile is typically assumed in simple analytical models \citep{Silk:98, Fabian:99, King:03, King:05}, the relevant equations of motion can be straightforwardly adapted for a Hernquist halo by taking the expression for the enclosed mass $M_{\rm halo}(<R)$ in Eq.~\ref{shellmom1} and Eq.~\ref{shellene1} as

\begin{equation}
M_{\rm halo} (<R) \,=\, M \frac{r^2}{(r + a)^2} \, ,
\end{equation}

where $M$ is the total mass and $a$ is the scale length of the halo.
The halo's total mass density is given by

\begin{equation}
\rho_{\rm halo} \,=\, \frac{M}{2 \pi} \frac{a}{r (r + a)^3} \, .
\label{rhohern}
\end{equation}

\begin{figure*}
\centering \includegraphics[scale = 0.6]{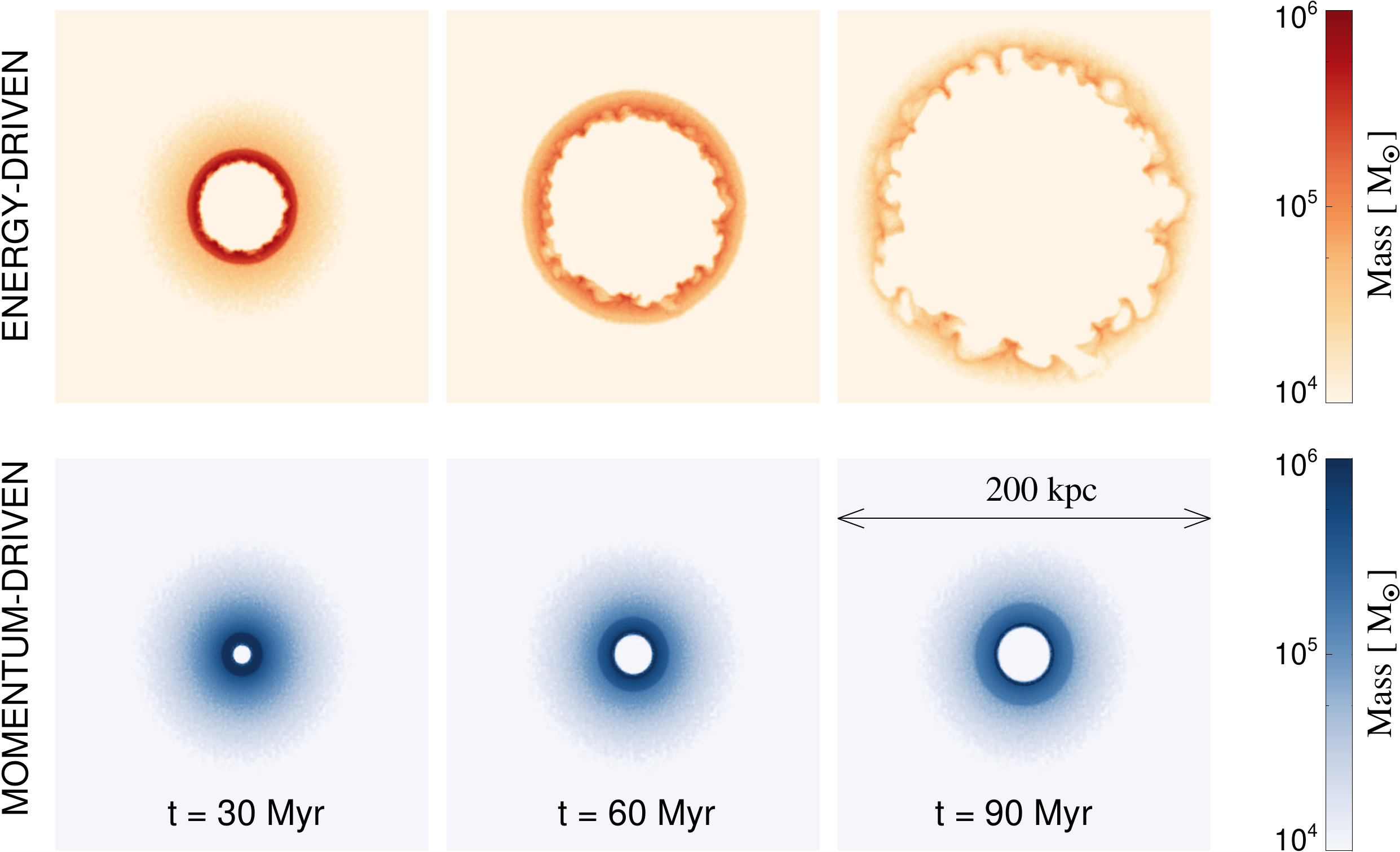}
\caption{Projection of gas mass along a slab of thickness $\, = \, 10 \, \rm kpc$ and length $\, = \, 200 \, \rm kpc$ for the energy-driven outflow model (top row) and for the momentum-driven outflow model (bottom row) at different simulation times for a black hole of mass $10^8 \, \rm M_\odot$ emitting at its Eddington luminosity. Injection of energy or momentum at the centre of the halo drives a shock that compresses almost all intervening gas into a propagating shell. For matching black hole mass, the energy-driven outflows propagate more rapidly than the momentum-driven outflows due to their higher efficiency in converting AGN luminosity to kinetic luminosity.}
\label{shellpanel}
\end{figure*}

The Hernquist potential has the advantage that its total mass converges as $r \,\rightarrow\, \infty$ (removing the need for a truncation at an arbitrary radius) while closely resembling the Navarro-Frenk-White profile \citep{Navarro:97}. 
Moreover, solutions to Eqs.~\ref{shellmom1} and \ref{shellene1} converge to a unique solution at large radii for a Hernquist halo \citep{McQuillin:12}, facilitating a comparison with analytical solutions.
For our numerical experiments, we select a halo of total mass $M \,=\, 10^{12} \, \rm M_\odot$, which we populate with $10^6$ resolution elements, resulting in an initial gravitational softening length of $0.5 \, \rm kpc$. 
Note that in {\sc AREPO}, gas has an adaptive gravitational softening set by the cell size. 
The minimum cell size in our simulations reaches $\approx 7 \, \rm pc$.
As in \citet{King:03, King:05}, we assume that a fraction $f_{\rm gas}$ of the total mass $M$ is in gas such that the gas density is $\rho_{\rm gas} \,=\, f_{\rm gas} \rho_{\rm halo}$.
We choose $f_{\rm gas} \,=\, 0.17$ in all our tests.
Finally, we choose the halo to have a concentration of $10$ as expected for a $10^{12} \, \rm M_\odot$ halo at $z \,=\, 0$ \citep[e.g][]{Maccio:08}.
This choice of parameters fixes the virial radius\footnote{$R_{\rm vir}$ defined as the radius enclosing a mean density $200$ times that of the cosmic mean.} to $R_{\rm vir} \,=\,162.62 \, \rm kpc$ and the scale length to $a \,=\, 28.06 \, \rm kpc$.

Initial conditions were generated by randomly sampling the halo's density structure out to a very large radius of $100 a \,=\, 2608 \, \rm kpc$ in order to ensure its density profile to be as close a match as possible to its analytical expression (Eq.~\ref{rhohern}). 
Starting from these initial conditions, the halo was then evolved non-radiatively for a period of $10 \, \rm Gyr$ to minimise halo relaxation effects.
A black hole sink particle was then introduced at the centre of the halo and its position was fixed at this location for the entire duration of the various simulations.
For the black hole mass, we explored values ranging from $5 \times 10^7 \, \rm M_\odot$ to $3 \times 10^8 \rm \, M_\odot$, in order to probe solutions for black hole masses below, equal and above the critical $M_{\sigma}$ value for the halo\footnote{For any halo with a peaked circular velocity ($v_{\rm c}$) profile, it can be shown that $M_{\rm \sigma} \,\approx\, \left( \frac{f}{0.17} \right) \left( \frac{v_{\rm c}}{200 \, \rm km \, s^{-1}} \right)^4 10^8 \, \rm M_\odot$ \citep{McQuillin:12}} respectively. 
In order to establish a close comparison with the models of \citet{King:03, King:05}, it is assumed that the AGN constantly emits at its Eddington luminosity for all simulations.
In Section~\ref{seccooling}, we relax this assumption.

\subsection{Energy-driven limit}

In the energy-driven regime, the hot shocked inner wind expands as it does `PdV' work on its surroundings (see Section~\ref{subsecene}).
Its thermal energy changes according to Eq.~\ref{shellene1}, where it was assumed that the kinetic energy of the inner wind is fully converted into thermal energy across the reverse shock.
We reproduce this behaviour in our simulations by letting a fraction $\epsilon \,=\, \frac{\eta}{2} \,=\, 0.05$ of the energy emitted by the AGN couple thermally with its neighbouring gas cells.
The algorithm first identifies the nearest $64$ gas cell neighbours, so defined if they are located within an SPH-kernel of the black hole particle.
Energy is then injected at a rate $\epsilon L_{\rm Edd}$ into the neighbouring cells in a kernel-weighted fashion, so that gas cells closer to the black hole receive a higher amount of thermal energy. 
For the simulations presented in Section~\ref{seccomp1}, note that $L \,=\, L_{\rm Edd}$ always. 
In Section~\ref{seccooling}, we explore varying the AGN lightcurve.
In contrast with typical implementations of this widely used AGN model, we do not prevent gas cells from heating above a maximum temperature. 
This procedure is usually adopted in order to prevent gas particles/cells from heating to temperatures at which relativistic effects become relevant and for which our treatment of hydrodynamics breaks down. In the context of this study however, imposing a temperature ceiling would effectively limit the energy injected by the AGN and would introduce an obvious source of disagreement with analytical models.
Our implementation is otherwise very close to previous implementations of AGN thermal feedback \citep{Springel:05, DiMatteo:05, Sijacki:07, DiMatteo:08, Sijacki:09, Booth:09, Dubois:13b, Dubois:13}.
In Appendix~\ref{appendixa}, we show that in the context of this study, this model is largely insensitive to the selected number of gas cell neighbours.

\subsection{Momentum-driven limit}

In order to reproduce the momentum-driven limit of AGN outflows, it is necessary to ensure that gas is accelerated by the impinging ram-pressure of the (unresolved) inner wind rather than `PdV' work of the shocked inner wind gas, which now has negligible thermal pressure. 
Instead of injecting thermal energy, we impart momentum kicks to the nearest neighbour gas cells directly at rate $L_{\rm Edd}/c$ in the same spirit as previous implementations of mechanical feedback \citep{Ostriker:10, Nayakshin:10, Debuhr:11, Choi:12, Choi:13, Choi:14}.
Note that the magnitude of the effective velocity kick is kernel-weighted, in order to ensure that gas closest to the black hole is expelled most efficiently.
We have also run a set of test simulations, in which we varied the number of gas cells into which momentum is injected (see Appendix~\ref{appendixa}). 
For a large number of neighbours, we have found that the numerical and analytical solutions converge.
The numerical results overshoot with respect to the analytical solution if the selected number of neighbours is too small.
In this case, gas cells effectively receive larger velocity kicks.
In this regime, agreement between numerical and analytical solutions can only be restored by increasing the time-stepping accuracy.

\section{Comparison of hydrodynamic simulations with the analytical wind model}
\label{seccomp}

\subsection{Simulations without radiative cooling}
\label{seccomp1}

\begin{figure*}
\includegraphics[scale = 0.5]{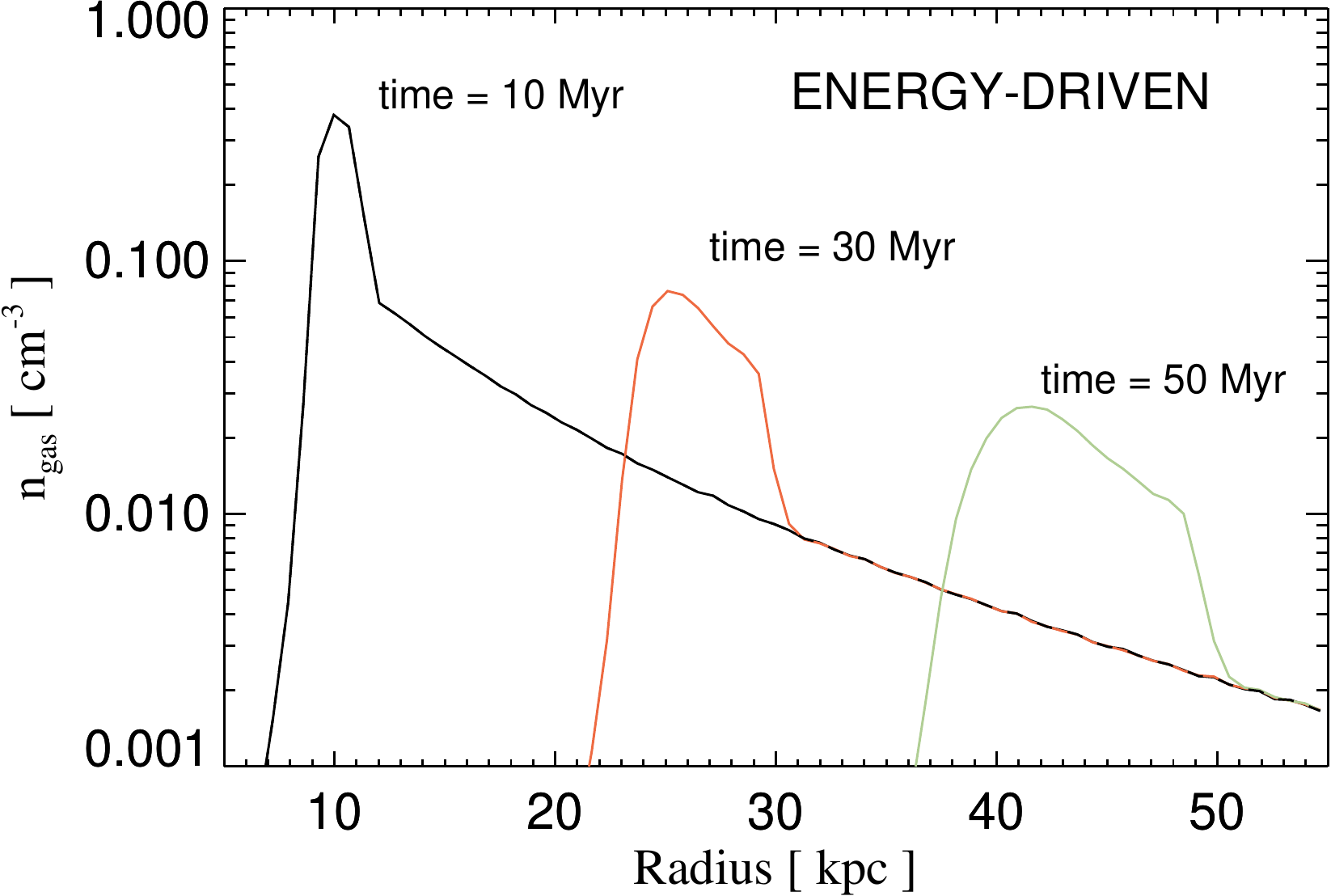}
\includegraphics[scale = 0.5]{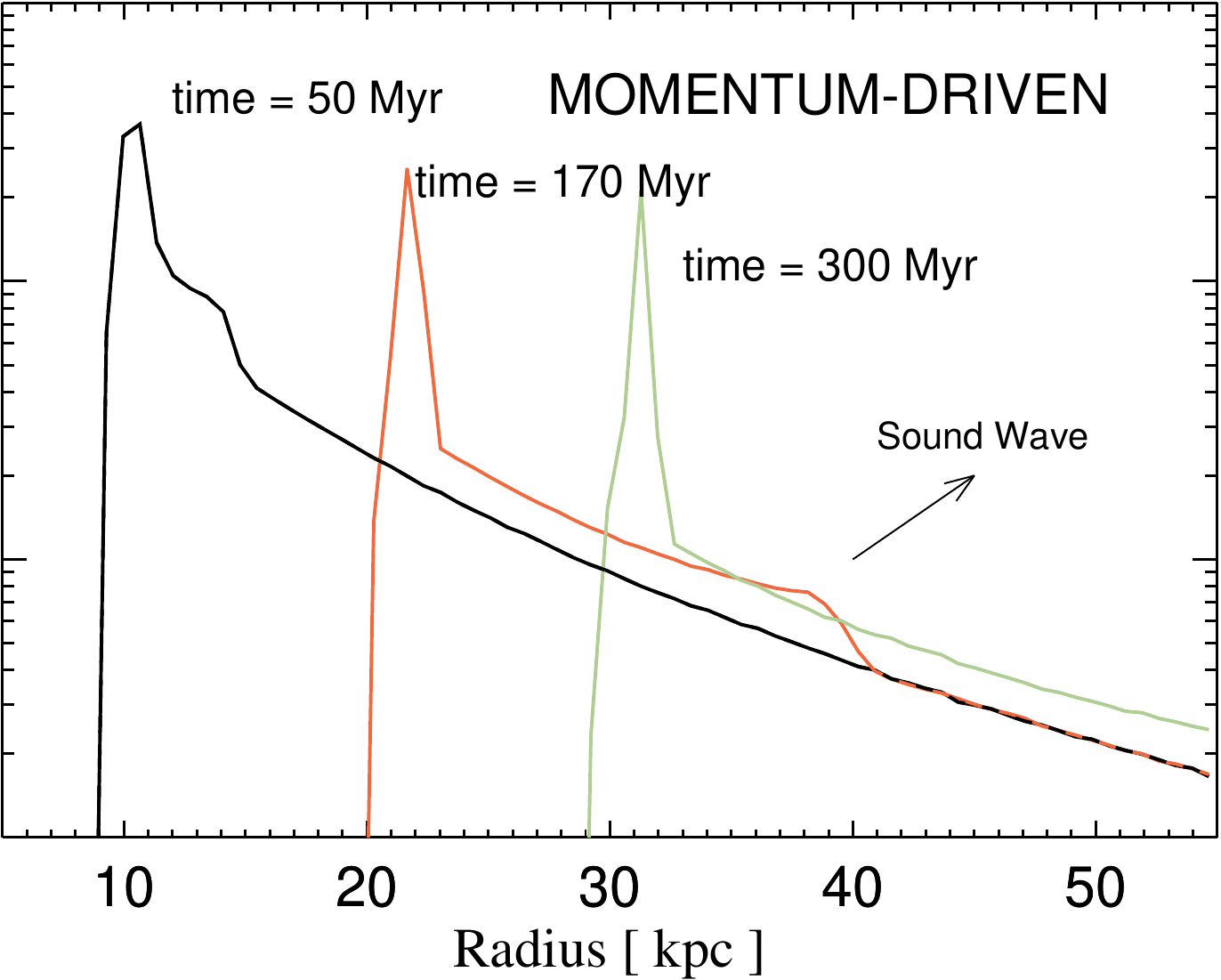}
\caption{Time sequence of gas density profiles in the numerical models for an energy-driven (left hand plot) and momentum-driven (right hand plot) outflow for a black hole mass of $10^8 \, \rm M_\odot$ emitting at its Eddington luminosity. Note that approximately matching radial distances are traversed in much less time by the energy-driven outflow than the momentum-driven outflow. The higher velocity of energy-driven outflows is a consequence of the higher conversion factor of AGN luminosity to kinetic luminosity of the outflowing shell. Note also that for the momentum-driven outflow, the speed of the outflowing shell quickly decelerates into the subsonic regime, generating a forward ripple travelling at the speed of sound.}
\label{thinshell}
\end{figure*}

A time sequence of the blast waves generated in the numerical models for energy and momentum-driven outflows detailed above is illustrated in Fig.~\ref{shellpanel}, where the gas mass was projected along a thin slab of thickness $10 \, \rm kpc$.
As in the theoretical picture proposed in \citet{Silk:98}, \citet{Fabian:99} and \citet{King:03}, a thin shell propagates outwards, carrying with it almost all intervening gas.
For matching black hole mass, the energy-driven outflow propagates more rapidly across the halo than a momentum-driven outflow in agreement with the considerations made in Section~\ref{secenevsmom}. 
Another feature that becomes clear in Fig.~\ref{shellpanel} is the development of Rayleigh-Taylor instabilities in the energy-driven outflow as hot gas rises buoyantly through under-densities in the shocked ISM component \citep[see also][]{King:10b, Zubovas:14}. These instabilities ultimately lead to a pronounced departure from spherical symmetry in the outflowing shell.

\begin{figure} 
\centering
\includegraphics[scale = 0.52]{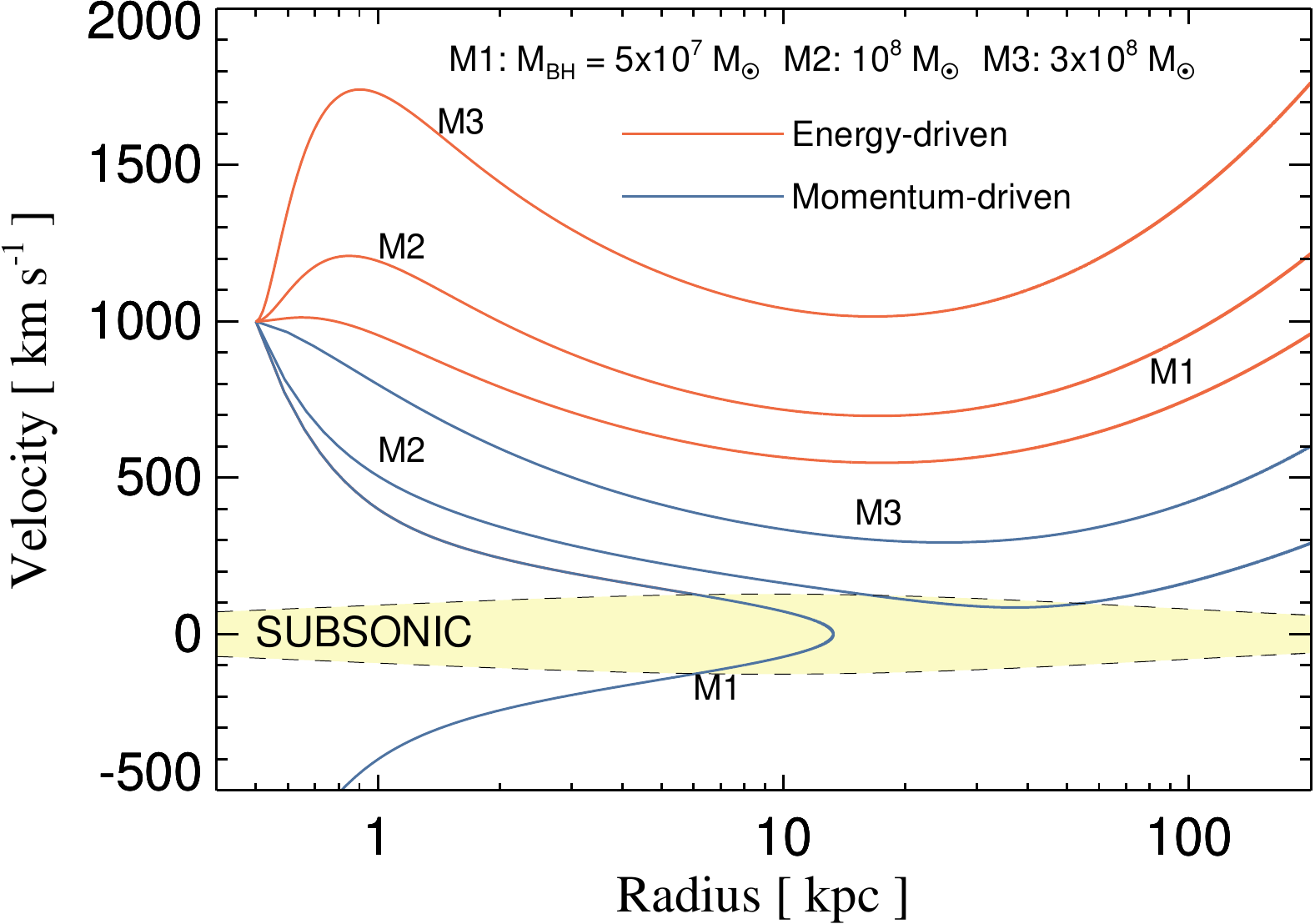}
\caption{Analytical energy- (red lines) and momentum-driven (blue lines) solutions to Eq.~\ref{shelleq}. The local speed of sound is shown as a dashed line separating the sub- and supersonic regimes of the solutions. Momentum-driven solutions for black hole masses $\lesssim 10^8 \, \rm M_\odot$ partially exist in the subsonic regime, where Eq.~\ref{shelleq} is invalid. The behaviour of our simulated solutions deviates from the analytical predictions in this regime quantitatively, but do not change the key conclusion that there is a critical black hole mass, below which momentum-driven outflows stall and above which they can freely expand.}
\label{solutions_analytic}
\end{figure}

The propagation of the shell is represented quantitatively in Fig.~\ref{thinshell}, where a time sequence of gas density profiles is shown for both energy- and momentum-driven models.
At large radii, where gas has not yet been perturbed by the outflow, the profiles follow the Hernquist profile very closely.
Strong deviations to the Hernquist profile occur at intermediate radii, where the gas density rises above the levels corresponding to a Hernquist potential at the same radius. 
This region is occupied by the shocked medium, swept up into a shell, and stretches all the way to the lowest radii, where the gas density then drops.
In the momentum-driven outflow, note that the density profile has a very sharp peak in the innermost radii of the shell.
This region is occupied by gas which is highly compressed after receiving a radial velocity boost and mimics the thin inner wind (isothermal) shock present in the momentum-driven outflow (see Fig.~\ref{outflowstructure_momene}).  
For a black hole mass of $10^8 \, \rm M_\odot$, the thickness of the shell increases and becomes comparable with its distance to the black hole at $r \,\approx\, 10 \rm \, kpc$. 
As shown later, this approximately matches the location at which the shell starts propagating subsonically\footnote{Note that we have explicitly verified in our simulations that the shell always remains thin for solutions in which it always moves supersonically.} (see Fig.~\ref{solutions_analytic}). 
In this regime, the forward shock weakens and eventually propagates as a sound wave.
At $t \approx 300 \, \rm Myr$ for instance, the forward disturbance satisfies $\Delta \rho/\rho \sim 0.6$, clearly indicating that it can no longer be described as a strong shock.
Due to the shape of the Hernquist profile, note also that the density of the shell of shocked matter drops with distance from the black hole for both energy- and momentum-driven outflows.
This is important when we later consider the cooling of the shocked medium, which, due to its dependence on $\rho_{\rm gas}^2$, is particularly efficient in the central regions.

\begin{figure*}
\includegraphics[scale = 0.38]{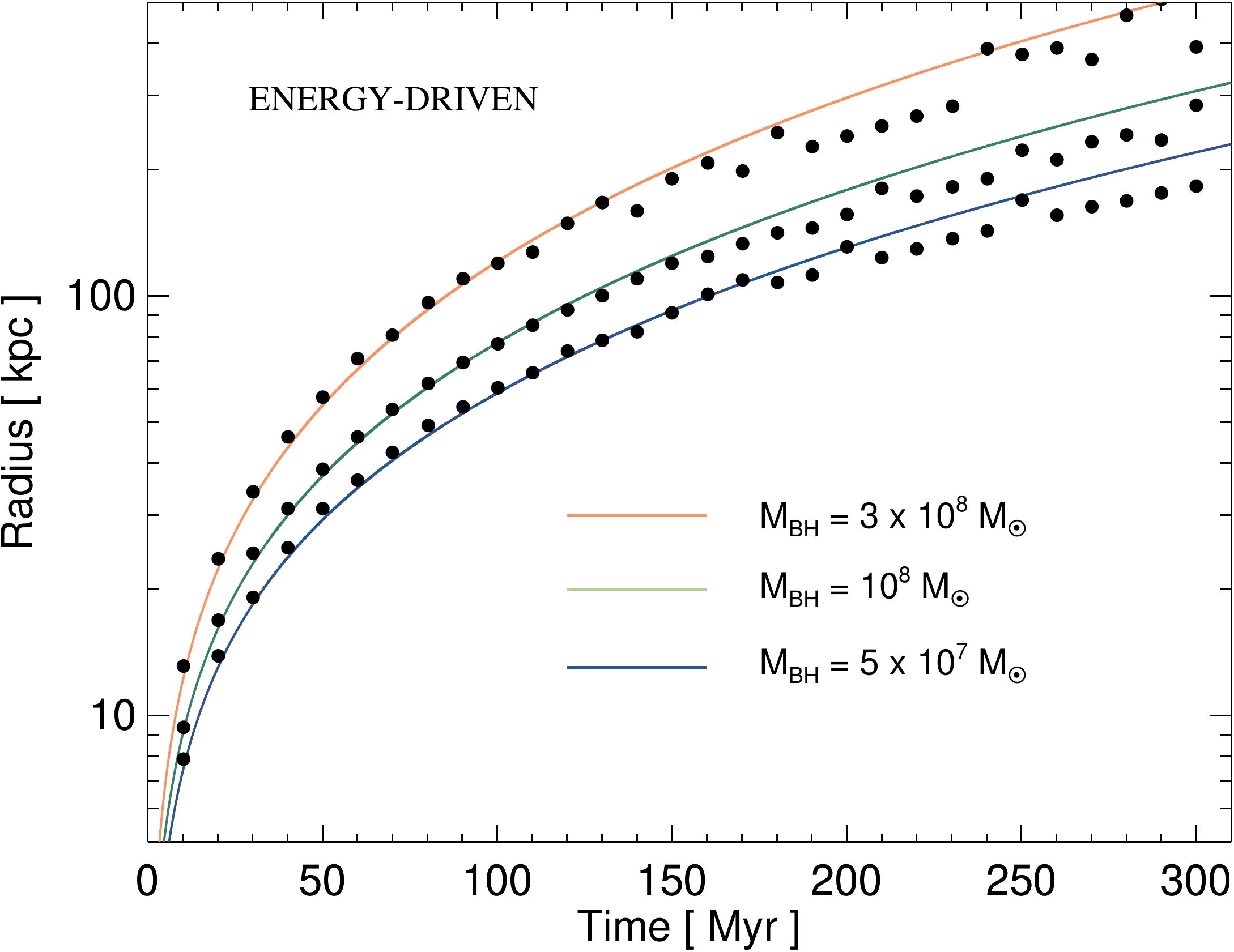}
\includegraphics[scale = 0.38]{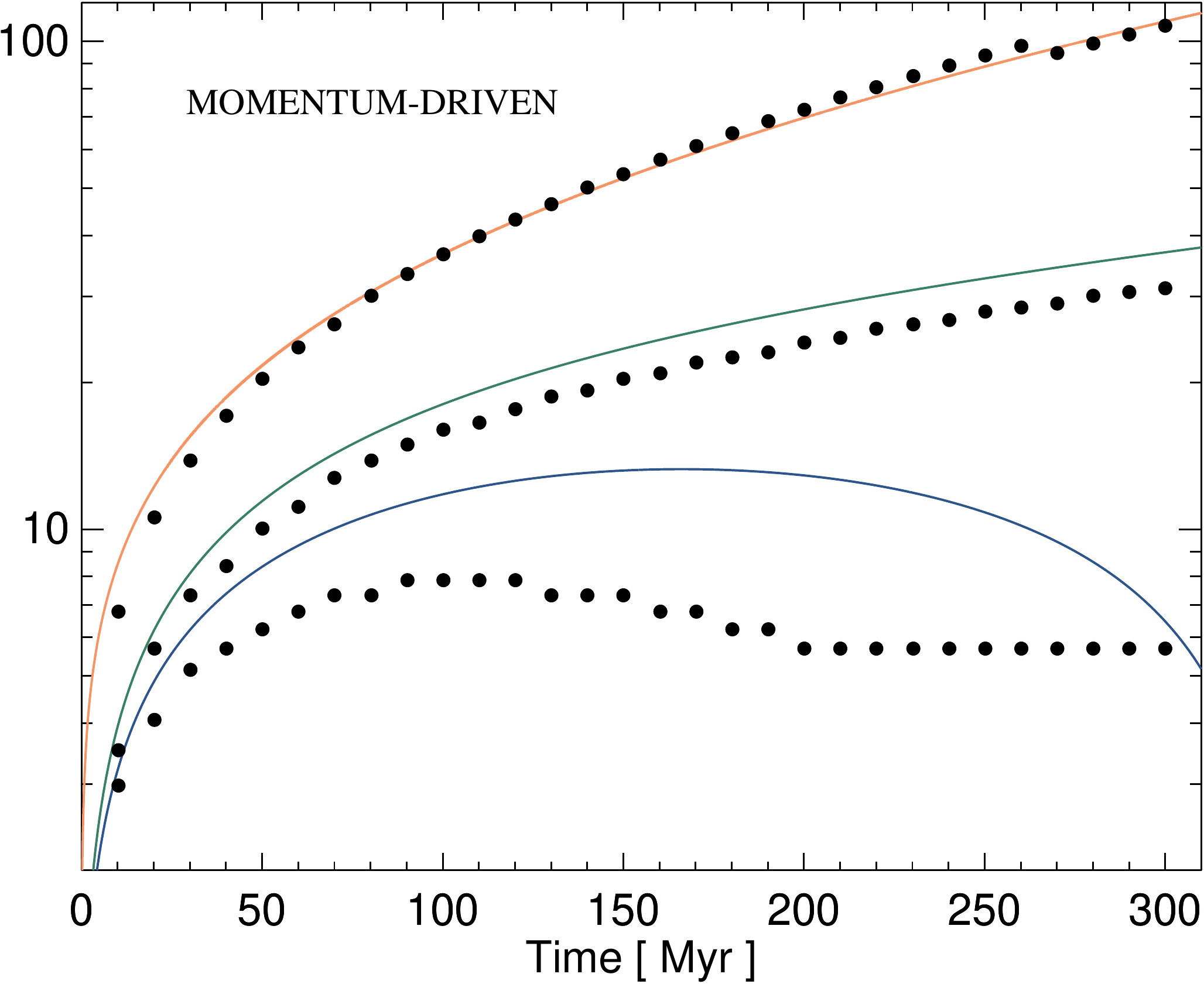}
\caption{Radius of the expanding shell as a function of time for the energy- and momentum-driven models for three different black hole masses (filled circles). Solid lines represent the corresponding analytical solutions. The close match between analytical and numerical results indicate that the chosen sub-grid models provide an adequate approximation to the energy- and momentum-driven solutions. For the momentum-driven solution, the agreement is poorer for lower mass black holes. The source of disagreement is the back-reaction of the ambient medium on the outflow as it enters the subsonic regime, which is accounted for in our hydrodynamical simulations but neglected in the equations presented in Section~\ref{enemomout}, which assumes the forward shock is strong (see text for more details).}
\label{trackshell}
\end{figure*}

In order to establish a close comparison of the predictions of our numerical models with those presented by \citet{King:03, King:05}, we solved Eq.~\ref{shellmom1} and Eq.~\ref{shellene1} numerically for $M_{\rm BH} \,=\, 5 \times 10^7 \,,\, 10^8\,,\, 3 \times 10^8 \, \rm M_\odot$ and a Hernquist halo as described in Section~\ref{hernhalo}.
We then tracked the location of the contact discontinuity as a function of time in our simulated outflows and compared it with predictions based on the equations described in Section~\ref{kingsmodel}.
The results of this comparison are shown in Fig.~\ref{trackshell}, where we plot the estimated location of the contact discontinuity in our various simulations as a function of time against the analytical predictions (solid lines) for a matching Hernquist halo and identical black hole masses. According to Fig.~\ref{trackshell}, the propagation of both energy- and momentum-driven outflows is in close agreement with analytical predictions throughout most of the simulation time. 
In energy-driven outflows, our simulations systematically underestimate the location of the contact discontinuity at later times when compared to analytical models.
This occurs as a consequence of the formation of Rayleigh-Taylor instabilities at the interface between the shocked inner wind and the shocked medium components \citep[see also][]{King:10b}.
Since the swept up material is no longer confined to a spherically symmetric shell, when tracking the contact discontinuity, our method preferentially picks up the location of overdense clumps that lag behind in the Rayleigh-Taylor unstable flow.

For the momentum-driven outflows shown on the right hand side in Fig.~\ref{trackshell}, agreement between numerical simulations and analytic solutions is less successful for the lower black hole masses, though close for $M_{\rm BH} \,=\, 3 \times 10^8 \, \rm M_\odot$.
In order to establish the source of this disagreement, we show the outflow velocity as a function of radius according to the various analytical solutions in Fig.~\ref{solutions_analytic}.
We also show the sound speed profile in the simulated Hernquist halo as a dashed line. 
For the momentum-driven solutions with $M_{\rm BH} \,=\, 5 \times 10^7 \, \rm M_\odot$ and $M_{\rm BH} \,=\, 10^8 \, \rm M_\odot$, the speed of the outflowing shell becomes comparable to the speed of sound at $r \,\approx\, 5 \, \rm kpc$ and $r \,\approx\, 15 \, \rm kpc$, respectively.
The assumption of a strong forward shock is however implicit in Eq.~\ref{shelleq}, which should otherwise not neglect the confining pressure of the ambient gaseous medium\footnote{Note that \citet{Faucher-Giguere:12} include it in their analytic treatment.} \citep[see e.g.][]{Weaver:77, Dyson:97}. 
As shown in Fig.~\ref{solutions_analytic}, the assumption of a strong forward shock is valid for all energy-driven solutions for the black hole masses considered but is unsuitable in the momentum-driven limit for black hole masses $M_{\rm BH} \lesssim 10^8 \, \rm M_\odot$. 
In this regime, the shock weakens (see Fig.~\ref{thinshell}) and the outflow becomes confined by the pressure of the ambient medium, causing it to decelerate.
Note, however, that this finding does not alter the core of the conclusions of \citet{King:03, King:05} qualitatively. 
Our results show that the shell of swept-up material still stalls when the black hole mass is lower than a critical value.

\begin{figure*}
\includegraphics[scale = 0.5]{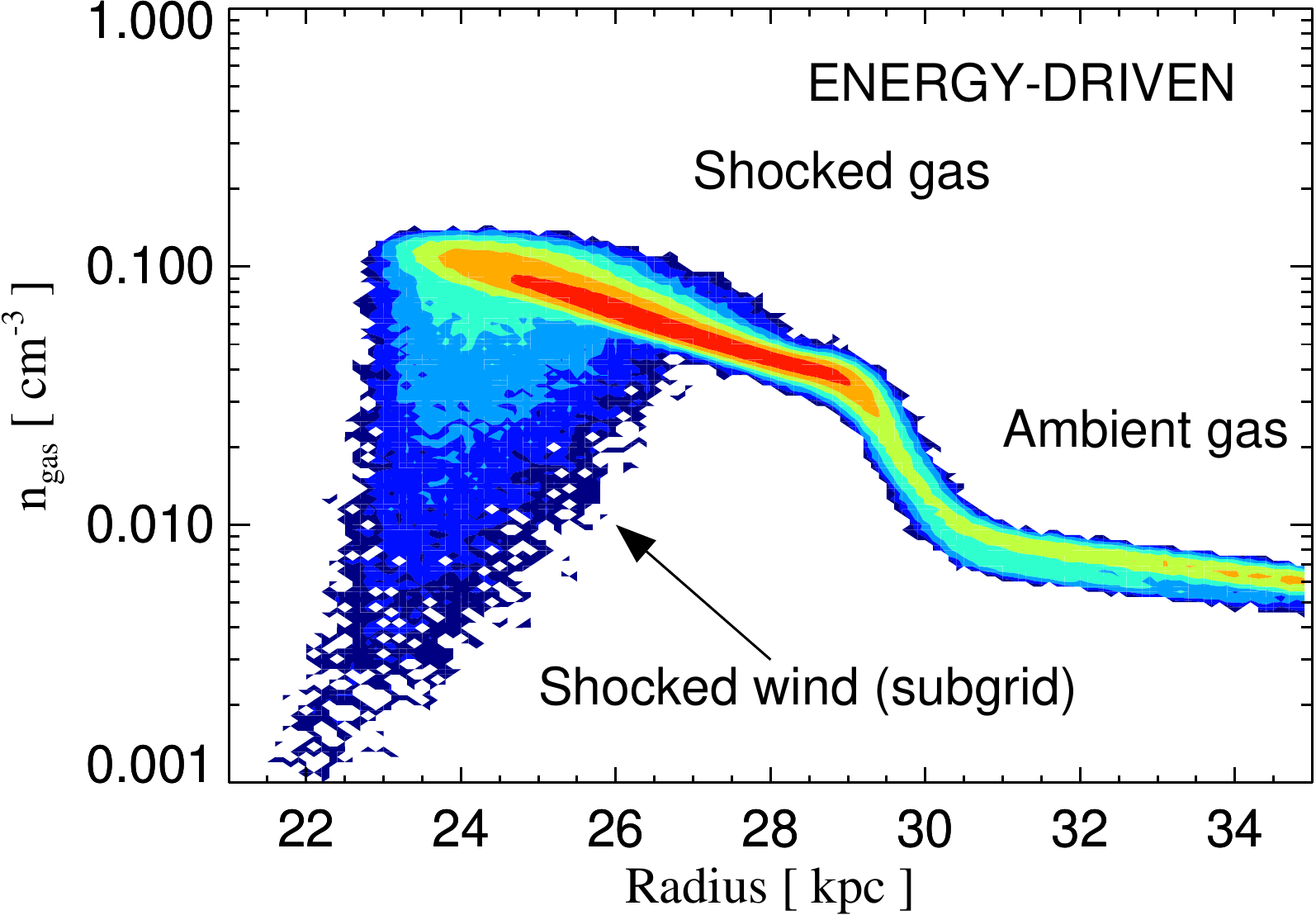}
\includegraphics[scale = 0.5]{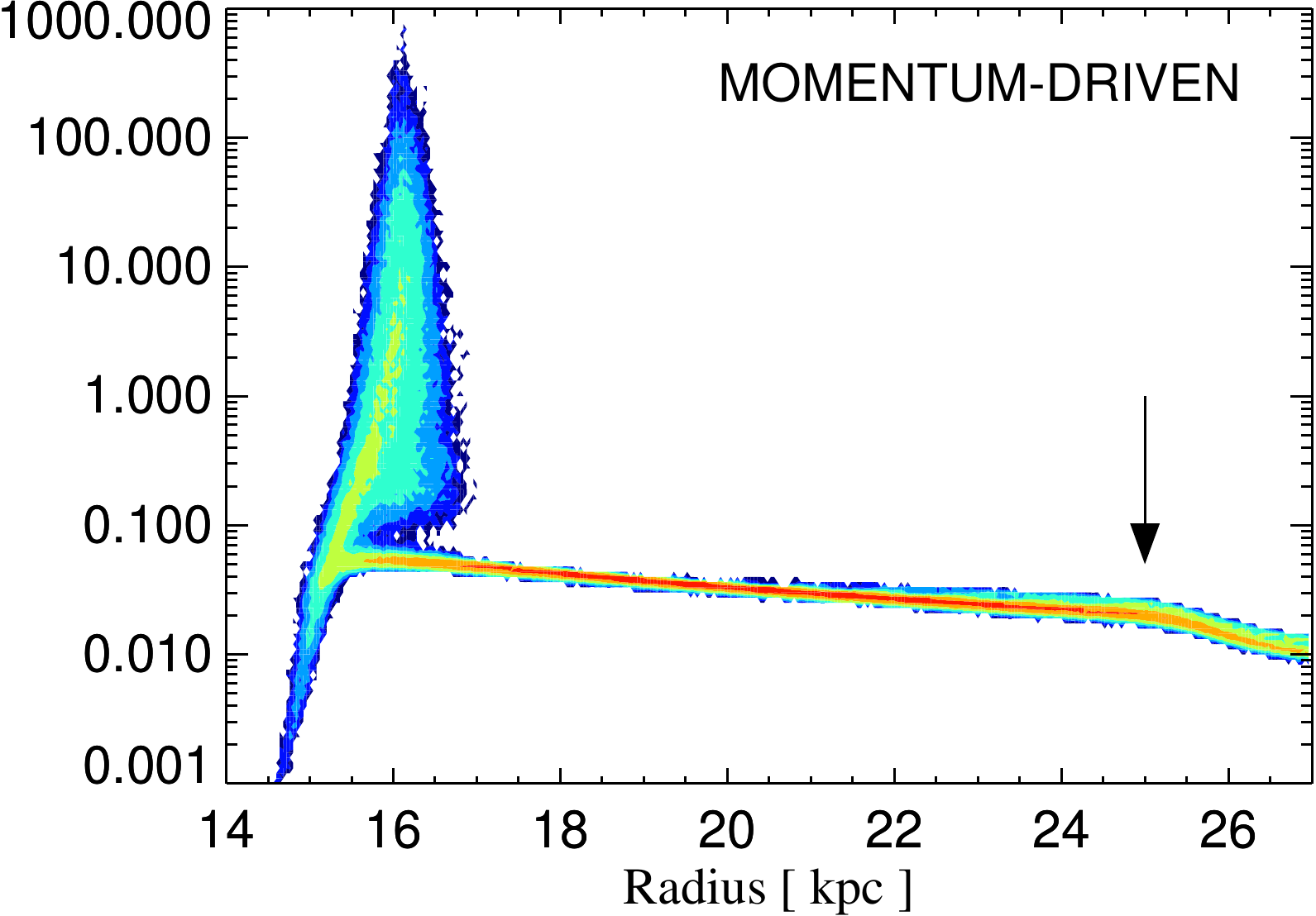}
\caption{The structure of the outflow shown as a two dimensional histogram of the gas density as a function of radius for the energy-driven outflow model at $t \,\approx\, 30.2 \, \rm Myr$ (left) and for the momentum-driven outflow at $t \,\approx\, 100.2 \, \rm Myr$ (right) for a black hole mass of $10^8 \, \rm M_\odot$. The outflow can be separated into three distinct zones: (a) a shocked wind phase which is modelled in a `subgrid' fashion by adding energy or momentum for energy- and momentum-driven outflows respectively, (b) a shocked medium phase, which is thick in the energy-driven outflow and (c) the ambient gas phase. In the innermost regions of the momentum-driven outflow, the shocked wind phase consists of a thin and highly compressed layer instead of a hot bubble, as in the energy-driven case. The arrow indicates the location of the forward (weak) shock.}
\label{shockfig}
\end{figure*}

The structure of the shock in our energy-driven and momentum-driven models is illustrated in Fig.~\ref{shockfig}, which shows two-dimensional histograms of gas density against radial distance from the black hole.
The colour coding reflects the local gas mass. 
Fig.~\ref{shockfig} shows that the flow patterns can be clearly separated into different phases.
For the energy-driven outflow, lower radii are sparsely populated by a diffuse hot component, which is generated in a `subgrid' fashion in our simulations by directly injecting energy into gas cells surrounding the black hole.
The adiabatic expansion of this component drives the forward shock containing the bulk of the outflow.
Note that it is because the shell of dense gas lies on top of the very low density hot bubble that Rayleigh-Taylor instabilities arise in the outflow (see Fig.~\ref{shellpanel}).
The shock structure looks very different for the momentum-driven outflow shown on the right hand side of Fig.~\ref{shockfig}, where the mildly shocked gaseous medium is separated from the inner void by a thin region of highly compressed gas at which momentum is being injected.
Because the shell lies on top of an even denser layer of gas (the `subgrid' realisation of the cooled shocked inner wind), no Rayleigh-Taylor instabilities arise in this case\footnote{We have checked that this is true for the entire duration of all our simulations for momentum-driven solutions.}.
This finding is in good agreement with the analytical arguments of \citet{King:10b}, which suggest that energy-driven shells should always be Rayleigh-Taylor unstable, whereas momentum-driven shells for black hole masses close or much higher than $M_{\rm \sigma}$ should be Rayleigh-Taylor stable.

\begin{figure*}
\includegraphics[scale = 0.5]{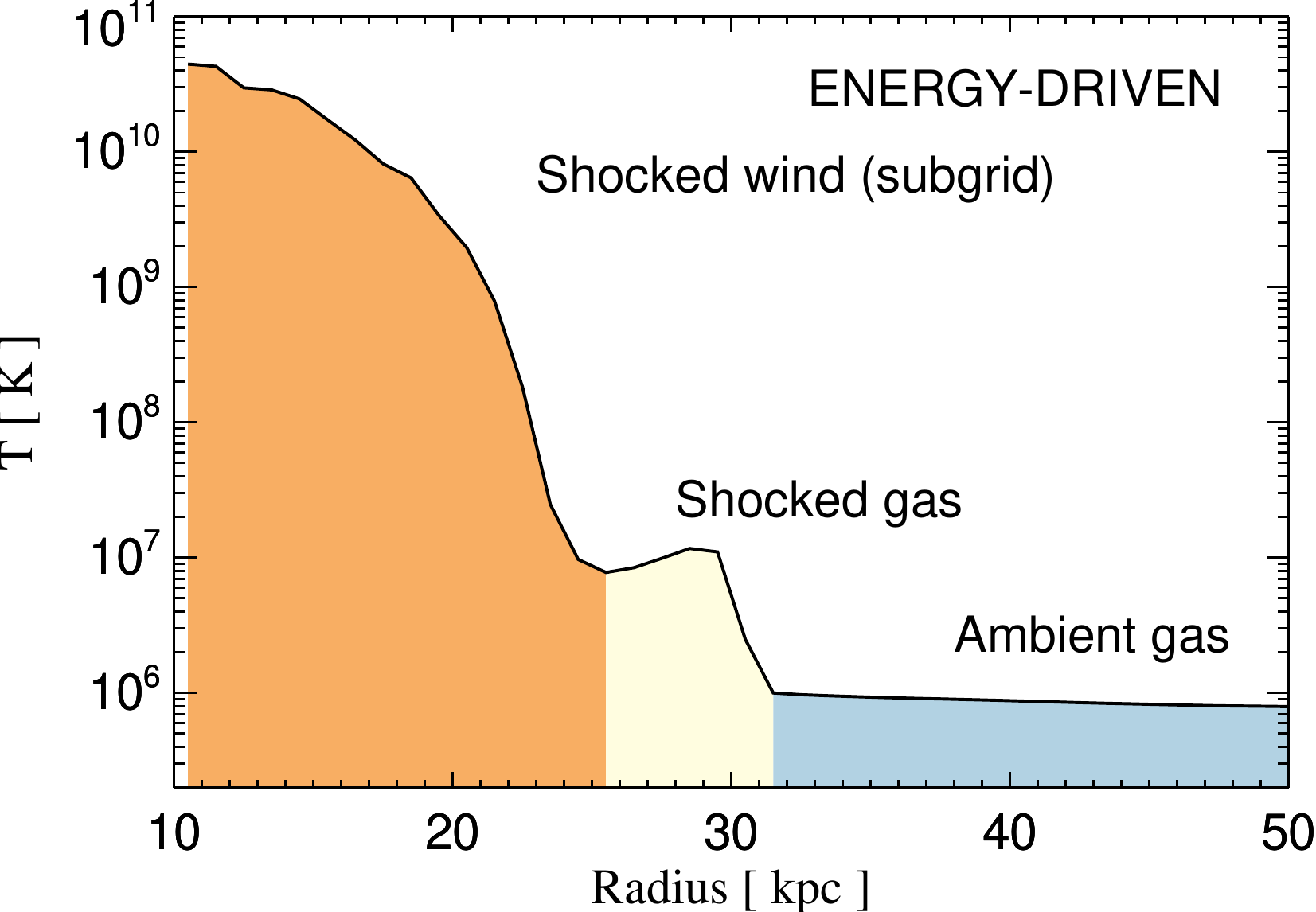}
\includegraphics[scale = 0.5]{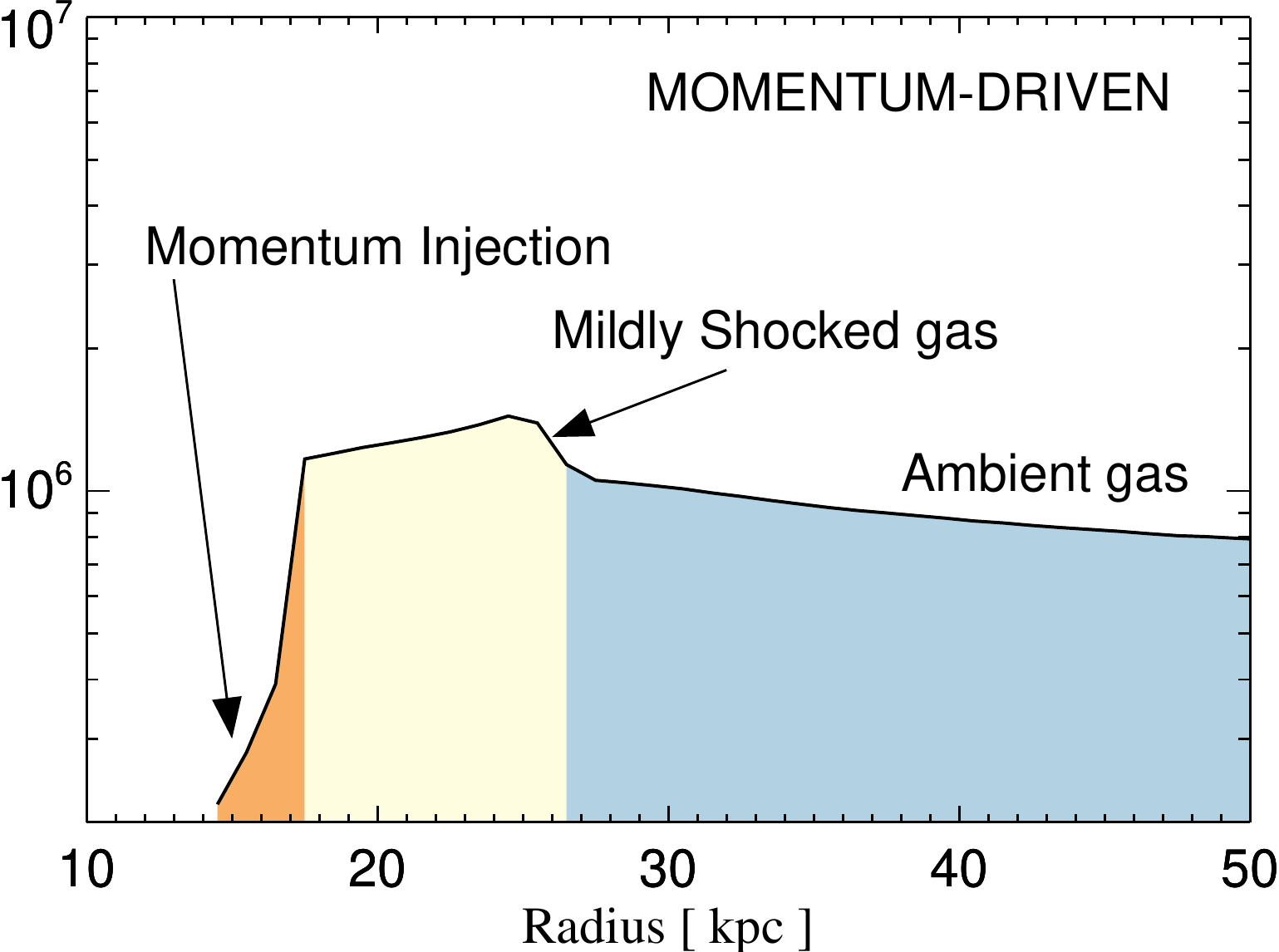}
\caption{Temperature profile of gas in the energy-driven outflow at $t \,\approx\, 30.2 \, \rm Myr$ (left) and in the momentum-driven outflow at $t \,\approx\, 100.2 \, \rm Myr$ (right) for a black hole with mass $10^8 \, \rm M_\odot$. The region directly affected by the `subgrid' model is shaded in orange. For energy-driven outflows, our `subgrid' model generates a very hot bubble that drives an outflow as it expands adiabatically, while for momentum-driven outflows, our `subgrid' model evacuates the central regions entirely by imparting velocity kicks to the nearest neighbouring gas cells of the central black hole. The temperatures of the different regions are an indicator of the cooling processes that are likely to dominate in each phase if radiative cooling is taken into account. The shocked wind gas in the energy-driven outflow should cool efficiently via inverse Compton cooling close to the AGN (see text), whereas the shocked medium component should only cool via standard two-body processes such as Bremsstrahlung or metal-line cooling.}
\label{outflow_temp}
\end{figure*}

Temperature profiles for energy- and momentum-driven outflows are shown in Fig.~\ref{outflow_temp}.
For the energy-driven outflow, three main regions can again be identified.
In order of increasing radial distance from the AGN, these are the shocked wind phase at temperatures of $\approx 10^{10} \, \rm K$, the shocked medium phase at temperatures of $\approx 10^7 \, \rm K$  \citep[in agreement with analytical estimates for an energy-driven outflow of speed $\sim 1000 \, \rm km \, s^{-1}$ as in][]{Zubovas:14} and the roughly isothermal ambient medium at the virial temperature $\lesssim 10^6 \, \rm K$.
For the momentum-driven outflow, the temperature of the mildly shocked gaseous medium increases only slightly and drops towards the centre of the halo as it becomes increasingly poorly populated.

\subsection{Cooling of the forward shock}
\label{seccooling}

Taking $n_{\rm e} \,\approx\, 0.05 \, \rm cm^{-3}$ (assuming a fully ionised primordial gas) and $T \,\approx\, 10^7 \, \rm K$ for the shocked medium phase as in our simulated energy-driven outflow, Eq.~\ref{freefree} gives a free-free cooling timescale of $t_{\rm ff} \,\approx\, 10^8 \, \rm yr$.
Only if the cooling time $t_{\rm cool}$ is comparable to the flow time $t_{\rm flow}$, can the shell of shocked material fragment into clumps of colder material.
Once the gas cools to $\approx 10^4 \rm \, K$ and hydrogen is no longer ionised, further cooling must proceed through molecular and metal cooling processes, not included in our numerical simulations.
The characteristic timescales for these processes are however expected to be very short \citep[$\sim 10\--100 \, \rm yr$, see][]{Zubovas:14} and validate the approximation taken in this study that gas at $\approx 10^4 \rm \, K$ traces molecular gas.
Observed molecular outflows appear to have momentum fluxes $\sim 20 \frac{L}{c}$ \citep{Sturm:11, Cicone:14}, suggesting that they must be interpreted as energy-driven within the picture in which AGN energy/momentum and interstellar gas couple hydrodynamically \citep{King:11, Faucher-Giguere:12, Zubovas:14}.
We accordingly focus solely on simulations of energy-driven outflows in this section.

Our simulation setup is identical to that described in Section~\ref{hernhalo}, but we now also include radiative cooling.
We assume the cooling function presented in \citet{Katz:96}, which is suitable for a primordial mixture of hydrogen and helium but probably underestimates the cooling of the shock-heated medium likely to be significantly enriched with metals and dust due to efficient star formation. 
The presence of metals and dust in the outflowing gas should greatly decrease the cooling times.
In order to bracket the likely efficiency of radiative cooling, we therefore run an additional simulation with the same cooling function, increased by a factor of $10$.
For temperatures in the range of $\sim 3 \times 10^4 \-- 10^7 \, \rm K$, the factor of $10$ increase results in a cooling function similar to that of gas enriched to solar metallicity \citep{Sutherland:93}.

\begin{figure}
\includegraphics[scale = 0.45]{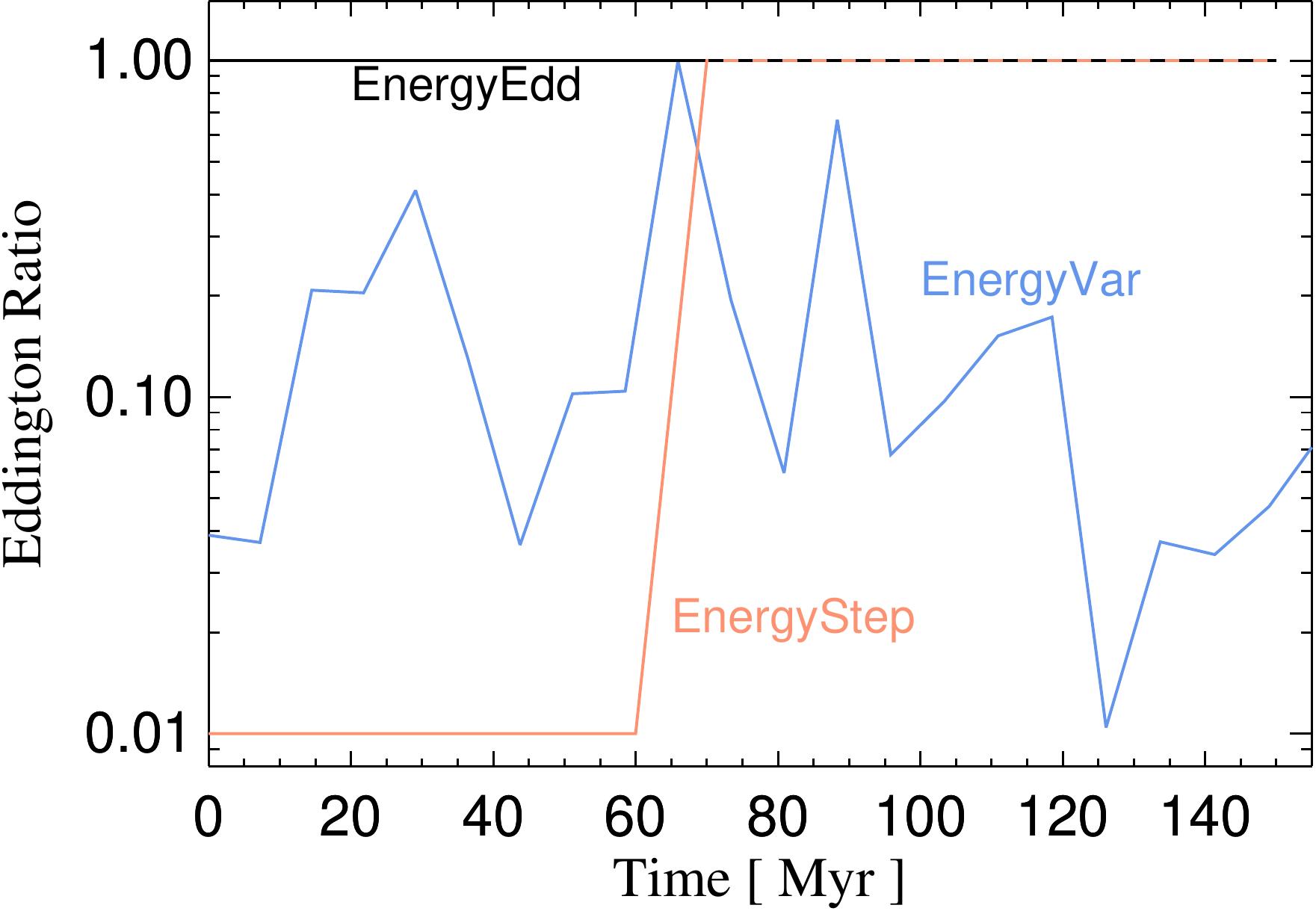}
\caption{Eddington ratio as a function of time for the AGN light curves considered: ($1$) the AGN emits at its Eddington rate throughout the entire simulation (black), ($2$) the AGN initially emits at a hundredth of its Eddington rate, but rises sharply to its Eddington limit about half-way through the simulation and ($3$) the AGN luminosity curve varies with time as predicted by cosmological simulations \citep{Costa:14}.}
\label{lightcurves}
\end{figure}

In our various numerical tests, we also explore relaxing the assumption that the AGN constantly emits at its Eddington luminosity.
The ratio $t_{\rm cool}/t_{\rm flow}$ is sensitive to the mass of the black hole and to the luminosity of the AGN that drives the outflow since $\dot{E} \,\propto\, L \,\propto\, M_{\rm BH}$.
In order to account for the (degenerate) effects of lowering the black hole mass and/or AGN luminosity, we consider three different AGN light curves as shown in Fig.~\ref{lightcurves}.
The first case assumes the AGN to emit at a constant luminosity $L_{\rm Edd}$ as before; in a second case we let the AGN luminosity rapidly rise from a value $L_{\rm Edd}/100$ to $L_{\rm Edd}$ and, in the third case, we adopt a variable light curve. 
In this last scenario, the chosen light curve is taken directly from a high resolution cosmological simulation that self-consistently follows black hole accretion \citep{Costa:14}. 
In the cosmological simulations, variability results from alternating accretion and outflow episodes, that respectively boost and suppress the AGN luminosity.
For brevity, we shall name the simulations as `EnergyEdd', `EnergyStep', `EnergyVar' and `EnergyEddCooling' for simulations of energy-driven outflows adopting the identically named lightcurves in Fig.~\ref{lightcurves} respectively. EnergyEddCool is identical to EnergyEdd, but is simulated using the cooling function increased by a factor of $10$.  
Note that in this section, we implicitly assume that an (unresolved) AGN inner wind is launched even if the Eddington ratio is low. We caution, however, that it is unclear whether an AGN inner wind can be launched if $L < L_{\rm Edd}$ for instance if the driving mechanism is radiation pressure on UV lines \citep[e.g.][]{Proga:04, Proga:05}.

\begin{figure}
\includegraphics[scale = 0.43]{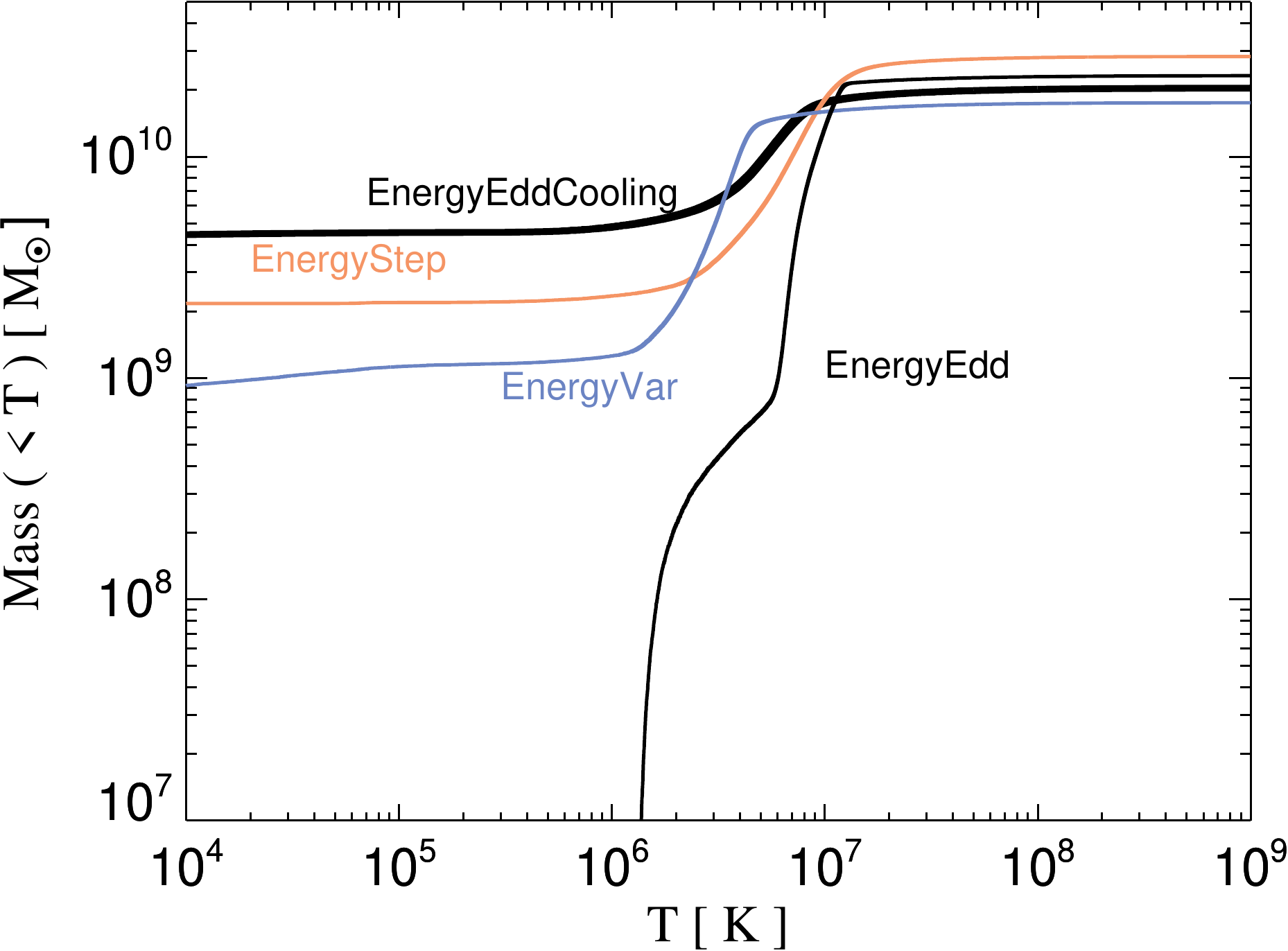}
\includegraphics[scale = 0.43]{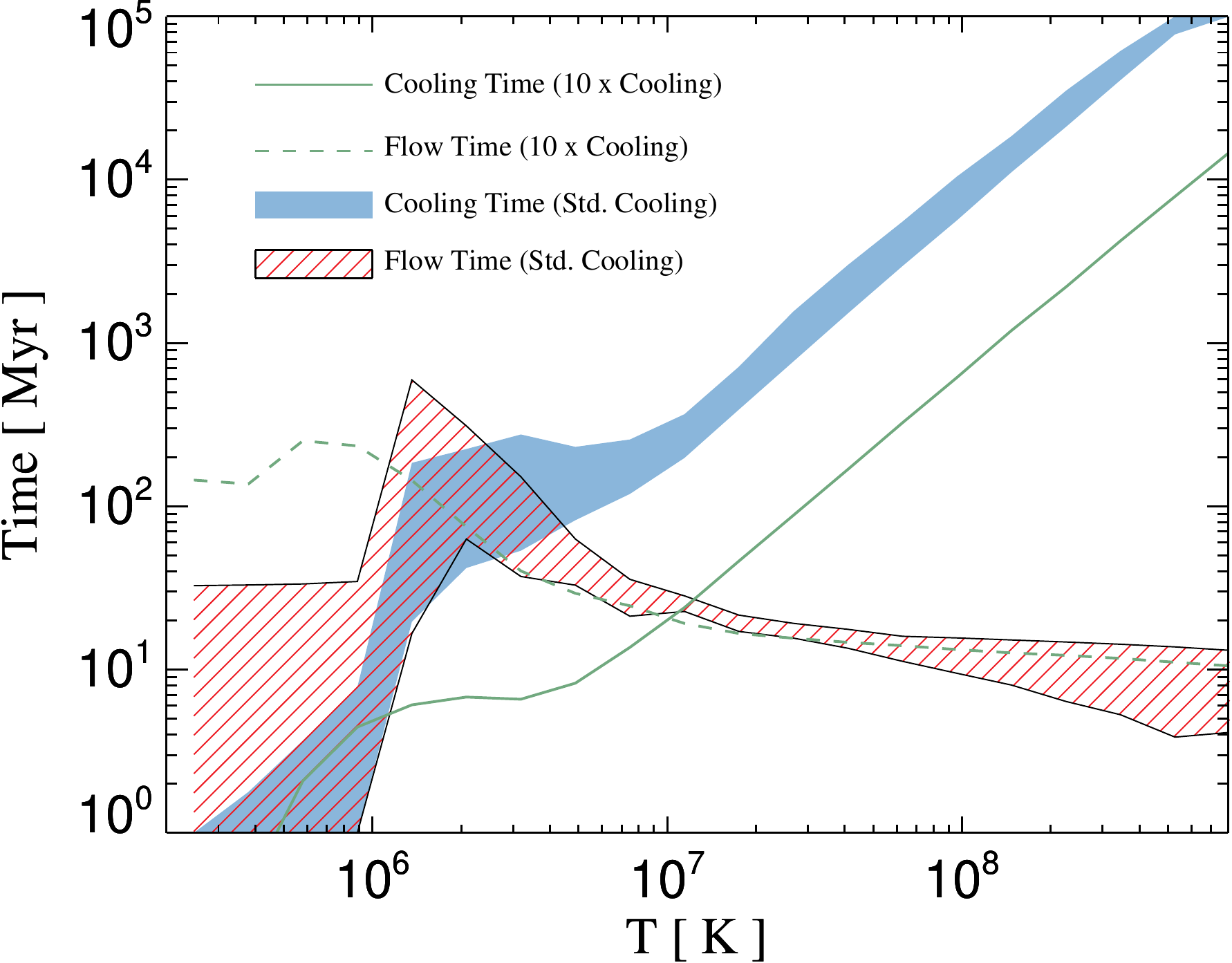}
\caption{The origin of multi-phase structure in outflowing material for energy-driven outflows with radiative cooling. The plot at the top shows the cumulative mass in outflowing gas as a function of temperature in the innermost $20 \, \rm kpc$ of simulations EnergyEdd, EnergyStep, EnergyVar (primordial cooling) and EnergyEddCooling ($10$ times more efficient cooling) for the light curves shown in Fig.~\ref{lightcurves} at a time where the bulk of the outflow is located at about $10 \, \rm kpc$ from the AGN. Large quantities of cold gas ($M_{\rm cold} \,\gtrsim\, 10^9 \, \rm M_\odot$) at $T \,\leq\, 5 \times 10^4 \, \rm K$ form for EnergyStep, EnergyVar and EnergyEddCooling, where cooling times are comparable with the flow times. The plot at the bottom shows the flow time (estimated as $= \, r/v_{\rm r}$) (solid lines) and the cooling time $= \, E_{\rm th}/\Lambda$ as a function of temperature in the different simulations. Since they yield similar results, the region enclosed by the maximum and minimum cooling times and maximum and minimum flow times are shaded in blue and by red lines respectively for simulations EnergyEdd, EnergyStep and EnergyVar. Green lines show the cooling (solid) and flow (dashed) times for EnergyEddCooling.}
\label{coolingprops}
\end{figure}

\begin{figure*}
\centering \includegraphics[scale = 0.7]{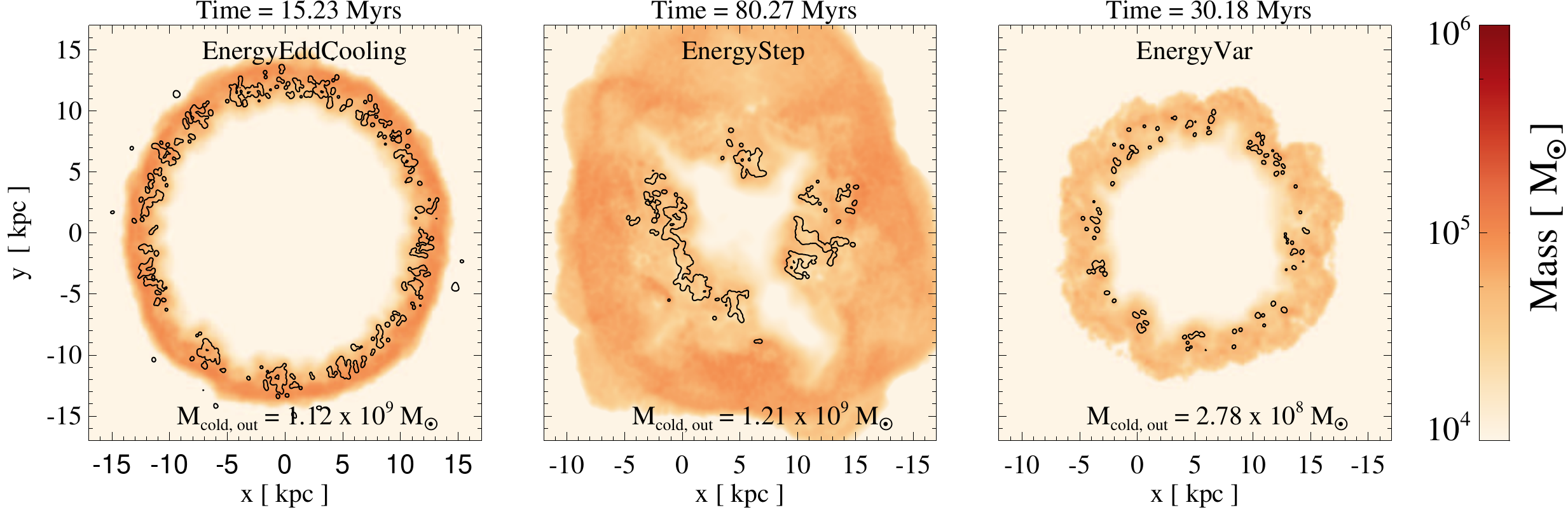}
\caption{Projection of gas mass along a slab of thickness $5 \, \rm kpc$ and length $R_{\rm vir}$ for three different simulations using the energy-driven outflow model and the different light curves presented in Fig.~\ref{lightcurves}. Note that due to the absence of any significant amount of outflowing cold gas in the simulation using standard cooling, here we show results for the run with the cooling function $10$ times increased. We show the output for snapshots corresponding to the time at which the outflow has reached radii of about $10 \, \rm kpc$ in each simulation. Black contours enclose regions of outflowing cold gas with $T \, \leq \, 5 \times 10^4 \, \rm K$. In all cases, large amounts of cold gas form in the outflow.}
\label{panel_cold}
\end{figure*}

\begin{table*}
\centering
\setlength{\tabcolsep}{5pt}
\begin{tabular}{lcccccccc}
\toprule 
Simulation & time & $L/L_{\rm Edd}$ & $M_{\rm cold} $ & $\langle v\rangle$ & $v_{\rm max}$ & $\dot{M}$  & $\dot{p}/(L/c)$ & $\dot{E}_{\rm k}/L$ \\
           & $\rm [Myr]$ &  & $\rm [M_{\odot}]$ & $[\mathrm{km \, s^{-1}}]$ & [$\mathrm{km \, s^{-1}}]$ & $\mathrm{[M_{\odot} \, yr^{-1}]}$ & & \\
\\ \midrule 
EnergyEddCooling  & $15.23$ & $1$     & $4.5 \times 10^9$ & $680.9$ & $907.4$ & $2474.0$  & $26.0$ & $0.026$  \\ 
EnergyStep        & $80.27$ & $1$     & $2.2 \times 10^9$ & $685.0$ & $707.1$ & $724.1$ & $4.7$  & $0.003$ \\ 
EnergyVar         & $30.18$ & $0.37$  & $1.1 \times 10^9$ & $340.1$ & $462.1$ & $260.1$ & $3.3$  & $0.002$ \\ 
\bottomrule
\end{tabular}
\caption{Properties of entrained cold gas ($T \,\leq\, 5 \times 10^4 \, \rm K$) in simulations EnergyEddCooling, EnergyStep and EnergyVar at a time when the bulk of the outflow is at a distance of about $10 \, \rm kpc$ from the AGN. From left to right, we provide values for the simulation time, the Eddington ratio of the AGN, followed by the total mass, the average speed, maximum speed, the outflow rate, the momentum flux normalised to $L/c$ of the AGN and the kinetic luminosity normalised to $L$ for outflowing cold gas. Results for simulation EnergyEdd are not included, because there is no cold gas present in this simulation at any time. The highest outflow rate, momentum boost and kinetic luminosity is obtained for simulation EnergyEddCooling, where cooling is most efficient, while AGN energy input is also very high.}
\label{table1} 
\end{table*}

Fig.~\ref{coolingprops} illustrates the impact of radiative cooling in our four numerical experiments.
At the top of the panel, we show the cumulative mass of outflowing gas in the innermost $20 \, \rm kpc$ of the halo as a function of temperature, at a time when the outflowing shell is located at a distance of about $10 \, \rm kpc$ from the AGN. 
For simulation EnergyEdd (thin black line), the lowest temperature in outflowing gas is about $10^{6} \rm \, K$, two orders of magnitude above the minimum temperature achievable due to primordial cooling.
In this particular case, it does not seem possible to cool the shocked medium gas to temperatures at which molecules can form and the outflow is solely composed of hot ionised gas.
However, in all other numerical tests, large amounts ($\gtrsim 10^9 \, \rm M_\odot$) of gas with $T \lesssim 5 \times 10^4 \, \rm K$ form.
We here term as `cold' any gas with temperatures $T < 5 \times 10^4 \, \rm K$, though none of our conclusions is very sensitive to this threshold as most of the `cold gas' has in fact reached the temperature floor of $10^4 \, \rm K$.
As we have seen, gas at such temperatures would likely cool rapidly to much lower temperatures.
Due to the very high cooling rates for gas in the shocked medium phase, simulation EnergyEddCooling produces $4 \times 10^9 \, \rm M_\odot$ of $10^4 \, \rm K$ gas in the outflow.
Large amounts of cold material ($\approx 2 \times 10^9 \, \rm M_\odot$) also form in simulation EnergyStep, where less efficient energy input from the AGN leads to a lower outflow speed and hence a lower $t_{\rm cool}/t_{\rm flow}$.
Finally, simulation EnergyVar, which has an identical cooling function as simulations EnergyEdd and EnergyStep, but an AGN luminosity intermediate between these two, correspondingly contains less cold material than EnergyStep but more than EnergyEdd.
Note also that for every curve shown at the top of Fig.~\ref{coolingprops}, there is a characteristic temperature at which the cumulative mass rises very sharply. In order of increasing temperature, this occurs first for run EnergyVar, followed by EnergyEddCooling, EnergyStep and finally EnergyEdd. 
This sharp rise in mass traces the uncooled component of the shell of swept-up interstellar gas.
Its temperature is sensitive to the outflow speed, which determines the temperature of the gas that passes the forward shock, and on the efficiency of cooling.

At the bottom of Fig.~\ref{coolingprops}, we explore the efficiency of cooling by plotting both $t_{\rm cool}$ and $t_{\rm flow}$ as function of outflowing gas temperature for all four simulations.
We shade the region enclosed by the maximum and minimum cooling times in blue and the region enclosed by the maximum and minimum flow times in red for simulations EnergyEdd, EnergyStep and EnergyVar. Results for simulation EnergyEddCooling are shown as green lines as labelled on the plot.
Note that for all simulations, cooling times rise extremely rapidly with increasing temperature.
In all cases, $t_{\rm cool} \geq 10^9 \rm \, yr$ for gas with $T \geq 10^8 \, \rm K$.
The hot bubble that drives the outflowing shell is therefore largely unaffected by cooling and the interpretation of the `subgrid' model as realising the effects of a shocked wind phase remains valid.
Cooling is efficient whenever $t_{\rm cool} \lesssim t_{\rm flow}$ and this can be seen to occur for simulations with standard cooling for outflowing gas at temperatures of $\approx 1 \-- 5 \times 10^6 \, \rm K$. For simulation EnergyEdd, both timescales become comparable at $T \,\approx\, 10^6 \, \rm K$, just above the virial temperature, while most outflowing gas has $T \,\approx\, 10^7 \, \rm K$ (see Fig.~\ref{coolingprops}).
Note also the slight bump in the flow time for all simulations at $T \,\approx\, 10^6 \, \rm K$ for EnergyEdd, EnergyStep and EnergyVar and at $5 \times 10^5 \, \rm K$ for EnergyEddCooling.
In all cases, this traces the gas in the ambient medium which is currently passing through the forward shock. 
Whereas in the simulations with only primordial cooling, this gas is still at the virial temperature (see Fig.~\ref{outflow_temp}) and the bump therefore occurs at $T \,\approx\, 10^6 \, \rm K$, in EnergyEddCooling, cooling is so efficient that it affects even the less dense ambient medium. The temperature of the gas that passes the forward shock in this case is therefore slightly lower, explaining the behaviour seen in Fig.~\ref{coolingprops}.

Fig.~\ref{panel_cold} illustrates that the distribution of cold gas in the outflow is clumpy.
In this panel, the location of cold clouds is shown as black contours plotted against the outflowing gas mass projected along a thin slab of thickness $5 \, \rm kpc$ for simulations EnergyEddCooling, EnergyStep, EnergyVar at a time when the outflow has reached approximately $10 \, \rm kpc$ in spatial extent.
Gas preferentially cools in the over-densities seeded by Poisson noise in the initial conditions of the gaseous halo, which leads to the formation of the observed clumpy structure.
Note that a departure from spherical symmetry in the outflow is present in all cases.
This is particularly evident in EnergyStep, where the very low luminosity of the AGN has not been able to drive gas to more than $5 \rm \, kpc$ away from the black hole, while the shocked medium has expanded adiabatically, leading to the diffuse appearance of the hot outflowing component.
The presence of Rayleigh-Taylor instabilities, now accentuated by cooling in the shocked medium phase, is also clearly visible in Fig.~\ref{panel_cold} as hot and cold gas mix in the interface between shocked medium and shocked wind.
Fig.~\ref{panel_cold} also shows the mass of cold gas found in the projected slab. 
In agreement with our previous discussion as well as Fig.~\ref{coolingprops}, the largest quantities of cold outflowing gas are found in EnergyEddCooling, followed by EnergyStep and EnergyVar.

We quantify various properties of the simulated energy-driven (cold) outflows in Table~\ref{table1}. 
Even though average outflow speeds are in the range $\approx 340 \-- 680 \, \rm km \, s^{-1}$, maximum speeds can reach $\sim 900 \rm \, km\,s^{-1}$ for sufficiently high AGN power, as is the case for EnergyEddCooling. 
In this case, outflow rates are very high ($\sim 2500 \, \rm M_\odot \, yr^{-1}$), momentum-boost factors reach $\approx 26 \frac{L}{c}$ (from Eq.~\ref{momboost}, this value is similar to what would be expected by taking $\epsilon_{\rm th-k} \,\approx\, 0.5$) and kinetic luminosities are $\approx 0.03 L$, in reasonable agreement with observational estimates for outflows driven by AGN powered by black holes of similar mass \citep[e.g.][]{Cicone:14}. 
Note than in order to obtain simultaneously high momentum boost factors and high kinetic luminosities, both high outflow rates and high outflow speeds are required.
In simulations EnergyStep and EnergyVar, outflow speeds are high ($\approx 700 \, \rm km \, s^{-1}$), but outflow rates are much lower than in EnergyEddCooling (only $\approx 260 \-- 700 \, \rm M_\odot \, yr^{-1}$). 
Both the momentum-boost and the kinetic luminosity are comparatively low in these cases ($\dot{p}\,\approx\, 3 \-- 5 L_{\rm Edd}/c$ and $\dot{E}_{\rm k}\,\approx\,0.002 \--0.003 L$).
In particular, the relative inefficiency of the energy-driven outflow in EnergyStep despite the high AGN power ($L \,=\, L_{\rm Edd}$) highlights the importance of efficient cooling. 
In this particular case and in general, metal-line cooling is probably crucial in generating quantities of cold material entrained in the outflow as large as observed.
More efficient cooling leads to higher cold outflow rates and to higher momentum-boosts and kinetic luminosities in improved agreement with observed outflows.

\section{Cosmological simulations}
\label{seccosmo} 

\subsection{Feeding black holes from the cosmic web}

In order to achieve efficient growth, supermassive black holes require a steady inflow of gas into their haloes at a high rate. 
Large amounts of gas can be brought in to the central galaxy as a result of a merger with another galaxy or as a consequence of filamentary inflow from the cosmic web \citep{DiMatteo:08, Sijacki:09, DiMatteo:12, Khandai:12, Dubois:12, Dubois:13, Costa:14, Feng:14}.
Inflowing material slows down the propagation of outflows \citep[see e.g.][]{Nayakshin:10} as additional work to overcome the inflow is required and will reduce the efficiency of AGN feedback.
Note also that a large portion of the ISM with which the luminous energy from the accreting black hole couples is cold unlike in the simple cases considered in Sections~\ref{enemomout} and \ref{secnumimp} and is distributed anisotropically around the AGN. 
Cold gas cools radiatively more efficiently and can lead to lower conversion efficiency between AGN luminosity and outflow kinetic energy, whereas an anisotropic mass distribution will cause the efficiency of feedback to be direction dependent.
A full description of the growth and evolution of supermassive black holes and the propagation of outflows launched by AGN activity therefore ultimately requires a detailed account of their cosmological context. 
In this section, we do this by incorporating the `subgrid' prescriptions that successfully reproduce the \citet{King:03, King:05} models of energy- and momentum-driven winds in their idealised settings as presented in Section~\ref{secnumimp} into fully cosmological simulations. 
In order to highlight complications introduced by the cosmological environment of AGN, we compare the outflows driven in the cosmological simulations with those launched in an isolated halo with a gravitational potential equivalent to the spherically averaged (cosmological) halo.
We assume a $\Lambda$CDM cosmology with a set of parameters identical to \citet{Costa:14}, i.e. $h \,=\, 0.73$, $\Omega_{\rm m}^{\rm 0} \,=\, 0.25$, $\Omega_{\Lambda}^{\rm 0}\,=\,0.75$, $\Omega_{\rm b}^{\rm 0} \,=\, 0.041$ and $\sigma_{\rm 8} \,=\, 0.8$. Due to the cosmological nature of the simulations, units are given in comoving coordinates in this section. 
Units of distance are accordingly prefixed with  `c' for `comoving'.

\subsection{Numerical Setup}

The numerical setup of the cosmological simulations presented in this study is very similar to that of \citet{Sijacki:09} and \citet{Costa:14} and here we only briefly review its main features.
We employ output from the Millennium simulation \citep{Springel:05b}, which follows the collisionless dynamics of dark matter in a cubic cosmological volume of side length $500 h^{-1} \, \rm Mpc$, in order to generate initial conditions for a QSO hosting halo at $z\,\approx\,6$.
The selected halo\footnote{The chosen halo corresponds to region `O$6$' in \citet{Costa:14}.} grows to a virial mass $M_{\rm 200} \,\approx\, 3 \times 10^{12} h^{-1} \, \rm M_\odot$ by $z \,\approx\, 6$ and is thus a suitable host for a high redshift QSO \citep{Sijacki:09, Angulo:12, Costa:14}.
The zoom-in technique \citep[see][for a more detailed description]{Sijacki:09} is adopted in order to simulate a small region centred on the selected halo at high resolution.
Resolution is gradually degraded with radial distance from the selected halo in order to accurately represent the large-scale cosmological tidal field, while maintaining computational efficiency.
The high resolution region is then populated by gas cells and is evolved from a starting redshift $z \,=\, 127$ down to $z \,=\,6.2$.

The hydrodynamics of gas is followed using {\sc AREPO}.
Gas in our simulations is assumed to consist of an optically thin primordial mixture of hydrogen ($76 \%$) and helium ($24 \%$). 
The N-body dynamics of dark matter is now followed self-consistently using a {\sc TreePM} algorithm. 
In order to isolate the effects of AGN-driven outflows from other feedback effects,  we deliberately limit the physical processes included in our simulations to radiative cooling \citep[according to][]{Katz:96} and star formation \citep[following][]{Springel:03}. 
Star-forming gas is thereby assumed to lie on an effective equation of state that accounts for the (self-regulated) balance of cold cloud formation via a thermal instability and subsequent evaporation in a hot component pressurised by supernova explosions. The resulting pressure gradients lead to dynamical stabilisation of the star forming gas, but do not trigger outflows.
Note further that supernova-driven outflows are not incorporated in our simulations. Supernova-driven outflows have nevertheless been found not to launch efficient outflows from the very deep potential well of the quasar host halo. Instead, as they suppress star formation in less massive galaxies surrounding the QSO, supernova-driven outflows generate a surplus of infalling material that causes more efficient black hole growth and hence more dominant AGN-driven outflows in the QSO host halo \citep[see][]{Costa:14}.
In order to avoid resorting to further `subgrid' models, black hole accretion is also not followed explicitly. 
Instead we seed a $10^8 h^{-1} \, \rm M_\odot \,\approx\, 1.37 \times 10^8 \, \rm M_\odot$ black hole sink particle at the centre of the most massive halo when this reaches a mass of $M_{\rm 200} \,=\, 10^{12} h^{-1} \, \rm M_\odot$ at $z \,\approx\, 7.8$.
Our choice of black hole mass is in close agreement with that predicted by self-consistently following black hole growth from a $10^5 h^{-1} \, \rm M_\odot$ seed black hole in cosmological simulations for the same halo and at the same redshift \citep{Sijacki:09, Costa:14}.
Note that in a $M_{\rm 200} \,\approx\, 10^{12} \, \rm M_\odot$ halo at this redshift, gas cooling is very efficient and the bulk of the baryons cannot form a quasi-static `atmosphere'.
At the time of black hole seeding, the galaxy thus has a stellar mass of $\approx 2.2 \times 10^{10} h^{-1} \, \rm M_\odot$ and a star formation rate of $\approx 149.7 \, \rm M_\odot \, yr^{-1}$ (estimated within the galaxy's half mass radius). 
We estimate the $1$D velocity dispersion of star particles within the half mass radius of the galaxy hosting the black hole to be $\approx 168.3 \, \rm km \, s^{-1}$. 
For this velocity dispersion, a black hole mass of $1.37 \times 10^8 \, \rm M_\odot$ agrees well with the expected value if the galaxy were to lie on the observed $M_{\rm BH} \-- \sigma$ relation \citep[$\approx 1.5 \times 10^8 \, \rm M_\odot$, ][]{Kormendy:13}.
The black hole mass is then kept fixed at this value for the entire duration of the simulations.
We set the dynamical mass of the black hole to a value hundred times the mass of the dark matter particle ($m_{\rm DM} \,=\,6.75 \times 10^6 h^{-1}\, \rm M_\odot$) in order to prevent it from unphysically oscillating around the centre of the halo.
Note that the position of the black hole particle is then simply used as a tracer of the location at which AGN feedback energy or momentum is to be injected. 
We then carried out separate simulations employing either energy- or momentum-driven outflow models as described in Section~\ref{secnumimp}.

\begin{figure*}
\centering 
\includegraphics[scale = 0.6]{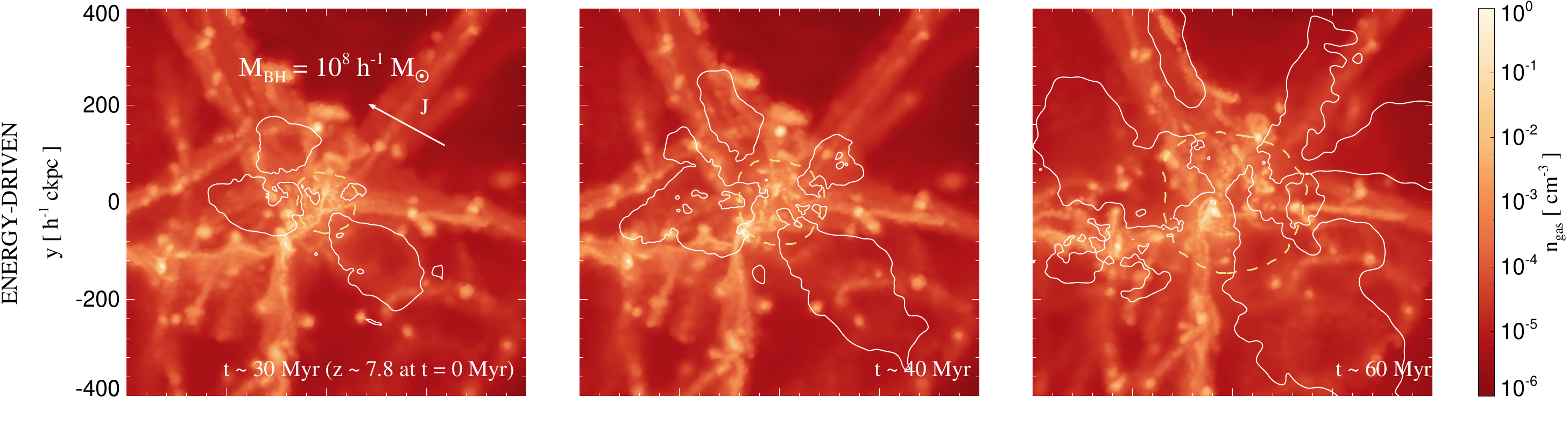}
\includegraphics[scale = 0.6]{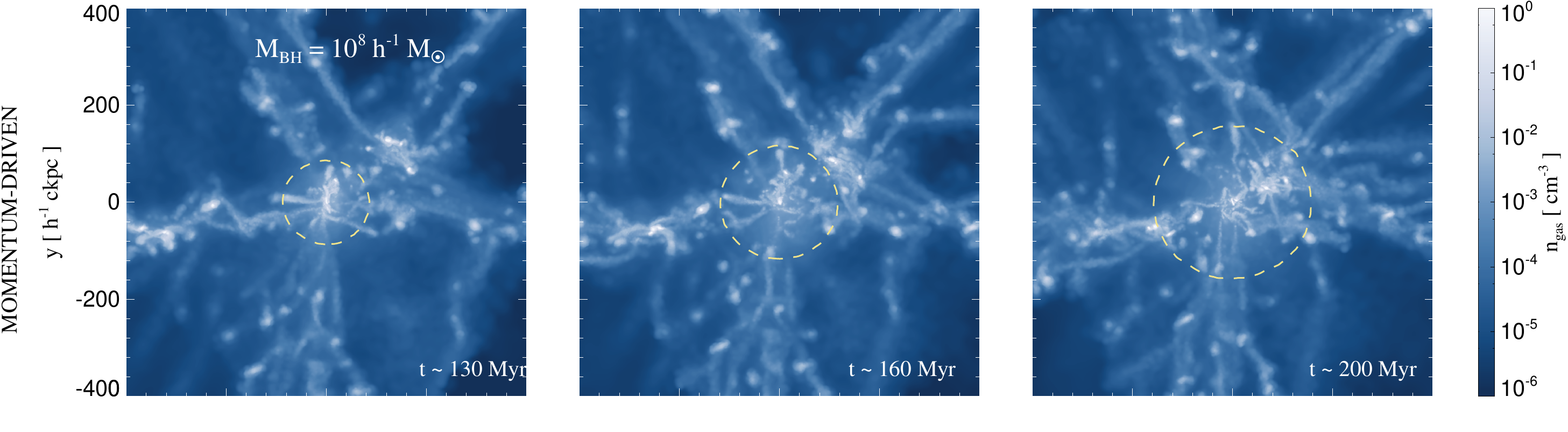}
\includegraphics[scale = 0.6]{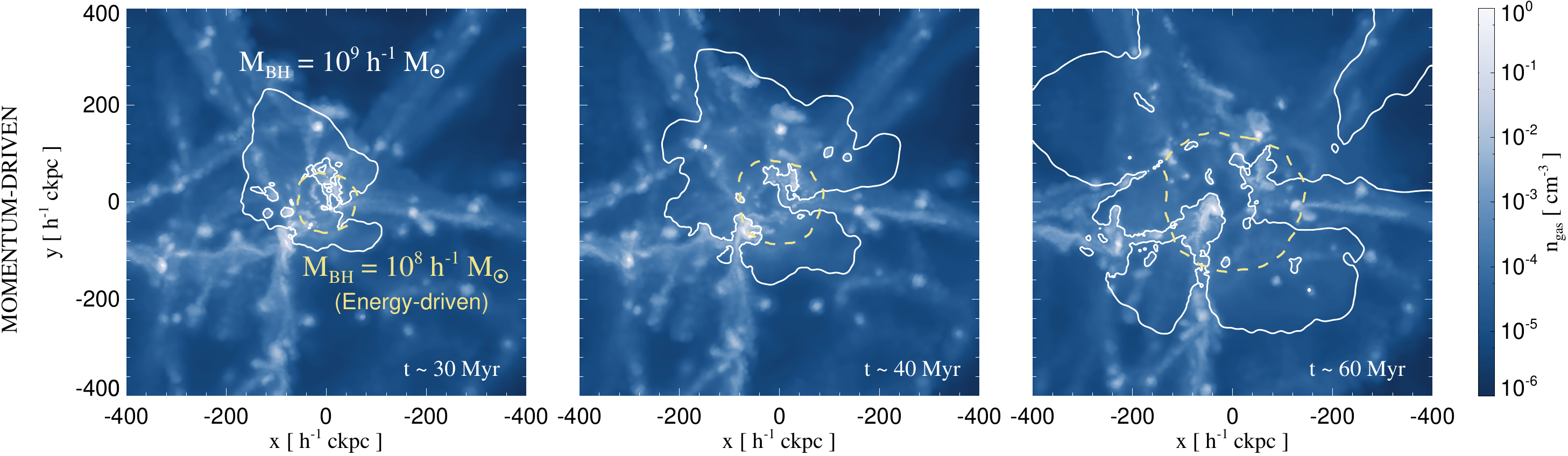}
\caption{The formation and evolution of an energy- (top row) and momentum-driven (two bottom rows) outflows in cosmological simulations. The AGN is powered by a black hole with $M_{\rm BH} \,=\, 10^8 h^{-1} \, \rm M_\odot$ ($M_{\rm BH} \,=\, 10^9 h^{-1} \, \rm M_\odot$ in the bottom row) and lies at the centre of a $\approx 10^{12} h^{-1}\, \rm M_\odot$ halo. The background rendering illustrates the gas density projected along a slab with thickness of $40 h^{-1} \, \rm ckpc$. White lines show contours for the radial velocity of outflowing gas in the cosmological simulations, whereas yellow dashed lines show identical contours for the outflow in an identical spherically averaged halo simulated in isolation. Contour levels are $\approx 880 \, \rm km \, s^{-1}$, $\approx 385 \, \rm km \, s^{-1}$ and  $\approx 880 \, \rm km \, s^{-1}$ for the top, middle and bottom panels respectively. Outflows driven in cosmological simulations for energy-driving with $M_{\rm BH} \,=\, 10^8 h^{-1} \, \rm M_\odot$ and momentum-driving with $M_{\rm BH} \,=\, 10^9 h^{-1} \, \rm M_\odot$ have a comparable spatial extent and are highly anisotropic when compared to the spherical outflows in the isolated haloes. Depending on direction, cosmological outflows propagate to lower or larger distances from the AGN than isolated halo outflows. In the case of momentum-driving with a $10^8 h^{-1} \, \rm M_\odot$ black hole, no outflow is observed in the cosmological simulation, whereas clearly present in the isolated halo simulation, clearly showing that a momentum input $\gg L_{\rm Edd}/c$ is required to launch large-scale momentum-driven outflows in cosmological environments.} 
\label{cosmo_outflow}
\end{figure*}

\begin{figure*}
\includegraphics[scale = 0.5]{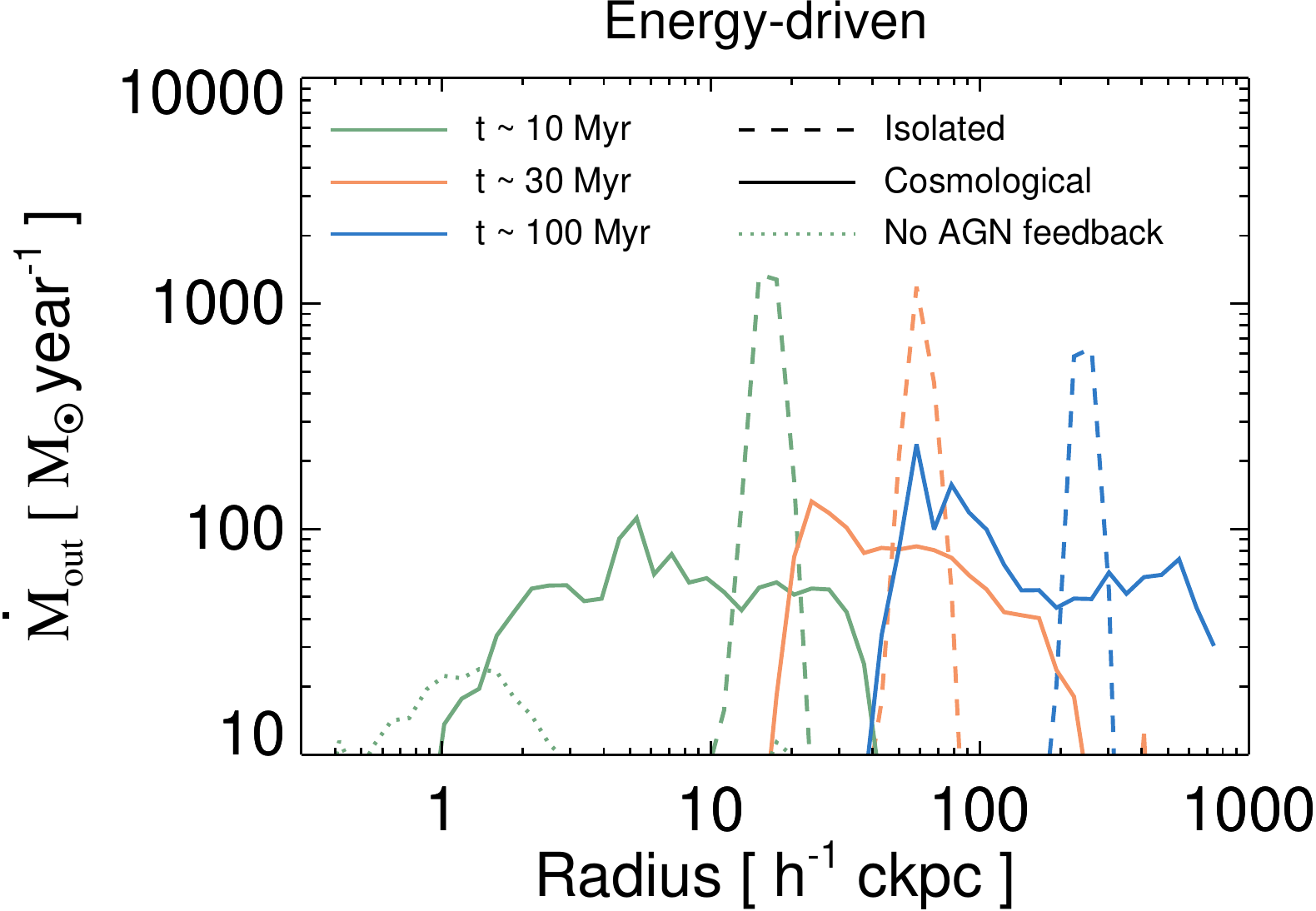}
\includegraphics[scale = 0.5]{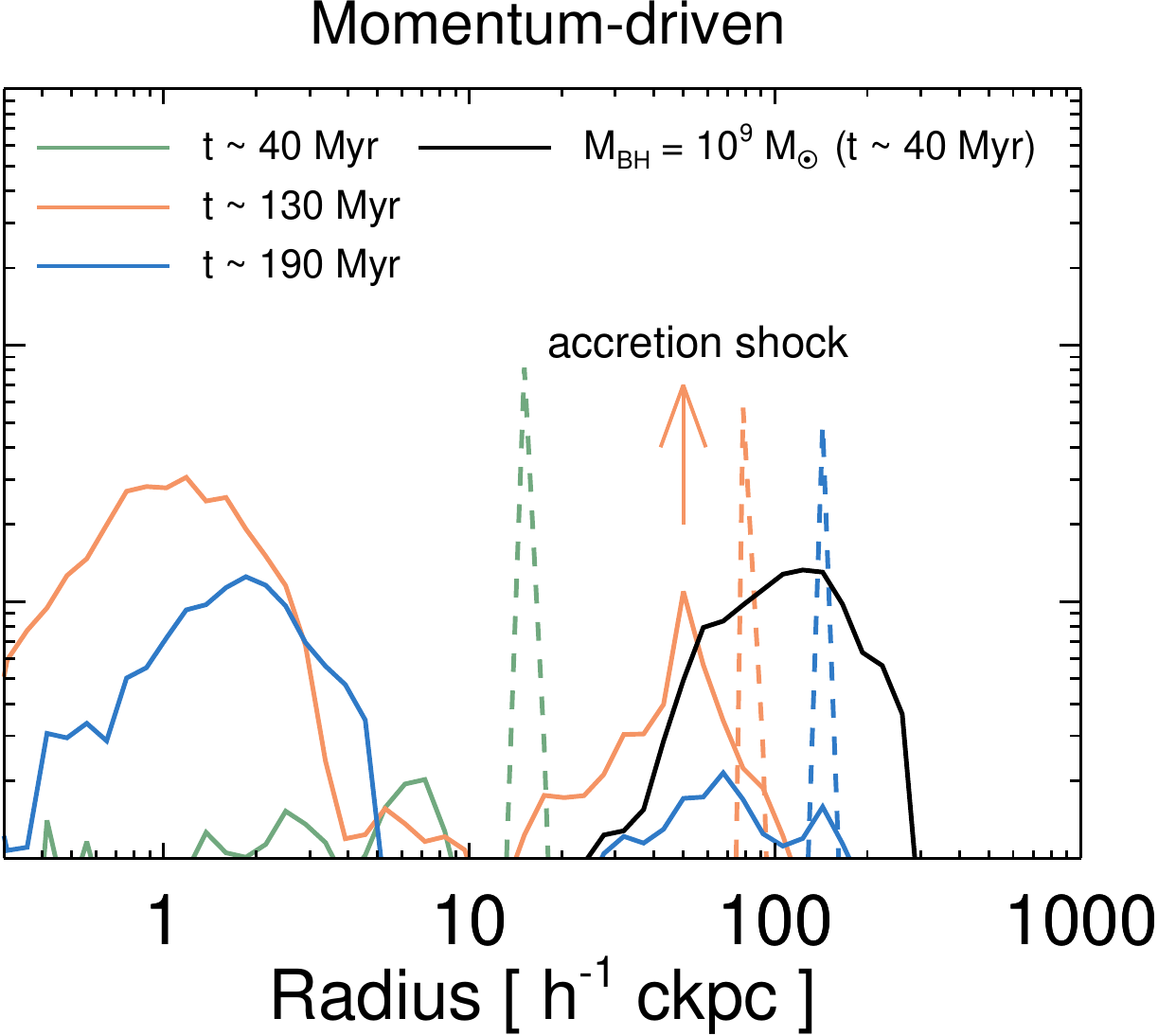}
\caption{Mass outflow rate profiles per logarithmic unit distance for the AGN in isolated halo simulations (dashed lines) and in cosmological simulations (solid lines) at different output times (colour-coded as labelled). Note that all the quantities are given in comoving units. Outflows are confined to an expanding shell in the isolated halo for both the energy- and momentum-driven cases. For the energy-driven outflow in the cosmological simulation, the outflow is spatially extended, but roughly located at similar distance. Note that no such outflow is present in the cosmological simulation of a momentum-driven outflow powered by a $M_{\rm BH} \,=\, 10^8 h^{-1} \, \rm M_\odot$ black hole, in contrast with the isolated halo case. Substantially higher masses (here shown for $M_{\rm BH} \,=\, 10^9 h^{-1} \, \rm M_\odot$ as a black solid line) are required to launch large-scale momentum-driven outflows with $\dot{p} \,=\, L_{\rm Edd}/c$. The enhancement in outflow rates seen for the momentum-driven outflow at $r \,\approx\, 50 h^{-1} \, \rm ckpc$ is due to an accretion shock and occurs simultaneously at the same spatial scales in simulations with and without AGN feedback.} 
\label{mrate_plots}
\end{figure*}

Initial conditions for the isolated halo were obtained by considering the spherically averaged profiles of dark matter, stars and gas in the cosmological halo at the time the black hole particle is seeded in its centre.
Mass density profiles were then separately produced for the gas component and for the collisionless components (dark matter and stars).
The collisionless component was treated as a static potential, as in the static halo models discussed in Sections~\ref{secnumimp} and \ref{seccomp}.
The gaseous component, whose distribution was obtained by sampling the (non-analytical) density profile, was further assumed to be in hydrostatic equilibrium.
Subsequently, a $10^8 h^{-1} \, \rm M_\odot$ black hole was seeded at the centre of the isolated halo and its position and mass were fixed for the duration of the simulations.
Note that since the density profiles used to construct initial conditions for the isolated halo were produced using the comoving units adopted in cosmological simulations, all quantities output in our isolated halo simulations must therefore also be interpreted as being given in comoving units.
For the gas particle mass and gravitational softening length in our isolated halo simulations, we adopt values identical to those taken for the cosmological simulations \citep[see][]{Costa:14}, i.e. $m_{\rm gas} \,=\, 1.32 \times 10^6 h^{-1}\, \rm M_\odot$ and $\epsilon_{\rm soft} \,=\, 1 h^{-1} \, \rm ckpc$ respectively. 
In {\sc AREPO}, the gravitational softening length is adaptive and is set by the cell size.
The typical minimum cell size in our cosmological simulations is $\sim 70 h^{-1} \rm \, pc$.
Simulations including radiative cooling and either energy- or momentum AGN feedback were then performed for this halo separately.
Note that our simulations for an isolated halo here differ from the simulations presented in Section~\ref{seccomp} only in resolution, the non-analytical nature of the potential and the fact that gas no longer exactly follows dark matter.

\subsection{Comparison with spherical isolated haloes}

Fig.~\ref{cosmo_outflow} shows a time sequence of the density field for our cosmological simulations for energy- (top panel) and momentum-driven (two bottom panels) outflow models. 
In contrast with the spherically symmetric galactic halo, the density field is very anisotropic and most matter is concentrated in thin filaments with an overall configuration that changes little over the time span covered in Fig.~\ref{cosmo_outflow}.
Each plot is centred on the galactic disc\footnote{This is only true at early times, since the outflow completely disrupts the disc after some time.} hosting the black hole, which is seen edge-on at $t \,\sim\, 20 \, \rm Myr$ as indicated by the projected angular momentum vector shown in the top left corner of the top panel.
The host galaxy is located at a prominent region of the density field at about the centre of the over-density and is continuously fed by filamentary infall of gas.
It is also in close proximity to another massive galaxy (about $100 h^{-1} \, \rm ckpc$ south-east at $t \,\sim\, 20 \, \rm Myr$) with which it eventually merges.
The location of the AGN host galaxy in a collapsing over-density favours the presence of copious amounts of gas that can fuel black hole accretion and is what ultimately enables the formation of a very massive black hole \citep{Sijacki:09, DiMatteo:12, Costa:14}.

\begin{figure*}
\includegraphics[scale = 0.5]{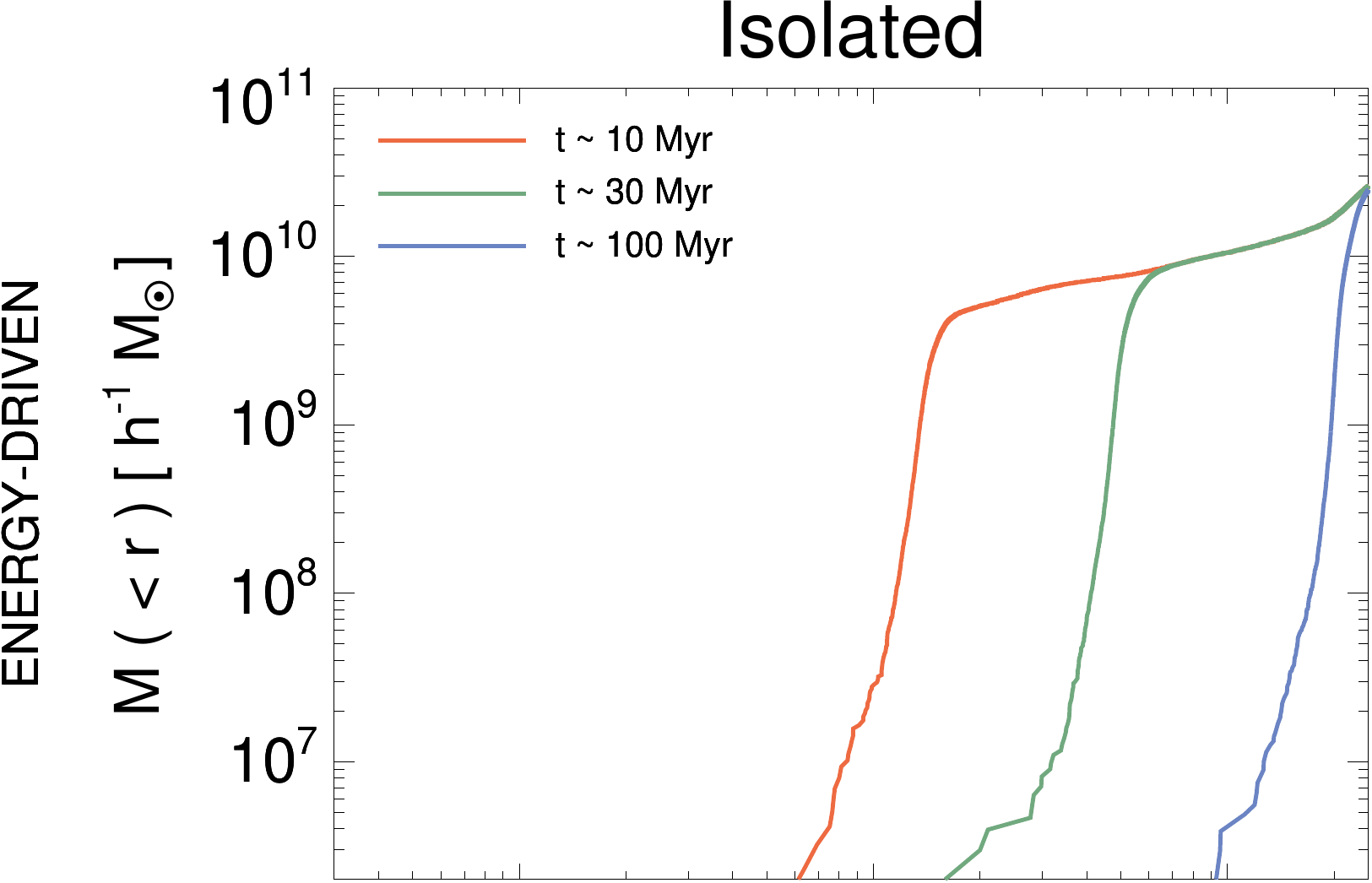}
\includegraphics[scale = 0.5]{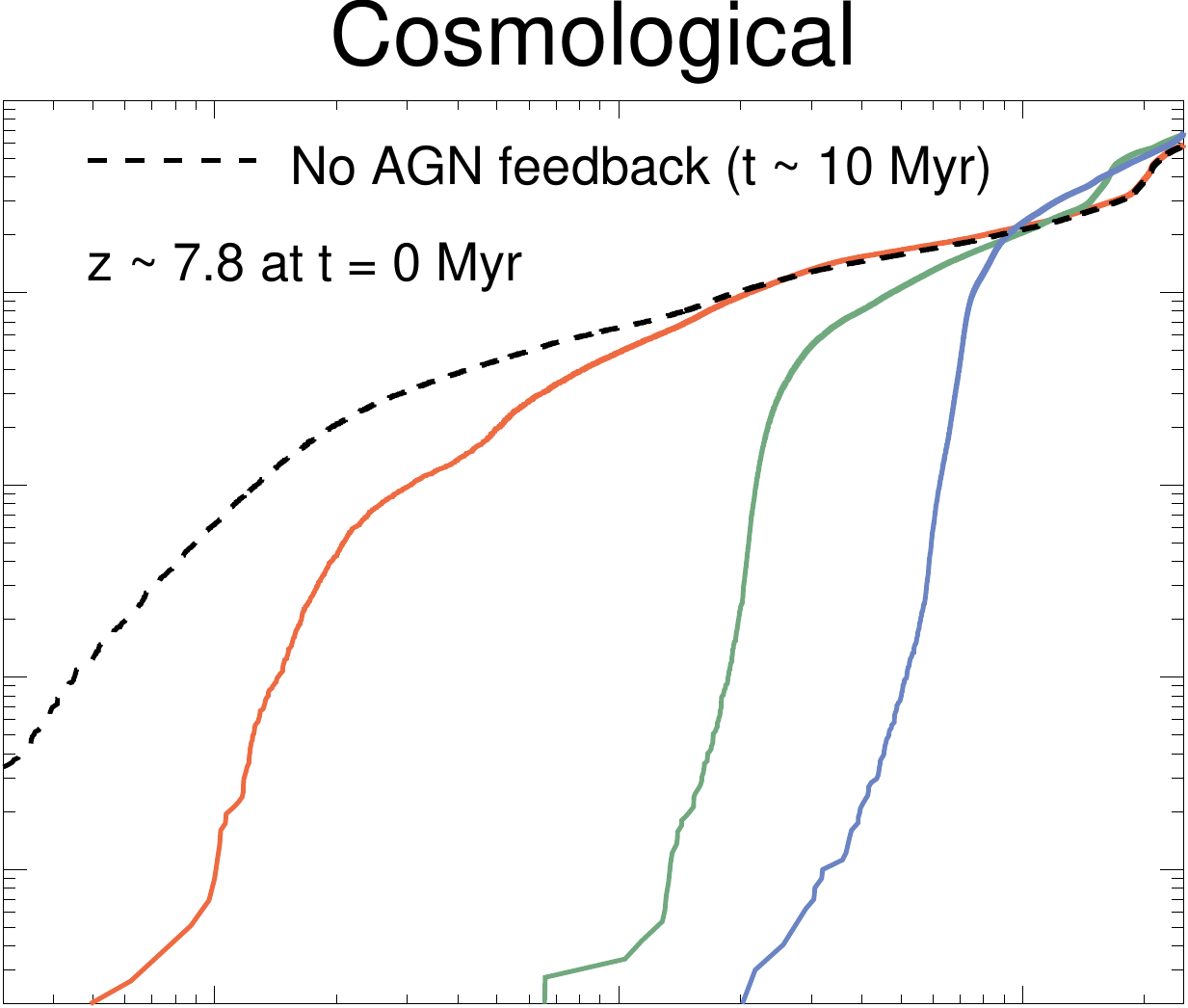}

\includegraphics[scale = 0.5]{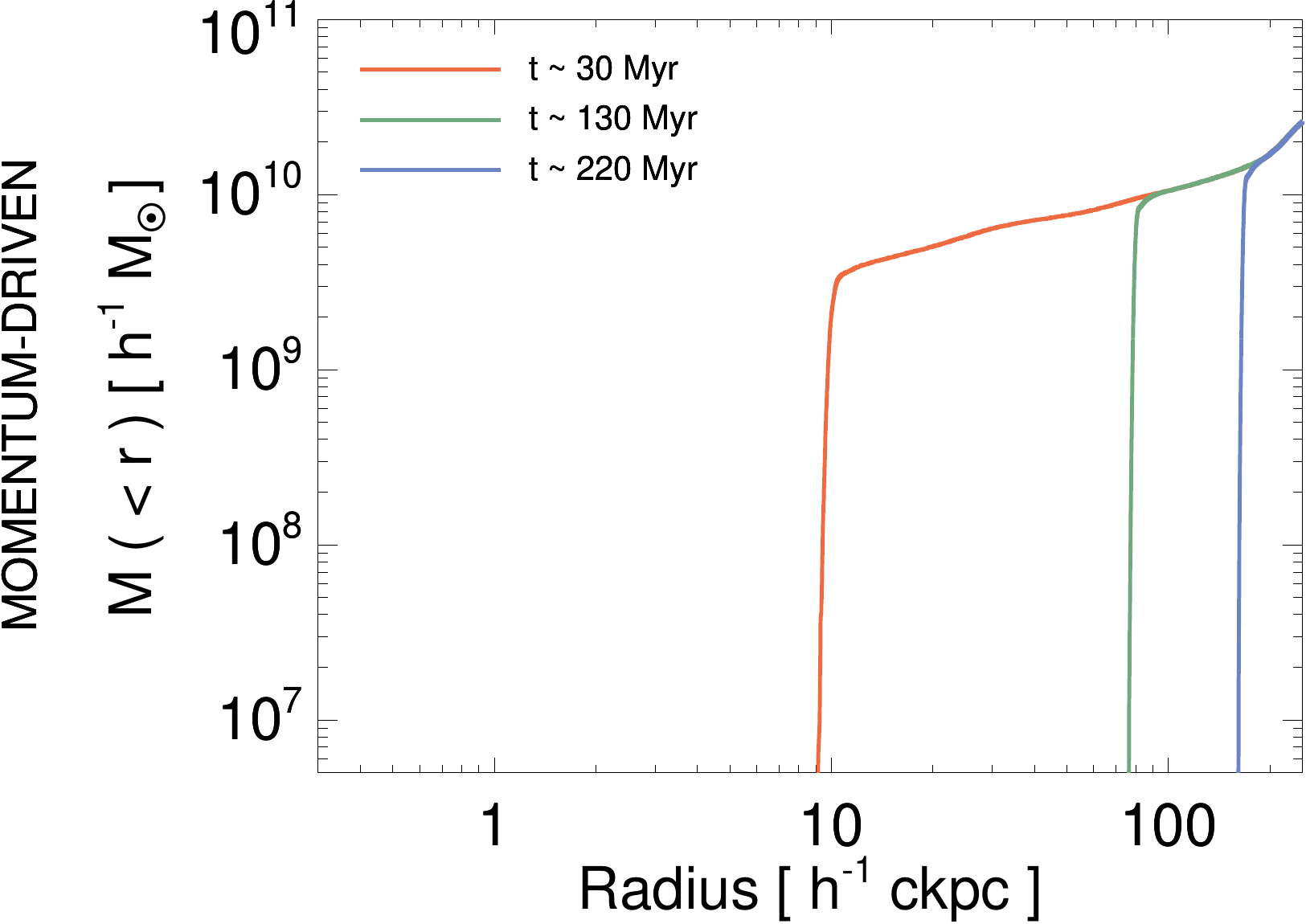}
\hspace{-0.03in}
\includegraphics[scale = 0.5]{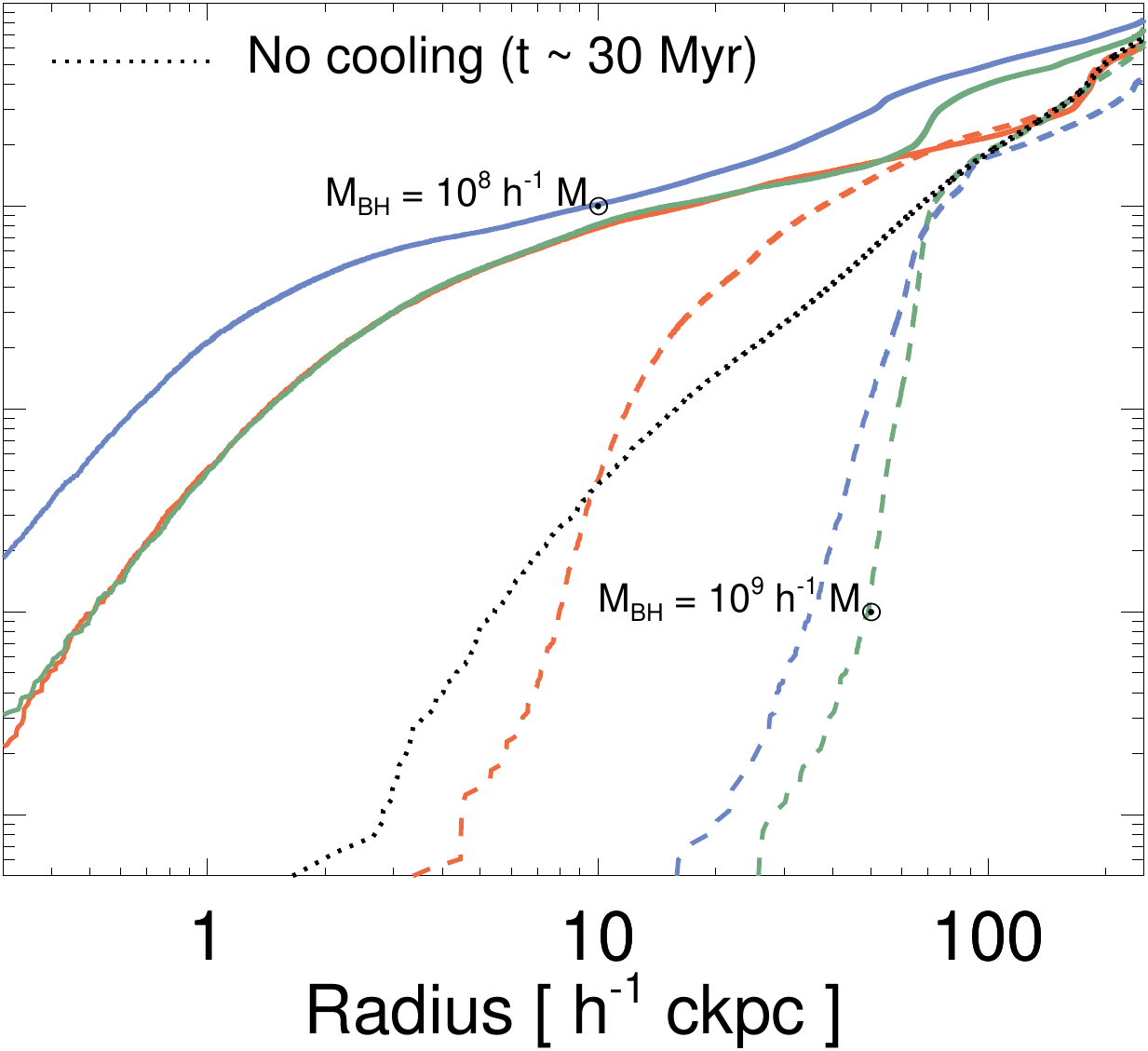}

\caption{Profile of total enclosed mass for energy-driven (top row) and momentum-driven outflows (bottom row). The left column shows results for isolated halo simulations and the right column for cosmological simulations. For energy-driven outflows, the innermost regions of the halo are emptied with similar efficiency in isolated halo and cosmological simulations, except at late times, when the outflow is more efficient in the isolated halo. Momentum-driving for a black hole mass of $10^8 h^{-1} \, \rm M_\odot$ yields very different results for isolated halo and cosmological simulations. For the latter, the total gas mass in the central regions grows due to efficient inflows, whereas a shell is efficiently driven outwards in the isolated halo. A black hole mass and $L_{\rm Edd}/c$ ten times higher yields results similar to energy-driven outflows for $M_{\rm BH} \,=\, 10^8 h^{-1} \, \rm M_\odot$ at matching times. The absence of radiative cooling also enables an outflow to efficiently propagate (dotted line). Note that all quantities are given in comoving units.} 
\label{enclosed}
\end{figure*}

Overplotted on the density fields are radial velocity contours (see figure caption for the contour levels), which trace the geometry and spatial extent of the AGN-driven outflows. 
Yellow contours refer to the outflows driven in the simulation for the isolated halo, while white contours refer to the cosmological simulations. 
For the latter, energy-driven outflows are highly anisotropic and in striking contrast with the outflowing spherical shell driven in the isolated halo simulation.
Initially, the cosmological outflow propagates in directions parallel to the angular momentum vector of the galaxy hosting the luminous AGN because it is confined by the galactic disc \citep[see also][]{Zubovas:11}.
At later times, the central regions of the halo are disrupted and the outflow becomes more spherical, though still more pronounced in regions of low density and less prominent in the direction of the filaments \citep[see also][]{Costa:14}.
Note that whether the outflow driven in cosmological simulations has travelled to higher or lower distances from the black hole than in the isolated halo now depends on direction.
As we shall see, the fact that the outflow driven in cosmological simulations travels more inefficiently than for the isolated halo along certain directions means that outflows driven in cosmological simulations are generally less efficient at clearing gas away from the centre of the halo.

For the equivalent cosmological simulation with momentum injection for $M_{\rm BH} \,=\, 10^8 h^{-1} \, M_\odot$ (second row), there is no discernible outflow.
Note that in the isolated halo case, the same black hole mass however gives rise to a momentum-driven shell that efficiently evacuates the central regions of the halo, indicating that a momentum input $\gg L_{\rm Edd}/c$ is required to drive a large-scale outflow in a cosmological setting \citep[cf.][]{Silk:10}. 
For momentum injection at a rate $L_{\rm Edd}/c$, a larger black hole mass is required, as shown in the bottom row of Fig.~\ref{cosmo_outflow}, where results for the cosmological simulation are shown for times matching those of the energy-driven outflow, but for a black hole mass of $M_{\rm BH} \,=\, 10^9 h^{-1} \, M_\odot$.
Results for the isolated halo are given as in the first row, i.e. for a black hole mass of $10^8 h^{-1} \, \rm M_\odot$ and for the energy-driven outflow model.
Note that the momentum-driven outflow now extends over spatial scales comparable to the energy-driven outflow (at matching times for a black hole mass ten times lower), suggesting that both outflows are now similarly efficient. 
In this case, the momentum-driven outflow is also highly anisotropic and, as in the energy-driven outflow, inefficient in directions of filamentary infall.

Fig.~\ref{mrate_plots} shows the outflow rate as a function of radius for simulations of energy- (left) and momentum-driven (right) outflows for a $10^8 h^{-1} \, \rm M_\odot$ black hole in the cosmological (solid lines) and isolated halo (dashed lines) simulations.
Here, all gas with positive radial velocity is taken into account.
While for the isolated halo, this velocity cut ensures that only gas which flows out due to AGN feedback is being traced, for the cosmological simulations, results must interpreted as upper limits because motion of gas which is not associated with AGN feedback will also be inevitably included.
Consistent with Fig.~\ref{cosmo_outflow}, Fig.~\ref{mrate_plots} shows that in cosmological simulations, the energy-driven outflow is spatially extended and anisotropic and not adequately described by a shell-like geometry.
The outflow rate levels are significantly above those occurring in the absence of an AGN (dotted line), but at any location lower than in a shell-like configuration.
Fig.~\ref{mrate_plots} also shows that the peak outflow rates in cosmological simulations systematically trail behind the outflowing shell in the isolated case.
This inefficiency of energy-driven outflows in cosmological simulations compared to the isolated halo case is moderate, however, and does not prevent the outflow from effectively clearing the centre of the halo.

The right hand side of the panel in Fig.~\ref{mrate_plots} shows the momentum-driven outflow case.
In the isolated halo, a momentum input $L_{\rm Edd}/c$ is clearly sufficient to drive a large-scale outflow, since a shell can be seen to propagate outwards.
Note that the free expansion of a momentum-driven shell is expected in this case since the black hole and its host galaxy lie on the $M_{\rm BH} \-- \sigma$ relation (see Section~\ref{enemomout}).
Results for the cosmological simulations confirm that, as seen in Fig.~\ref{cosmo_outflow}, there is no significant bulk outflow and the gas instead piles up in the central regions of the halo.
The associated increase in outflow rates in the central regions is due to an increase in the speed of random motions of gas in these regions.
On the same plot, we show the outflow rates resulting from a momentum-driven outflow for a black hole mass of $10^9 h^{-1} \, \rm M_\odot$ at $t \sim 40 \, \rm Myr$.
The outflow that forms in this case has reached a distance of $\approx 100 h^{-1} \, \rm ckpc$ from the black hole.
The peak outflow rate is $\gtrsim 100 \, M_\odot \, \rm yr^{-1}$ at $\approx 120 h^{-1} \, \rm ckpc$, similar to results for an energy-driven outflow due to a black hole mass of $M_{\rm BH} \,=\, 10^8 h^{-1}\, \rm M_\odot$ at a similar simulation time of $t \,\sim\, 30 \, \rm Myr$.

Fig.~\ref{enclosed} shows the total enclosed mass as a function of radius for the isolated halo (left column) and cosmological (right column) simulations. 
Results for the energy-driven outflow are given in the first row and for the momentum-driven outflow in the second row.
Consistent with results presented in Fig.~\ref{mrate_plots}, the innermost regions of the AGN host galaxy are emptied with comparable efficiency in cosmological and isolated halo simulations for energy-driven outflows for a black hole mass of $M_{\rm BH} \,=\, 10^8 h^{-1} \, \rm M_\odot$ at early times.
We verified that the star formation rate of the AGN host galaxy accordingly drops to zero.
For the momentum-driven outflows, the total mass of gas in the central regions increases with time, leading to an extremely high star formation rate of $\approx 787 \, \rm M_\odot \, yr^{-1}$ by $t \,\approx\, 130 \, \rm Myr$ in the AGN host galaxy.
For a black hole mass of $M_{\rm BH} \,=\, 10^9 h^{-1} \, \rm M_\odot$ however, the now efficient outflow is in fact comparable to the energy-driven outflow for a black hole mass ten times lower at matching times and the central regions of the galaxy are cleared.

From Fig.~\ref{cosmo_outflow}, we had seen that even when able to propagate out to large distances, outflows are always far more (if not totally) inefficient along directions of filamentary infall.
Thus, a fundamental requirement for efficient AGN feedback is the outflow's ability in preventing the central regions from being replenished by these inflows.
We next verified that in the absence of such (cold) inflows, a momentum input rate of $L_{\rm Edd}/c$ is sufficient to clear the innermost regions of the galaxy.
For this purpose we performed another cosmological simulation with momentum injection from a black hole mass of $10^8 h^{-1} \, \rm M_\odot$, but artificially suppressed the cooling function from the time at which the black hole is seeded.
This procedure guarantees that the density distribution of this and the previous simulation with standard primordial cooling is identical at the time at which momentum injection starts taking place. 
Suppressing cooling equates to reducing the efficiency at which the galaxy is replenished with fresh material.
The dotted black line in Fig.~\ref{enclosed} shows that momentum-driving now efficiently expels gas from the inner regions of the galaxy.

\begin{figure}
\includegraphics[scale = 0.5]{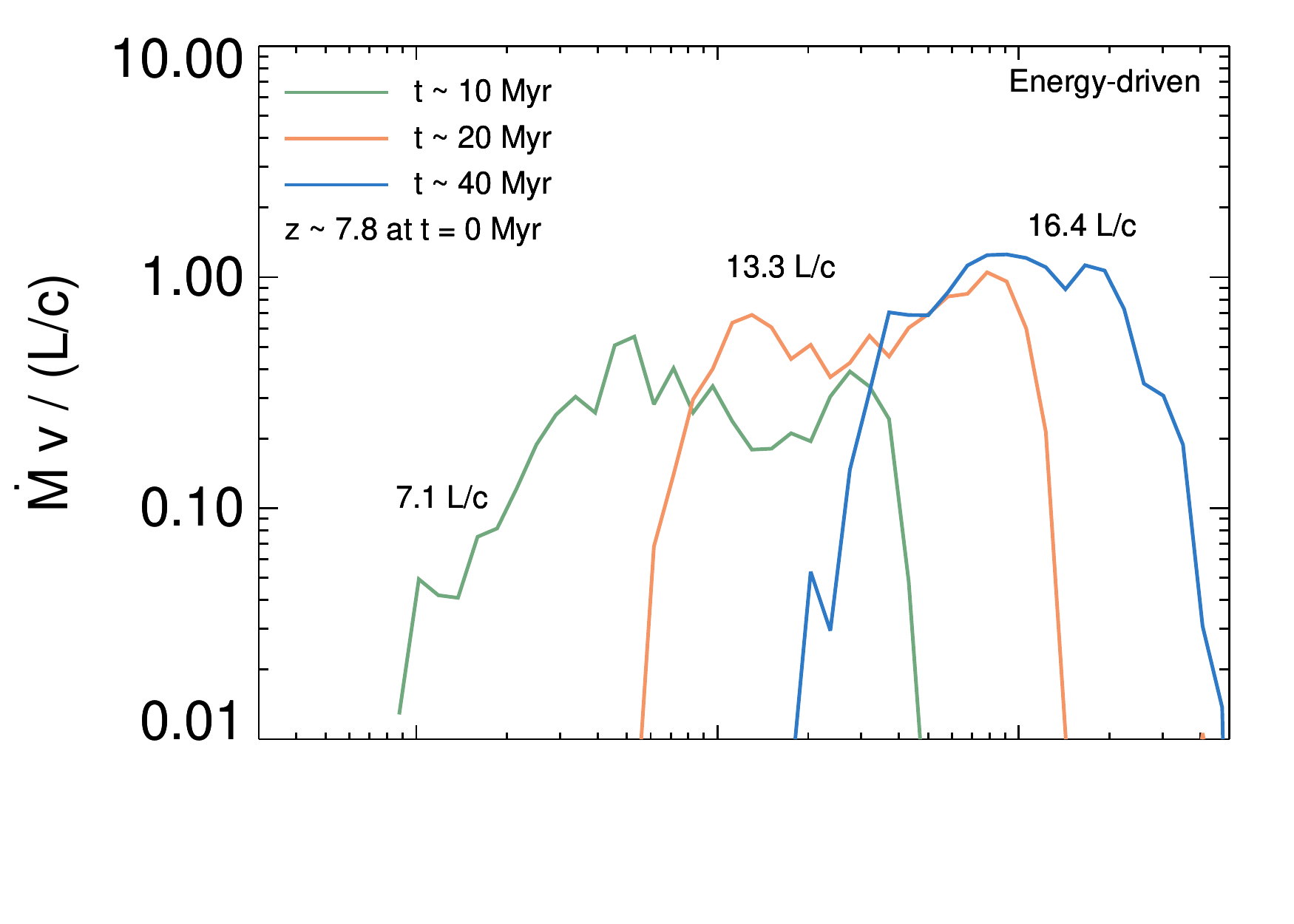}
\includegraphics[scale = 0.5]{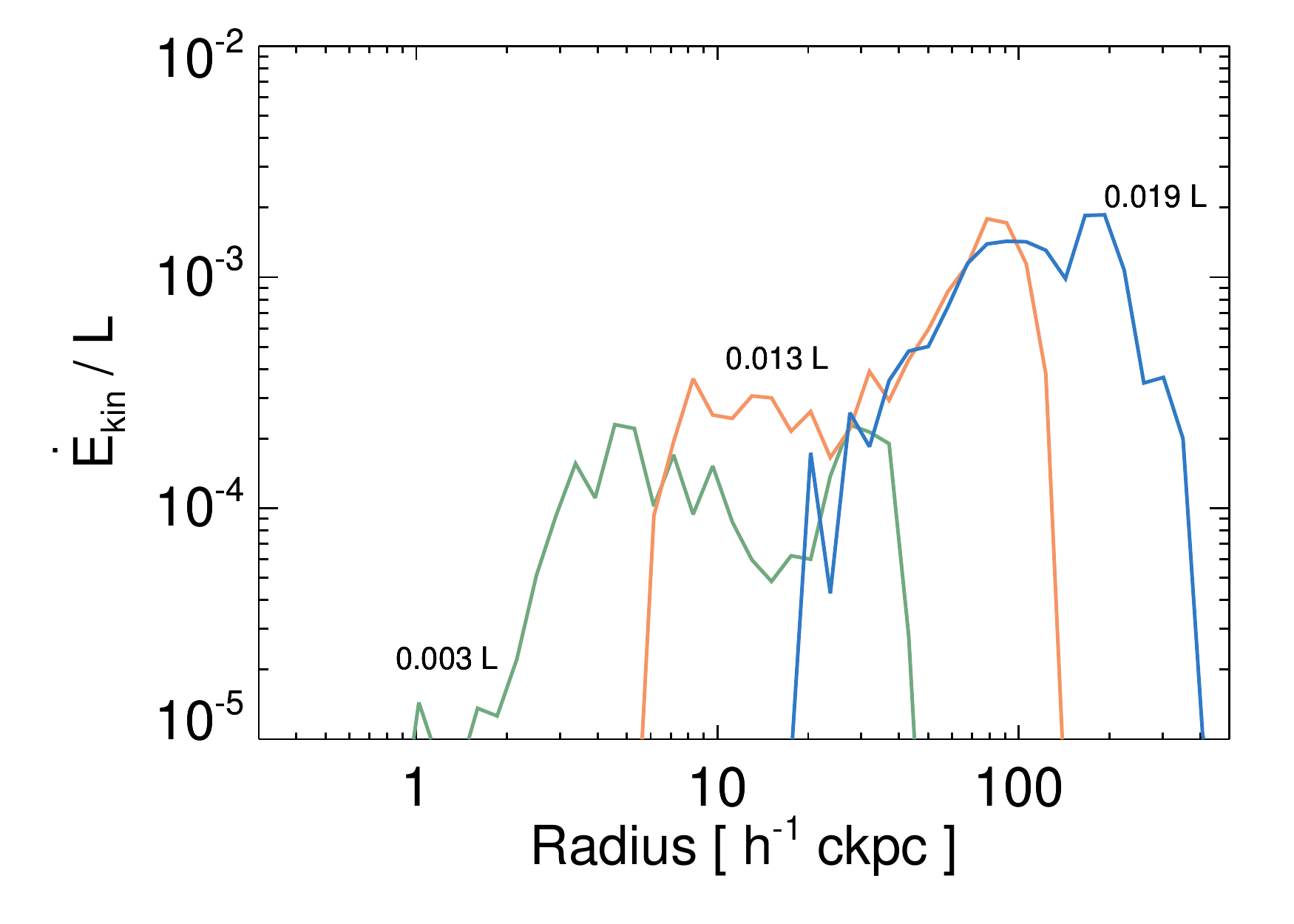}
\caption{Momentum flux and kinetic luminosity per logarithmic unit distance for an energy-driven outflow in the cosmological simulations at different times. Integrated quantities are labelled next to the corresponding curves. Note that, since units are given in comoving coordinates, these results imply that an outflow with momentum flux $\approx 13 L_{\rm Edd}/c$ and kinetic luminosity $\approx 0.01 L$ builds up at scales of $100 h^{-1} \, \rm ckpc \,\lesssim\, 20 \, \rm kpc$. The magnitude of the momentum fluxes, kinetic luminosities and spatial scales agree well with observed AGN-driven outflows and indicate that the energy-driven model is a viable mechanism for AGN feedback in cosmological simulations.}
\label{mdotv}
\end{figure}

The results presented in this section indicate that if the AGN energy/momentum couples to the ISM of the host galaxy hydrodynamically via an inner wind, an efficient large-scale outflow must be energy-driven already at scales of $\approx 70 h^{-1} \, \rm cpc$ (minimum cell size).
In the cosmological simulations, a momentum flux of $5 \-- 10 L_{\rm Edd}/c$ is required to efficiently disrupt the central disc and revert the filamentary inflows from which it is fed, while $L_{\rm Edd}/c$ is sufficient to drive an unbound shell for the spherically averaged isolated halo.
In Fig.~\ref{mdotv}, we verify that the momentum fluxes and kinetic luminosities of energy-driven outflows resulting from a $10^8 h^{-1} \, \rm M_\odot$ black hole in our cosmological simulations are comparable to observed values.
Both quantities can be seen to rise with time as thermal energy is converted into kinetic energy of the outflow.
The net momentum fluxes and kinetic luminosities, estimated by integrating over the curves shown in Fig.~\ref{mdotv} are of the order of $\sim 7 \-- 16 L_{\rm Edd}/c$ and $0.003 \-- 0.02 L_{\rm Edd}$, respectively, at scales $\lesssim 100 h^{-1} \, \rm ckpc \,\approx\, 17 \, \rm kpc$ in good agreement with observed outflows \citep[e.g.][]{Maiolino:12, Cicone:14}.
 
\subsection{Implications for the origin of the $M_{\rm BH} \-- \sigma$ relation}
We have seen that in a realistic cosmological context, a much larger amount of momentum input is required to suppress the infall of (cold) gas than is available  in the momentum driven model suggested by \citet{King:03} based on spherical static haloes. 
This is not surprising as not only is the gas not static, but a significant fraction of the  gas falls in along well defined filamentary structures with a small covering factor which does not offer a large working surface for the outward streaming AGN-driven outflow. 
A significantly larger momentum input than $L_{\rm Edd}/c$ is therefore required to even temporarily interrupt the inflow. 
Most of the injected momentum and energy escapes to large radii with little effect on the majority of the inflowing gas.
Note that similar conclusions are drawn by \citet{Nayakshin:14, Zubovas:14b} and \citet{Bourne:14} for a clumpy gaseous halo.  
Our cosmological simulations thereby suggest that the self-regulation of the black holes is not due to a complete ejection of the gas from the surrounding galactic haloes, but due to a temporary suppression/reduction of the inflow rate of cold gas.  
It is thereby not obvious at what radius the feedback loop limiting the black hole mass is closed. 
If cooling is efficient, the momentum inflow rate of infalling gas and the 
energy input rate required to eject the inflowing matter scale with $\sigma^4$
and $\sigma^5$, respectively. 
The requirement of stopping the inflow and partially ejecting the gas should thus lead to a self-regulated $M_{\rm BH} \-- \sigma$ relation with a slope somewhere in between these two cases, as is indeed observed. 
As discussed in considerable detail by \citet{Kormendy:13}, such a self-regulated $M_{\rm BH} \-- \sigma$ relation is then almost certainly modulated by the late-time merging of then gas-poor galaxies, especially at the high-mass end. 
Note also, that $\sigma$ in real galaxies is only weakly dependent on distance to the centre.
The scale at which the feedback loop is closed should therefore have only a modest effect on an $M_{\rm BH} \-- \sigma$ relation resulting from such self-regulation. 
This may explain why a wide range of feedback implementations in simulations of widely varying resolution appear to be able to reproduce at least approximately the slope of the observed $M_{\rm BH} \-- \sigma$ relation.
The main conclusion from our study regarding the $M_{\rm BH} \-- \sigma$ relation is, however, that the observed AGN-driven winds must become energy-driven at sufficiently small galactic scales such that the outflow driven by the inner AGN 
wind is still sufficiently fast to convert about $5\%$ of the initially liberated energy into kinetic energy of the outflow. 
As we have explicitly shown there is then energetically no problem to realise 
momentum fluxes of $5-30 L_{\rm Edd}/c$ in the outflow as observed.
Note that the picture would be the same if a similar amount of energy was injected by a different means than via thermalisation of a fast inner AGN wind. A possibility to be explored further here is Compton heating in phases in which the AGN spectral energy distribution is particularly hard.
Implications of our findings on future AGN feedback models are discussed in Appendix~\ref{appendixb}.

\section{Main conclusions and summary}
\label{secconc}

\begin{itemize}

\item
We have confirmed that simplified analytical models of energy- and momentum-driven AGN outflows in a spherically symmetric isolated halo without cooling can be accurately reproduced in numerical simulations for similar assumptions.
Our simulations, based on energy and momentum injection into the black hole's nearest neighbour cells, give rise to expanding shells, with speeds and structure in close agreement with analytical solutions.

\item
The velocity of the propagating shells in the hydrodynamical simulations are found to fall below those of the analytical solution when the shells move (sub)sonically through the halo's gaseous medium. 
For a Hernquist halo with total mass $10^{12} \, \rm M_\odot$, this occurs for momentum-driven outflows for black holes with mass $\lesssim 10^8 \, \rm M_\odot$ at a distance of a few $\rm kpc$ from the centre of the halo.
This behaviour does however not alter the key conclusion of \citet{King:03} that there is a critical mass below which the momentum input from the AGN cannot fully expel material from the centre of the halo.
The necessary black hole mass to launch an unbound shell in the simulated halo was found to be that which is required to place the black hole and its host halo on the observed $M_{\rm BH} \,\--\, \sigma$ relation.

\item
We have performed simulations of energy-driven outflows including radiative cooling in order to investigate the formation of cold material entrained in hot shocked gas in the outflowing shell.
If the cooling time is comparable with the outflow time of the shell, large amounts of gas cool out of the expanding shell of shocked material.
The radius at which this occurs depends sensitively on the outflow rate as well as the cooling properties of the gas.
By artificially varying the AGN light curve (previously assumed to always correspond to the Eddington luminosity of the AGN) and the efficiency of cooling, we found that the amount of cold material can vary from no material able to cool, e.g. if the AGN emits at its Eddington limit but cooling is inefficient, to $\sim 10^9 \, \rm M_\odot$ if either the AGN luminosity is sub-Eddington or if cooling is very efficient.
In all cases, cold material entrained in the (energy-driven) outflows attains a specific kinetic energy close to the available specific thermal energy. 

\item
In cases in which only small quantities of cold gas form, the total momentum flux in cold gas is only a factor of a few times higher than the momentum flux of the AGN radiation field. 
In order to generate cold outflows with very high momentum fluxes $\sim 20 \, L/c$ and kinetic luminosities $\sim 0.05 L$ as observed, efficient cooling and high AGN luminosities are simultaneously required. 
Only in a simulation including a cooling rate ten times higher than that for primordial cooling and an Eddington limited AGN, do we find high momentum-boost factors $\approx 26 L/c$ and high kinetic luminosities $\approx 0.03 L$.
This  suggests that efficient cooling, e.g. via metal-lines, is likely to play an important role in generating sufficient amounts of cold material in the otherwise hot outflows.

\item
When included into cosmological simulations that account for the environment of rapidly growing supermassive black holes  at $z \,\gtrsim\, 6$, the adopted `subgrid' prescriptions for AGN (energy and momentum) feedback give rise to highly anisotropic large-scale outflows.
The simulated outflows propagate most efficiently along paths of least resistance. At small scales, the outflows preferentially move along directions perpendicular to the host galactic disc and at large-scales, they avoid regions of filamentary gas infall.

\item
We have compared the efficiency of the AGN-driven outflows in cosmological simulations to our idealised spherically symmetric simulations by adopting the spherically averaged gravitational potential of the galactic halo in which the black hole is seeded in our cosmological runs.
For matching black hole mass, energy-driven outflows remove gas from the innermost regions of the AGN host halo with comparable efficiency in cosmological and isolated halo simulations. At large radii (i.e. late times) however, energy-driven outflows are unable to disrupt the dense filaments that continuously transport infalling cold gas into the main halo.

\item
The efficiency of momentum-driven outflows in cosmological simulations is drastically lower than in isolated halo simulations. 
In cosmological simulations, a momentum input of $L_{\rm Edd}/c$ is unable to revert the inflow of fresh material into the centre of the galaxy, where the total gas mass instead increases with time, while the same $L_{\rm Edd}/c$ input is sufficient to expell a momentum-driven shell for the isolated halo simulation.
A momentum input of at least  $5 \-- 10 L_{\rm Edd}/c$ is required to prevent the central (resolved) regions of the AGN host galaxy to be replenished by inflows. 
For this to be energetically feasible, the outflows have to become energy-driven at sufficiently small galactic scales that the outflow driven by the shocked inner AGN wind rapidly converts about $5\%$ of the energy released by the AGN into kinetic energy of the outflow. 
Models in which AGN energy and momentum couple to the ISM radiatively would require very high and probably unrealistic effective infrared optical depths of $\tau \approx 10$ at scales $\ga 10 \, \rm kpc$ in order to produce efficient feedback.
\end{itemize}

By comparing idealised numerical simulations of galactic outflows expanding in spherical gravitational potentials, carefully calibrated against analytical solutions, we have shown that in realistic cosmological environments, about a factor ten larger momentum flow is required for AGN feedback to efficiently  affect cooling-driven inflows and to self-regulate black hole growth.  
We have further demonstrated that such large momentum flow rates occur naturally if about 5\% of the expected accretion luminosity of AGN is thermalised in the galaxy host on scales of $\sim 0.1 \-- 1 \, \rm kpc$). 
The resulting energy-driven outflows contain substantial amounts of entrained cold gas which is cooling from the mostly very hot outflow in encouraging  agreement with observed molecular outflows. 
The most plausible mechanism for the input of thermal energy at these distances from the central engine is the thermalisation of the kinetic energy of the ultra-fast nuclear outflows observed in many AGN.
However, direct Compton heating during accretion phases resulting in exceptionally hard SEDs with Compton temperature $\ge 10^{7 \-- 8} \, \rm K$ also merits further study. 
With the dynamic range and resolution of cosmological simulations of AGN-driven outflows further improving, there is the exciting prospect of modelling the transition from ultra-fast nuclear to energy-driven galactic outflows and the effect of Compton cooling as well as Compton heating self-consistently in such simulations in the not too distant future. 

\section{Acknowledgments}

We thank the anonymous referee for thoroughly reading the manuscript and for useful suggestions on how to improve it.
We also thank Volker Springel, Cathie Clarke, Andrew King, Sergei Nayakshin, Andy Fabian, Martin Rees, Claudia Cicone and Roberto Maiolino for insightful comments and helpful discussions.
TC thanks Matt Young, Mike Curtis and Laura Keating for useful discussions and suggestions and Amanda Smith for assistance with designing Fig.~\ref{outflowstructure}.
All simulations presented in this study were performed using the Darwin High Performance Computer facilities based at the University of Cambridge, UK and at the DiRAC Complexity Cluster based at the University of Leicester, UK.
TC is supported by an STFC studentship.
MGH and TC acknowledge support from the FP7 ERC Advanced Grant Emergence-320596.

\bibliographystyle{mn2e} 
\bibliography{references}

\appendix

\section{Convergence properties of outflow models}
\label{appendixa}

All results presented in this paper rely on the usage of two `subgrid' models, whose goal is to realise energy- and momentum-driven AGN outflows at scales that can be accurately resolved in our simulations.
In this Appendix, we summarise results from several numerical tests performed on our `subgrid' models in order to test their robustness against changes in various numerical parameters.

\begin{figure*}
\includegraphics[scale = 0.5]{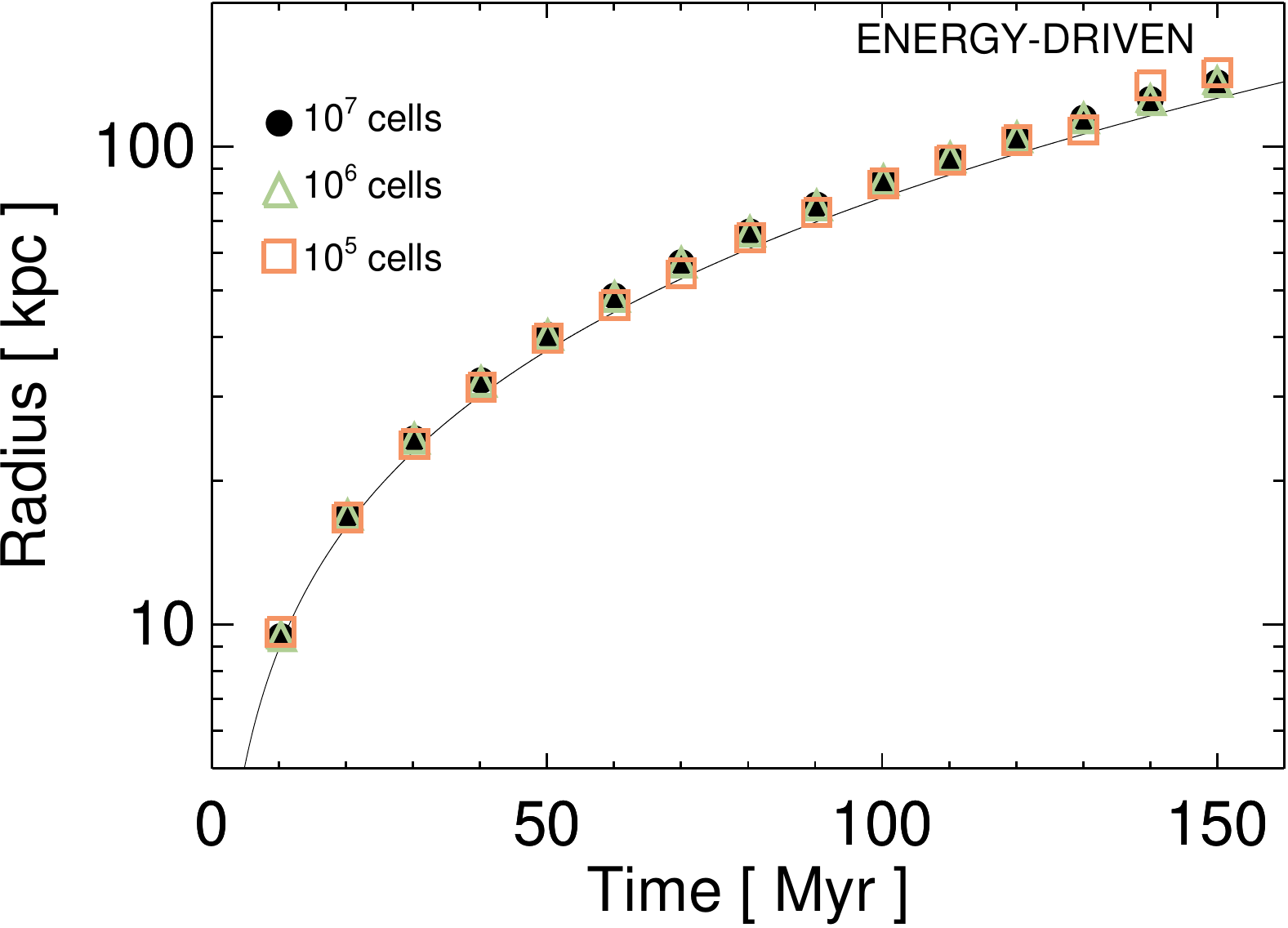}
\includegraphics[scale = 0.5]{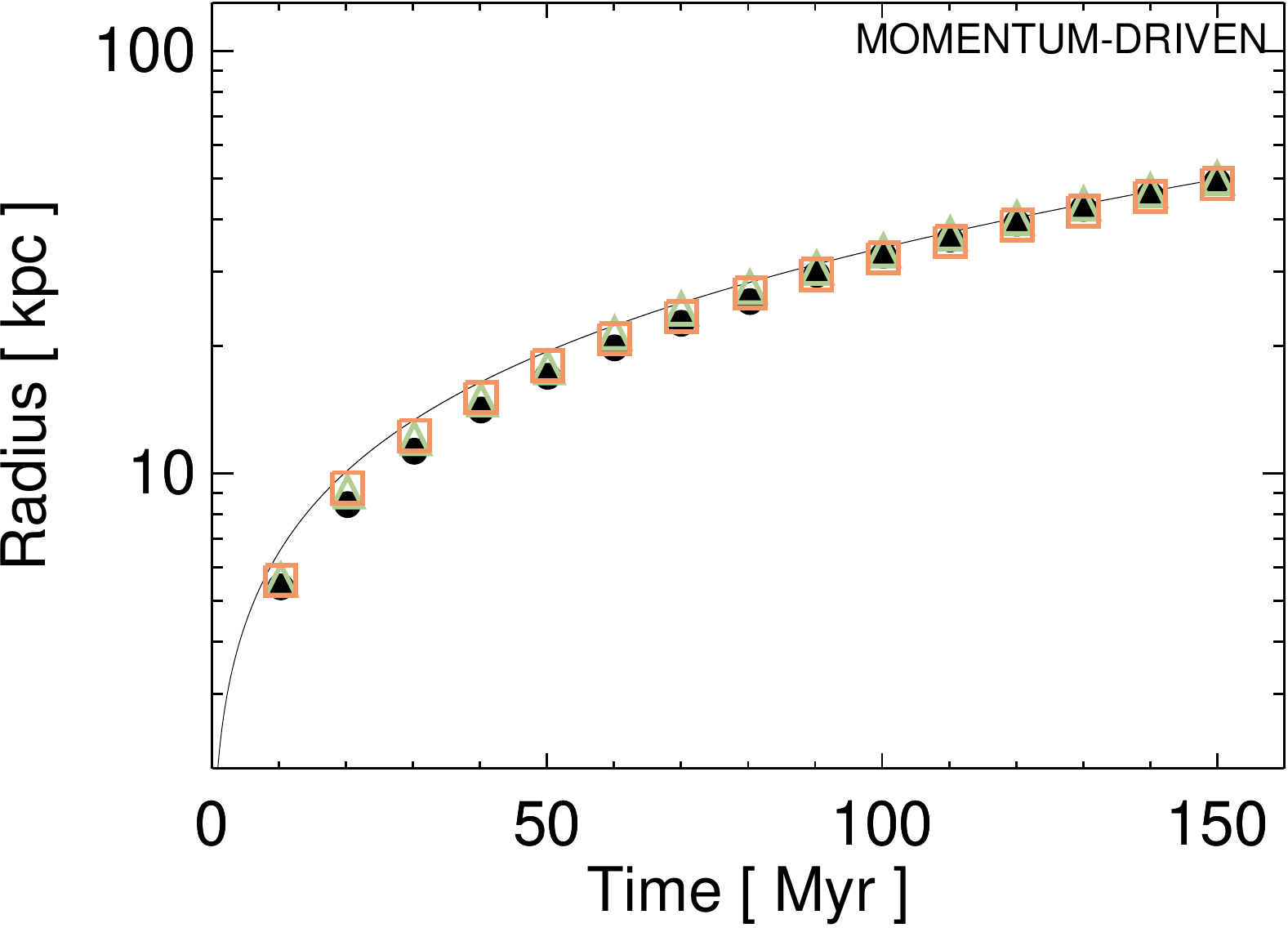}
\caption{Effect of varying the numerical resolution of simulations for energy- (left) and momentum-driven (right) outflows in a static Hernquist potential with total mass $10^{12} \, \rm M_\odot$ and black hole mass $M_{\rm BH} \,=\, 10^8 \, \rm M_\odot$. Our adopted `subgrid' models yield numerically converged solutions even for $10^6$ resolution elements. We conclude that both energy- and momentum-driven outflows are well converged in this study.}
\label{numconv}
\end{figure*}

\begin{figure*}
\includegraphics[scale = 0.5]{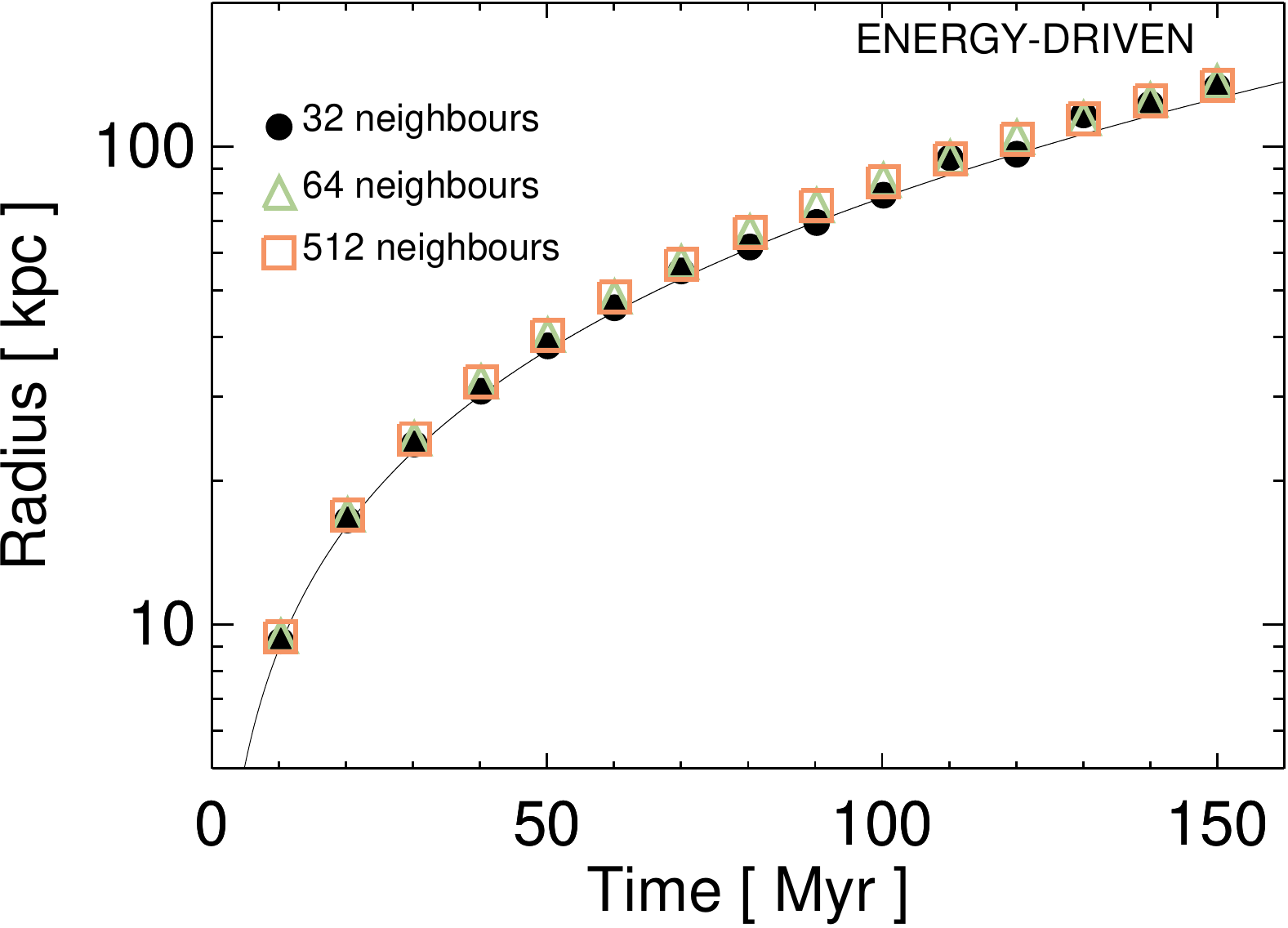}
\includegraphics[scale = 0.5]{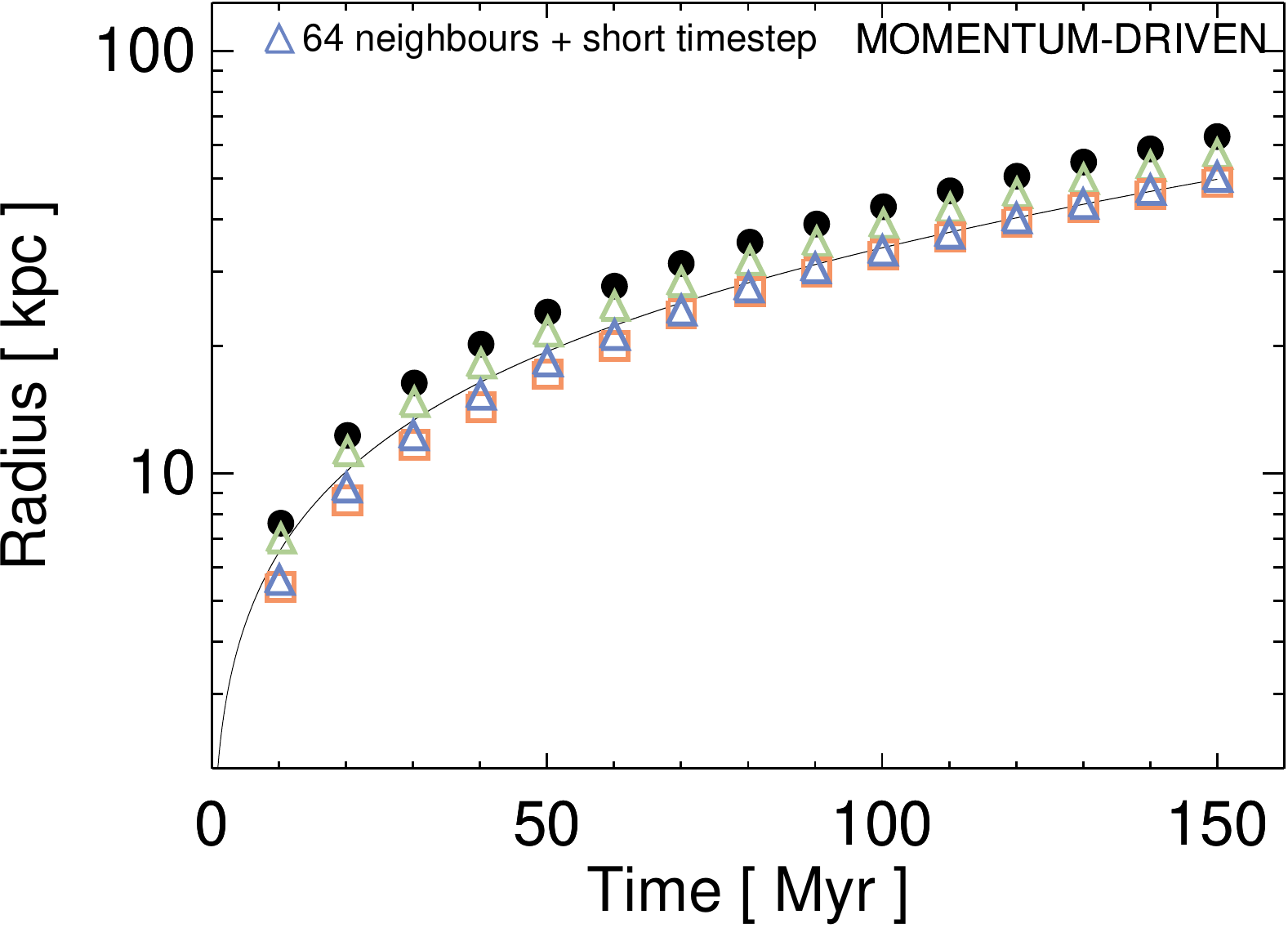}
\caption{Effect of varying the number of cell neighbours into which the black hole deposits energy (left) and momentum (right). Here, we use $10^6$ resolution elements. The numerical solution resulting from our energy-driven model is robust against changes in the number of neighbour cells to the black hole and in each case agrees with the analytical solution. Our momentum-driven model is more sensitive to the number of cell neighbours and diverges from the analytical solution for lower number of neighbours. In this regime, gas cells are kicked with very high speeds and a higher time-stepping precision is required in order to bring such numerical solutions to agreement with the analytical predictions, as shown with blue triangles for a simulation with $64$ cell neighbours but a maximum time-step five times lower.}
\label{ngbconv}
\end{figure*} 

In Fig.~\ref{numconv}, we examine the numerical convergence properties of our simulations. 
For this purpose, we performed three simulations for energy- and momentum-driven outflows for a static Hernquist potential with total mass $10^{12} M_\odot$ (i.e. the same halo as discussed in the main body of this study) and a black hole mass of $10^8 \, \rm M_\odot$. 
In this set of simulations, we varied the number of resolution elements with which the gaseous halo is sampled, keeping all other paramaters fixed.
We investigated simulations with $10^5$, $10^6$ and $10^7$ gas resolution elements with respective initial gravitational softenings of $1$, $0.5$ and $0.2 \, \rm kpc$.
For the energy- and momentum-driven outflows, numerical solutions presented in Fig.~\ref{numconv} are very well converged to the analytical solutions.
Energy- and momentum-driven outflows presented in this study should therefore be expected to be robust against changes in numerical resolution.

We also varied the number of gas cell neighbours into which energy or momentum is injected by the black hole in another set of simulations.
Note that we otherwise adopted the same halo (with $10^6$ resolution elements) for all simulations and that, as before, all other parameters were kept fixed in these experiments.
The plot on the left of the panel in Fig.~\ref{ngbconv} indicates that the numerical solution is robust to changes in the selected number of neighbours in the energy-driven model.
For simulations using $32$, $64$ and $512$, all curves fall on the analytical solution shown as a solid line.
Varying the number of gas cell neighbours in this case equates to varying the characteristic mass of gas into which thermal energy is deposited.
Note that such good convergence is expected to break down if cooling is very efficient, in which case providing the same amount of thermal energy to a smaller mass is likely to lead to more efficient feedback than if this energy is distributed over a larger mass and hence more efficiently radiated away.
On the right hand side of the panel, we investigated varying the number of cell neighbours for the momentum-driven model.
Note that here we use simulations for a black hole mass of $3 \times 10^8 \, \rm M_\odot$, for which the propagating shell is always in the supersonic regime and a good match to analytical solutions can be obtained.
Varying the number of neighbour cells in the momentum-driven model means varying the mass into which the same momentum is imparted.
As a result, gas cells are kicked with different speeds depending on the selected mass of black hole neighbours.
Fig.~\ref{ngbconv} shows that the numerical solution converges to the analytical solution for a high number of neighbours.
This agreement, however, becomes increasingly poor as the number of neighbours is decreased.
The latter case corresponds to increasing the magnitude of the typical velocity kicks imparted on the gas cells.
Increasing the time-stepping accuracy, however, brings the numerical solution back in agreement with the analytical solution as shown with blue triangles for the case of $64$ cell neighbours and maximum timestep five times lower.

\section{Energy versus Momentum injection in cosmological simulations}
\label{appendixb}

Assuming spherically symmetric outflows, in the absence of a plausible physical model for the injection of the required amount of momentum at galactic scales, the injection of energy should be the method of choice in cosmological simulations. On resolved galactic scales, hydrodynamical codes are able to handle all relevant physics properly.
Scaling the energy injected to the energy released by accretion and controlling it by a single parameter characterising the radiative losses on unresolved scales should therefore be the most consistent way of treating AGN-driven outflows.
Emphasis should then be given to a correct treatment of the relevant cooling processes which will in particular require modelling the metal enrichment of the ISM and CGM.
As discussed in Section~\ref{secnumimp}, injection of momentum on the other hand will correspond to amounts of kinetic energy which depend strongly not only on resolution, but also on the details of injection implementation.
Specifically, collimated bipolar outflows containing the same momentum input of $L/c$ may be much more effective at driving large-scale outflows.
Understanding how filamentary inflow will be affected in this case warrants detailed cosmological simulations in future work.
The choice of the preferred injection method may change towards momentum injection once the dynamic range of simulations has sufficiently increased to be able to resolve the fast inner AGN winds for which radiation driving with a constant momentum flux of $L/c$ and constant mass flow rate/mass loading is a plausible physical model.
Once this is possible, the goal should be to get the inner wind started with momentum injection and to simulate the change in cooling properties of the outflow with radius which triggers the transition from a likely radiatively/momentum-driven inner wind to a thermal/energy-driven galactic wind self-consistently.

Since we have placed emphasis on the efficiency of injecting AGN energy/momentum in driving large-scale outflows, note that we have simplified the physics of the black hole accretion that powers these outflows. Clearly, in the full picture, the input of energy/momentum by an AGN should be scaled to the instantaneous black hole accretion rate, which should ideally be computed self-consistently. In Section~$4.2$ we have explicitly verified that varying the AGN light curve sensitively regulates the energetics of the outflow and shapes its geometry as it may allow time for denser regions to cool and hence confine the outflow along paths of least resistance. These findings further highlight the need to follow black hole accretion self-consistently in cosmological simulations.

It is also worth emphasising that additional physics, which we have neglected in this study, such as super-Eddington accretion, magnetic fields and non-ideal plasma effects (viscosity and thermal conduction) may play an important role, which opens interesting additional avenues for future cosmological simulations.

\end{document}